\documentclass{aastex}
\usepackage{emulateapj5}



\newcommand{\lsim}{\mbox{\hspace{.2em}\raisebox{.5ex}{$<$}\hspace{-.8em}\raisebox{-.5ex}{$\sim$}
\hspace{.2em}}}
\newcommand{\gsim}{\mbox{\hspace{.2em}\raisebox{.5ex}{$>$}\hspace{-.8em}\raisebox{-.5ex}{$\sim$}\hspace{.2em}}}

\def\asca       {{\em ASCA}\/}
\def\suzaku       {{\em Suzaku}\/}
\def\chandra    {{\em Chandra}\/}
\def\xmm        {XMM-{\em Newton}\/}
\def\rosat      {{\em ROSAT}\/}
\def\sax        {{\em BeppoSAX}\/}

\def\iras        {{\em IRAS}\/}
\def\wmap        {{\em WMAP}\/}

\begin{document}

\title{\chandra\ studies of the X-ray gas properties of galaxy groups}

\author{
M.\ Sun\altaffilmark{1},
G.\ M.\ Voit\altaffilmark{1},
M.\ Donahue\altaffilmark{1},
C.\ Jones\altaffilmark{2},
W.\ Forman\altaffilmark{2},
A.\ Vikhlinin\altaffilmark{2}
}

\altaffiltext{1}{Department of Physics and Astronomy, MSU, East Lansing, MI 48824; msun@virginia.edu}
\altaffiltext{2}{Harvard-Smithsonian Center for Astrophysics,
60 Garden St., Cambridge, MA 02138}

\shorttitle{Properties of galaxy groups}
\shortauthors{SUN ET AL.}

\begin{abstract}

We present a systematic analysis of 43 nearby galaxy groups ($kT_{500} = 0.7 - 2.7$
keV or $M_{500} = 10^{13} - 10^{14} h^{-1}$ M$_{\odot}$, 0.012 $<z<$ 0.12), based
on \chandra\ archival data. With robust background subtraction and modeling,
we trace gas properties to at least $r_{2500}$ for all 43 groups. For 11 groups,
gas properties can be robustly derived to $r_{500}$. For an additional 12 groups,
we derive gas properties to at least $r_{1000}$ and estimate properties at $r_{500}$
from extrapolation. We show that in spite of the large variation in temperature
profiles inside 0.15 $r_{500}$, the temperature profiles of these groups are similar
at $>$ 0.15 $r_{500}$ and are consistent with a ``universal temperature profile.''
We present the $K - T$ relations at six characteristic radii (30 kpc, 0.15 $r_{500}$,
$r_{2500}$, $r_{1500}$, $r_{1000}$ and $r_{500}$), for 43 groups from this work and
14 clusters from the Vikhlinin et al. (2008) sample. Despite large scatter in the
entropy values at 30 kpc and 0.15 $r_{500}$, the intrinsic scatter at $r_{2500}$
is much smaller and remains the same ($\sim$ 10\%) to $r_{500}$. The entropy excess
at $r_{500}$ is confirmed, in both groups and clusters, but the magnitude is smaller
than previous \rosat\ and \asca\ results. We also present scaling relations for the
gas fraction. It appears that the average gas fraction between $r_{2500}$ and $r_{500}$
has no temperature dependence, $\sim$ 0.12 for 1 - 10 keV systems. The group gas
fractions within $r_{2500}$ are generally low and have large scatter.
This work shows that the difference of groups from hotter clusters stems from the
difficulty of compressing group gas inside of $r_{2500}$. The large
scatter of the group gas fraction within $r_{2500}$ causes large scatter in the group
entropy around the center and may be responsible for the large scatter of the
group luminosities. Nevertheless, the groups appear more regular and more like clusters
beyond $r_{2500}$, from the results on gas fraction and entropy.
Therefore, mass proxies can be extended into low mass systems. The $M_{500} - T_{500}$
and $M_{500} - Y_{\rm X, 500}$ relations derived in this work are indeed well behaved
down to at least 2 $\times 10^{13}$ h$^{-1}$ M$_{\odot}$.

\end{abstract}


\section{Introduction}

Most baryons in clusters reside in the hot ICM,
and most clusters are low-mass groups and poor clusters because the mass function
of virialized systems is steep.
Studies of galaxy groups are thus especially important for understanding the
gravitational and thermal evolution of the bulk of matter in the Universe.
In contrast to hot clusters, galaxy groups are systems where baryon physics
(e.g., cooling, galactic winds and AGN feedback) begins to dominate
over gravity. Groups are not just scaled-down version of massive clusters.
Cluster scaling relations (e.g., $L - T$ and entropy - $T$) show
deviations from the self-similar relations at the low-mass end (reviewed
in Voit 2005), reflecting the importance of baryon physics, which
is essential to shape the properties of the cluster galaxies
and ICM. Cooling sets a characteristic entropy threshold
in the ICM (Voit \& Ponman 2003) and is required to understand the cluster
scaling relations (Voit 2005). On the other hand, cooling has to be balanced
by feedback to preserve the observed cluster stellar fraction and the galaxy luminosity
function. The most important feedback may be the AGN outflows driven by the
central SMBH. AGN outflows can simultaneously explain the antihierarchical
quenching of star formation in massive galaxies, the exponential cut-off
at the bright end of the galaxy luminosity function, and the quenching of
cooling-flows in cluster cores (e.g., Scannapieco et al. 2005; Croton et al. 2006).
Along with SN winds, they act to suppress cooling and structure formation.
They pump thermal energy into the surrounding ICM as the host galaxy and the
central SMBH formed
and evolve. These imprints are recorded in the ICM and are reflected in
the cluster scaling relations.

There are two important ICM properties that directly reflect the role of baryon physics.
The first is ICM entropy.
With \rosat\ and \asca\ data, Ponman et al. (2003) showed that at 0.1 $r_{200}$,
the ICM entropy ($K$) obeys a simple relation with the cluster temperature ($T$),
$K \propto T^{\sim 0.65}$, which is different from the self-similar
relation ($K \propto T$). Ponman et al. (2003) and Voit \& Ponman (2003) demonstrated
that this $K - T$ relation at 0.1 $r_{200}$ can be understood with a simple model
involving a fixed cooling threshold, which may be related to galaxy formation.
However, the details of cooling and feedback that determine this threshold and its scatter
are still unclear. Finoguenov et al. (2002) and Ponman et al. (2003) showed that groups and 
clusters have significant excess entropy at $r_{500}$. Voit \& Ponman (2003) suggested
that smooth accretion (via e.g., galactic winds in subhalos) can boost entropy
higher than lumpy hierarchical accretion to produce the observed excess. However,
Borgani et al. (2005) show that galactic winds alone are not able to boost
entropy very much at radii beyond $r_{500}$. The AGN-like feedback algorithm in
Kay (2004) has a more substantial effect on entropy at large radii, which may
imply that radio outflows have an important impact on group properties.
Thus, it is important to robustly constrain the entropy around $r_{500}$ with
observations and to connect the dispersion with other group properties. The
Ponman et al. (2003) results rely on extrapolations based on assumed density
($\beta$-model) and temperature (polytropic-model) profiles, which may not be
adequate and may bias the derived ICM properties (e.g., Borgani et al. 2004; Vikhlinin
et al. 2006, V06 hereafter). These early results can be examined with \chandra\
and \xmm. There has been some work on the ICM entropy of groups with the \chandra\
or \xmm\ data (e.g., Mushotzky et al. 2003; Sun et al. 2003; Pratt et al. 2005,
2006; Mahdavi et al. 2005; Finoguenov et al. 2007). Our much larger sample allows
us to study the scatter in group properties. As the ICM properties are traced to at least
$r_{2500}$ for each group, the $K - T$ relation at $\gsim r_{2500}$ can be better
constrained from our studies.

Another important ICM property is the gas fraction. Groups have low gas
fractions within $r_{2500}$ (0.03-0.07, V06; Gastaldello et al. 2007, G07 hereafter). 
However, few groups have gas fraction measured to $r_{500}$. V06 and G07
derived gas fraction within $r_{500}$ for four low-temperature systems
($kT <$ 2.7 keV). The results, 0.06 - 0.15, span a large range. Gas fraction
in simulated groups and clusters is directly related to the strength of cooling and
star formation (e.g., Kravtsov et al. 2005), and a small gas fraction in groups
may imply efficient cooling and star formation. The enclosed gas fraction can
also be modified by the AGN feedback (e.g., Puchwein et al. 2008), thus
bearing the imprint of the feedback history. It is therefore of great interest to know how
gas fractions and total baryon fractions (gas + stars) of groups at $r_{500}$
compare with those of clusters.
The $f_{gas} - T$ relation is also important for cosmology, e.g., determining
cosmological parameters, and using the Sunyaev-Zel'dovich flux as a proxy for
the cluster mass (e.g., Vikhlinin et al. 2008, V08 hereafter).
Besides the science related to entropy and gas fraction, better determination of
the group gas properties are also essential to constraining the low-mass ends of
other important scaling relations (e.g., $M - T, M - Y_{\rm X}$ and $c - M$).
$M - T$ and $M - Y_{\rm X}$ relations are essential for using clusters to study
cosmology (e.g., Kravtsov et al. 2006; V08),
while the $c - M$ relation is important for understanding the formation and evolution
of the dark matter halos (e.g., Buote et al. 2007).

We started a systematic analysis of galaxy groups in the \chandra\ archive to better
constrain the ICM properties in groups and to better understand the difference
between groups and clusters. In this paper, we present the results on 43 galaxy
groups that appear relaxed beyond the central core. Our sample is larger
than in the previous work on 3 - 16 groups with \chandra\  and \xmm\ data
(Mahdavi et al. 2005; Pratt et al. 2005; Finoguenov et al. 2007; G07; Rasmussen \&
Ponman 2007). The data are homogeneously analyzed with results on temperature, entropy,
gas mass and total mass. The sample size also allows us to measure the scatter in
various interesting ICM properties.
We focus on the data reduction and the derived scaling relations in this
paper, while more extensive discussions with modeling and work on an extended
sample including irregular groups will be presented in future papers.
The group sample is defined in $\S$2. The data analysis is presented
in $\S$3, including spatial and spectral analysis. We especially discuss the
\chandra\ background and our method of background subtraction in this section and
Appendix. In $\S$4, we define four tiers of groups with different data coverage.
Different characteristic radii ($r_{2500}$, $r_{1500}$, $r_{1000}$ and $r_{500}$)
are reached in different tiers. We also define the system temperatures
in $\S$4 ($T_{500}$ etc.) and derive their empirical relations.
The group temperature profiles are discussed
in $\S$5, while $\S$6 is about the ICM entropy. In $\S$7, we discuss $M - T$ and
$M - Y_{\rm X}$ relations, gas fraction, concentration parameter, baryon fraction and fossil
groups in this sample. There are groups with signs of AGN heating and
groups with strong central radio sources in this sample. We discuss them in $\S$8.
Systematic errors are discussed in $\S$9. $\S$10 is the summary and conclusions.
We assumed H$_{0}$ = 73 km s$^{-1}$ Mpc$^{-1}$, $\Omega$$_{\rm M}$=0.24,
and $\Omega_{\rm \Lambda}$=0.76.

\section{The group sample}

The groups in the sample are mainly selected from the \chandra\ archive as of September, 2008. 
We also include proprietary data on three groups from the \chandra\ program 09800349
(PI: Vikhlinin). We started to collect groups from several previous
group samples constructed from \rosat\ observations: Mahdavi et al. (2000),
Mulchaey et al. (2003) and GEMS (Osmond, \& Ponman 2004). We have also searched
for low temperature systems ($T < 3$ keV) in the on-line database 
BAX\footnote{http://bax.ast.obs-mip.fr/}. However, most \rosat\ groups with
well-determined temperatures are quite nearby. The most distant group in GEMS is at
$z$=0.0282 (NGC~6338). As we want to constrain gas properties out to at least $r_{2500}$
(ideally $r_{500}$) for each system, many nearby systems in these \rosat\
catalogs are excluded. Thus, we have had to extensively examine the data in the
\chandra\ archive seeking groups with good observations out to these large radii.
Our selection criteria are: 1) full-frame Advanced CCD Imaging Spectrometer (ACIS)
data at the focal plane temperature of -120C (after Jan. 29, 2000);
2) $T_{500} \lsim 2.7$ keV and a global temperature constrained to better
than 15\%; 3) 0.015 $<z<$ 0.13 and group emission traced to at least $r_{2500}$
with the \chandra\ data;
4) group emission well centered around the cD and not significantly
elongated or disturbed beyond the group core.
$T_{500}$ is the temperature measured between 0.15 $r_{500}$ and $r_{500}$
(defined and discussed in $\S$4). The upper limit on $T_{500}$ is determined
from the $M_{500} - T_{500}$ relation in V06 to make sure that we are studying low-mass
systems with $M_{500}$ of $< 10^{14} h^{-1}$ M$_{\odot}$. We understand that
there is not a well-defined temperature boundary separating galaxy groups and clusters.
Many people may consider groups to be systems with temperatures of $<$ 2 keV.
Nevertheless, we refer to all the systems in our sample as groups for
convenient comparison with the clusters in V08. The requirement to well constrain
$T_{500}$ allows us to derive a temperature profile for each group.
The constraint on redshift enables a first cut to make sure $r_{2500}$ can
be reached. We used the $r-T$ relation in V06 as the first guess. It turns out
the derived $r-T$ relations from our work are close to V06's. With the $r_{2500} - T$
relation in V06, $r_{2500}$ = 11.6$'$ for an 1 keV group at $z=0.015$. An ACIS
pointing can reach this radius although the coverage is partial.
The only exception to the redshift requirement is NGC~1550 at $z=0.0124$. There
are two offset observations for this luminous group that allow us to reach $r_{1000}$.
Clearly some hotter systems at $z\gsim0.015$ may not meet our criterion to reach
$r_{2500}$ and are excluded after their temperatures are constrained.
We indeed have examined many more systems than the ones in our final sample.
With \chandra's superior angular resolution, almost all groups have substructures
around the center, and in some cases these features can be very significant and
striking, like the central 60 kpc of IC~1262 ($\sim 0.1 r_{500}$).
Nevertheless, they are included as long as the group emission is well centered
and regular at $\gsim 0.15 r_{500}$, where our main interests are.
Our science goals concerning gas entropy are also not affected by the dynamic state of the
group. Moreover, at least the $Y_{\rm X}$ mass proxy is still robust in unrelaxed systems
(Kravtsov et al. 2006).

The \chandra\ archival search not only includes targets in the cluster and galaxy
categories, but also targets in the AGN category, as many radio galaxies are in
group environments. There are also targets selected optically. Many faint
systems are included, as shown by the wide range of gas entropy at 30 kpc and
0.15 $r_{500}$ ($\S$6). Thus, our final sample is not much biased to the X-ray
luminous systems. The final sample includes 43 groups, listed in Table 1.
It is hardly a homogeneous or a complete sample. But the wide spread of the gas
densities and entropy values
at $r < 0.15 r_{500}$ indicates a wide luminosity range as cool cores in groups contribute
a large portion of their luminosities (See $\S$7.2).
If there were groups that are much fainter or poorer in gas than the faint systems in this sample,
it is difficult to constrain their gas properties with the current
X-ray instruments anyway. Because of the layout of ACIS, full
coverage at large radii (defined as $>$ 80\% coverage for the outermost spectral
extraction bin, as there is always area missing from extended and point sources and chip gaps)
is only achieved in 7 groups (Table 2).
However, the coverage at $r \lsim r_{2500}$ is much better. Previous analysis
for nearby clusters or groups with the \chandra\ data all had partial
coverage at large radii (e.g., V06; Rasmussen et al. 2007; V08) and our sample
is about 3 times bigger than these previous ones.

\section{The data analysis}

\subsection{\chandra\ observations \& calibration steps}

All observations were performed with ACIS.
Standard data analysis was performed which includes
the corrections for the slow gain change
\footnote{http://cxc.harvard.edu/contrib/alexey/tgain/tgain.html}
and Charge Transfer Inefficiency (for both the FI and BI chips). 
We investigated the light curve of source-free regions (or regions with a small
fraction of the source emission) to identify and exclude time intervals with particle
background flares, including weak flares.
The relevant information on the \chandra\ pointings is listed in Table 1.
We corrected for the ACIS low energy quantum efficiency (QE) degradation
due to the contamination on ACIS's Optical Blocking Filter
\footnote{http://cxc.harvard.edu/cal/Acis/Cal\_prods/qeDeg/index.html},
which increases with time and is positionally dependent.
The dead area effect on the FI chips, caused by cosmic rays, has also been
corrected. As the background subtraction is important for this project,
we present it in detail in the next section and Appendix.
We do not use any data on the S4 chip, because of the residual streaks
often seen at low levels (after running the CIAO tool ``DESTREAK'') and the
lack of the stowed background data.
The ``readout artifact'', seen in groups with very bright cores, is
also corrected (see e.g., V06). We used CIAO3.4 for the data analysis.
The calibration files used correspond to \chandra\ Calibration Database
(CALDB) 3.4.3 from the \chandra\ X-ray Center, released in March, 2008.
We are aware of a possible over-correction of the \chandra\ effective area
beyond the Iridium M-edge ($\sim$ 2 keV) in CALDB 3.4.3 and before
(see David's presentation in the 2007 \chandra\ Calibration Workshop)
\footnote{http://cxc.harvard.edu/ccw/proceedings/07\_proc/presentations/david},
which can bias \chandra\ temperatures to higher values, especially for hot
clusters ($T > 4$ keV). The difference between \chandra\ and \xmm\ temperatures
is also shown in Snowden et al. (2008). The calibration work to incorporate
this into the \chandra\ CALDB is ongoing. However, as shown in both
David's presentation and Snowden et al. (2008), the agreement between \chandra\ and
\xmm\ temperatures for $\lsim$ 4-5 keV systems is very good. The highest gas
temperature at any radii in our sample is $\sim$ 3.3 keV for several groups at the center,
while temperatures at large radii are much lower and mainly determined by
the iron L hump (instead of the continuum slope as for hot clusters). Thus, any
changes in our results from this correction should be smaller than the current
measurement errors and scatter.
In the spectral analysis, a lower energy cut of 0.4 keV (for the BI data)
and 0.5 keV (for the FI data) is used to
minimize the effects of calibration uncertainties at low energy.
The solar photospheric abundance table by Anders \& Grevesse
(1989) is used in the spectral fits. Uncertainties quoted in this paper are 1 $\sigma$.

\subsection{Determining the local background}

Proper background subtraction is important for the analysis in the low surface
brightness regions of the groups. A detailed discussion of the \chandra\ background
and the relevant dataset is presented in the Appendix.
We determine the local background based on the stowed background data
and modeling. The corresponding stowed background of each observation is
scaled according to the flux ratio in the 9.5 - 12 keV band (e.g., V05).
We adopt a 3\% uncertainty on the particle background (PB) normalization (5\% for BI
data in period E, see Appendix) in the error budgets.

The local cosmic X-ray background (CXB) can be modeled.
The best fit of the hard CXB component is determined with an absorbed power
law with a photon index of 1.5. The absorption is determined from spectral fits
to the group spectra, which involves iterations and is present in $\S$3.3.
The soft CXB component is adequately described by two thermal components
at zero redshift, one unabsorbed component with a fixed temperature of 0.1 keV and
another absorbed component with a temperature either derived from spectral fits
or fixed at 0.25 keV (see Appendix). Abundances of both components are fixed at Solar.
We can compare our CXB model with other work. V05 used the blank sky
background and corrected for differences between a local control field and
the blank sky background. As the scaling factors of the PB are
close to unity in their sample, V05 ignored the correction for the unresolved hard CXB.
However, the correction for hard CXB becomes more important after the middle of 2004 when
the PB flux is 30\% - 50\% higher than that in the blank sky
background (Appendix). V05 found that generally a single thermal component with a
temperature of $\sim$ 0.2 keV and solar abundance can fit the soft CXB excess or
decrement well. In regions with high RASS R45 values, a second thermal
component with a temperature of $\sim$ 0.4 keV and solar abundance is required.
Humphrey \& Buote (2006) and G07
also used two thermal components with solar abundances to describe the soft CXB. The
CXB temperatures are fixed at 0.07 keV and 0.2 keV. As discussed in the Appendix,
Snowden et al. (2008) used three thermal components to describe the soft CXB. Thus,
our model of the soft CXB only has less freedom than the model by Snowden et al. (2008).
However, our model balances the requirement of having enough components to fit the local
CXB and avoiding parameter degeneracy in generally low-statistics data. First, the two
$\sim$ 0.1 keV components will naturally be mixed in our energy band when the
absorption is low. Second,
statistics for most of our groups at large radii are not very good, so uncertainties
are mainly statistical errors. As long as the background uncertainties are folded
into the final error budgets, very detailed modeling of the soft CXB component
is not crucial. Third, we test different models of the soft CXB for groups in our sample
for which a large group-emission-free region is available. All spectra can be well
fit by our two-component model, while the three-component model by Snowden et al. (2008)
makes little or no improvement. In fact, the soft CXB can be well fitted by a single
thermal component in most cases. However, the abundance usually has to be free
and the best-fit value is usually very close to zero. Therefore, we conclude that
the two-component soft CXB model is adequate and also necessary for our analysis.

The \chandra\ observations in our sample can be classified into two categories, one with
regions free of group emission in the FOV, another with group
emission detected to the edge of the field.
Cool systems at higher redshifts (e.g., $z > 0.04$) usually have group-emission-free
regions in the off-center chips (e.g., S2 for ACIS-I observations,
I2/3 and S1 for ACIS-S observations). There are also two groups with adjacent
\chandra\ pointings for unrelated targets in coincidence (ESO~306-017 and A1692).
We took the following approach to look for group-emission-free regions.
The radial surface brightness profile is derived with the exposure
correction and background subtraction using the scaled stowed background.
Different pointings for the same group are combined.
The obtained surface brightness profile has group emission plus the CXB.
The region where CXB is more dominant than the group emission, if present, can be
determined from the flattened portion at the outer region of the surface brightness profile.
We used a power law plus a constant at large radii to determine whether a significant
group-emission-free region exists and the radial range of that region.
If such a region is found, we have a control field with only CXB.
In this work, the inner radius of the group-emission-free region is
$> 3'$ larger than the outermost radius for the group temperature profile.
This control field may be a single region on the S2 chip (for the ACIS-I data)
or two separate regions on the S1 and I2/I3 chips respectively (for the ACIS-S
data). Although we could simply fit the spectrum (or spectra) of this control
field to determine the local CXB, the statistics are generally not sufficient. Moreover,
the hard CXB in this control field may be larger than that in the outermost
bin for the group temperature profile, as the flux of the hard CXB depends on
point source excision. Thus, we fit the spectra of the control field and the 
outermost bin for the group temperature profile together to better constrain the
local CXB. The normalizations of the soft CXB in different regions 
are linked by the ratios of their covered solid angles of the sky, while the normalizations
of the hard CXB are not linked. Therefore, we can determine the local CXB in the
outermost bin for the group temperature profile. In this work, we assume that the
soft CXB is constant across the examined group area.
Accurate determination of the local CXB is only important for large radii.
The covered area at large radii tends to be in the same direction
from the group center so the assumption should be reasonable. There is a complication
that the hard CXB may be smaller in inner radial bins but it is a small effect
and can be corrected (also discussed later in this section).

However, many groups in our sample have group emission detected to the edge of the
field, so we have to fit the CXB components with the group emission together.
Since these groups are limited by the \chandra\ FOV, the group emission is generally
still significant near the edge of the field. We fit the spectra from the two outermost 
radial bins together, with the normalizations of the soft CXB components linked.
Generally in this work, we are conservative and do not
include the outermost bin in the temperature profile as the uncertainties are generally
large. However, there are a few groups that remain bright to the edge of the field (e.g.,
NGC~1550, A262 and MKW4). They all have S1 data, so the soft CXB component can be easily
separated from the group emission (e.g., Fig. 1), owing to the good response
of the BI chips at $<$ 1 keV. As the temperatures of the cool gas can be well
constrained from the iron hump centroid, group emission can be robustly separated from the
soft CXB even to the edge of the field.
Thus, we derive temperature profiles to the edge of the field for these groups.

As we separate the PB from the CXB, the absolute CXB flux can be
derived in each group field. In Table 2, we list the flux of the local background
components, for both the soft CXB and the hard unresolved CXB (in the outermost bin).
We expect the derived soft CXB is on average
correlated with the RASS R45 flux in the surrounding area (excluding the source region,
see Table 2), which is true as shown in Fig. 2.
One should be aware of the uncertainty in the R45 flux, as the RASS soft X-ray background
maps have 0.2 deg pixels and the various uncertainties combined are not small (e.g.,
variable SWCX emission).
There are also uncertainties related to cross-calibration and
cosmic variance of both the soft and hard CXB, so a detailed one-to-one comparison
is hardly meaningful.
The derived 2- 8 keV flux of the hard CXB component depends on the limiting flux
of the observation. As discussed in Appendix, there is an empirical
relation between the limiting flux for point sources and the average
2 - 8 keV flux of unresolved hard CXB (K07). We estimated the average limiting flux
of the group observations in the outermost spectral bin.
The regions we used to control the local background generally have an area of about one ACIS
chip. As shown in Hickox \& Markevitch 2006 (HM06 hereafter) and K07, the cosmic
variance in this angular scale (depending
on the two-point angular correlation function of point sources) is 20\% - 30\% and
the total hard CXB flux has $\sim$ 10\% uncertainty. Thus, the expected hard CXB flux
from the empirical relation is only meaningful in an average sense. We use the CIAO
tool MKPSF to generate several PSFs (at an energy of 1.4 keV) in the outermost
spectral bin. We then derive the 90\% enclosed power aperture and measure the 3$\sigma$
limits at these regions. Their average is taken as the estimate for the limiting flux
for point sources. Compared with the average growth curve determined by K07 (their Fig. 19), we have an estimate of the unresolved hard CXB, which is also listed in Table 2.
We can see from Table 2 that the general agreement is quite good, while the
uncertainties for the hard X-ray CXB are much larger than the difference.
It is also true that the limiting flux for point sources depends on the off-axis angle.
The \chandra\ limiting flux changes little within the central 6-7$'$ from the
aimpoint, but increases rapidly beyond 10$'$. This can be corrected from the slope
of limiting flux vs. unresolved hard CXB in K07 and the absolute flux in the
outermost bin. The correction is small, also because the errors of the hard CXB are not small.

Besides the 3\% uncertainty on the PB mentioned above (5\% for BI data in period E),
we also included the following error budgets and added them (and the PB uncertainty)
to the statistical uncertainties of temperatures, abundances and surface brightness
in quadrature. For the hard CXB, we set the photon index at 1.1 and 1.9 (e.g., HM06)
and repeat the analysis to estimate the local background in those situations. For the
soft CXB component, if the temperature of the hotter component cannot be derived from
the data and is fixed at 0.25 keV, we change it to 0.2 keV and 0.3 keV and repeat the
analysis to estimate the local background in those situations. Thus, we are
conservative in the uncertainties for the local CXB.

\subsection{Spectral analysis and deprojection}

Once the local CXB is determined, we proceed to derive the projected temperature
profile. The radial bins for temperature measurement are decided from the outermost
bin, by requiring the temperature to be constrained to better than 30\% (1 $\sigma$
with all error budgets) and $r_{out} / r_{in}$ = 1.3 - 1.75. After determining the
outermost bin, the inner bins are determined progressively with $r_{out} / r_{in}$ =
1.25 - 1.6. Point sources and sub-clumps are excluded. The absorption column is
determined from the MEKAL fit to the integrated spectrum between $0.1 r_{500}$
and $0.4 r_{500}$. In this process, we also tried the VMEKAL model with extra free
parameters of O, Ne, Mg, Si, S, Fe and Ni abundances, as a lower O/Fe ratio may
cause excess absorption (Buote et al. 2003). This process and the determination of
the local CXB are done in iterations to obtain the final value of the absorption.
If the derived absorption is consistent with the Galactic absorption from the HI
survey (Table 2) within 1 $\sigma$, the Galactic value is used. Only when both MEKAL
and VMEKAL fits show excess absorption, the X-ray absorption from the MEKAL fit
(always consistent with the VMEKAL fit within 1 $\sigma$) is adopted (Table 2).
About 44\% of groups show excess absorption relative to the Galactic value
and indeed we find \iras\ 100 $\mu$m enhancement in many of these groups (see Appendix
for discussions of some groups). We also examined the absorption variation with
radius in each group. Beyond the central 20 kpc, no significant absorption variation
is found for any group so we used a fixed absorption column for each group. In several
cases, we observed an absorption increase within the central 20 kpc, but this analysis
is complicated by the possible multi-phase gas around the center. Nevertheless,
it has little effect on any of our results. We also discuss the systematic error related to
absorption in $\S$9.

We used the MEKAL model to fit the spectra of the group emission. The free parameters
are temperature, abundance and normalization, once the absorption is determined.
For spectra around the center, we also include a component to account for the LMXB
emission from the cD galaxy. The LMXB component is represented by a power law with
an index of 1.7, within $D_{25}$ aperture obtained from HyperLeda.
The total LMXB luminosity is fixed from the $L_{\rm X} - L_{\rm Ks}$ relation
derived in Kim \& Fabbiano (2004), where $L_{\rm Ks}$ is the total $K_{\rm s}$ band
luminosity of the cD from 2MASS. We also assume that the LMXB emission follows
the $K_{\rm s}$ band light (see also Gilfanov 2004).

The projected group temperature profiles are shown in Fig. 3 - 5.
We used the algorithm determined by Vikhlinin (2006) to derive the deprojected
temperature profiles in a parametric way, which was first applied in V06. The required
inputs are the three-dimensional (3D) or deprojected profiles
of gas density and abundance, and the projected temperature profile.
V06 simply used the projected abundance profile as their sample is dominated
by hot clusters. However, the emissivity of $\lsim$ 2 keV plasma is sensitive
to the chemical abundances so the 3D abundance profile is required for our work.
We applied the non-parametric geometrical deprojection (summarized in e.g.,
Pizzolato et al. 2003; G07) to derive
deprojected abundances in wider radial bins, generally merging 2 - 3 adjacent
bins for the temperature profile. The deprojected abundance is an emission-weighted
average in each bin so an effective radius in each bin is required.
We define the effective radius as the emissivity-weighted radius in each bin,
where the plasma emissivity, $\varepsilon (T, Z, r)$, depends on the 3D profiles of
temperature, abundance and density.

Since the determination of the 3D profiles of temperature, abundance and density
depends on each other, iterations have to be done to derive the best-fit 3D profiles.
Usually at most three iterations are required to stabilize the best-fits of these
profiles.
We assumed the following 3D abundance profile:

\begin{eqnarray}
Z(r) = Z_{0} + Z_{1} exp (-(r/r_{Z})^{\alpha_{\rm Z}})
\end{eqnarray}

This simple form can fit the abundance profiles of all groups in the sample.
An example is shown in Fig. 6 that represents the best-constrained 3D abundance
profile in this sample (thus the most difficult to fit, as the errors are the smallest).
Most groups in this sample only have 3D abundances constrained
in 3 - 5 bins with larger errors so good fits can always be achieved with this simple
function. Once the best-fits of all profiles are achieved, we applied 1000 Monte Carlo
simulations to address the uncertainties of the 3D abundance profile. The simulation is
realized by scattering the 3D abundance profile according to the measurement
errors. In this process,
to be conservative, we also include an uncertainty of 10\% of the bin size on the
effective radius of each bin for the abundance profile.
Thus, besides the best-fit 3D abundance profile, we have 1000 simulated profiles to
cover the error ranges, which will be used to estimate
the uncertainties of the 3D temperature and density profiles.

We used the same form of the 3D temperature profile as used in V06:

\begin{eqnarray}
T(r) = \frac{T_{0}(r/r_{cool})^{a_{cool}}+T_{min}}{(r/r_{cool})^{a_{cool}}+1}
 * \frac{(r/r_{t})^{-a}}{[1+(r/r_{t})^{b}]^{c/b}}
\end{eqnarray}

The exceptions are A1139, A1238 and A2092, which are faintest 2 - 3 keV systems
in this sample so the errors on temperatures are large. Their 3D temperature
profiles are modeled as: $T(r) = T_{0} [1+(r/r_{t})^{a}]^{b/a}$, which fits their
temperature profiles very well (Fig. 4 and 5). 
The inner boundary of the fit is 5 - 20 kpc, depending on the quality of the fits
at the center. As we focus on properties at large radii, the detailed choice of the 
inner boundary (5 - 20 kpc) is not a concern.
Because the sizes of bins at large radii are not small, each bin is further divided into 6
sub bins. The temperature modeling is done at the center of each sub bin, only added up
(with the algorithm by Vikhlinin 2006) at the end to obtain the expected
temperature for each original bin of spectral analysis. In this way, the binsize effect
is minimized and we don't need to worry about the accurate determination of effective
radius for each radial bin, as most previous work had to address.
The uncertainty of the 3D temperature profile
is derived from 1000 Monte Carlo simulations. In each simulation, the measured projected
temperature profile is scattered according to the measurement errors, and a new simulated
abundance profile is input. The density profile is fixed at the best-fit value, as the
temperature error is the dominant error source.
The reconstructed 3D temperature profiles, with 1 $\sigma$ uncertainties from simulations,
are also shown in Fig. 3-5. There are groups where a central corona exists
(e.g., Sun et al. 2007), so naturally the temperature gradient at 5 - 20 kpc
is large in these cases (e.g., A2462, A160, ESO~306-107, HCG~51 and NGC~6269). Sometimes
the 3D temperature appears too low within 5 - 10 kpc radius (e.g., A160 and NGC~6269),
which however affects none of our results in this work as core properties are
excluded.

\subsection{Gas density}

We extract the surface brightness profile in the 0.7 - 2 keV band, as suggested by
V06 to avoid the 0.6 keV hump in the soft CXB. Point sources and chip gaps
are excluded. The scaled stowed background is subtracted so only the group emission
and the local CXB is left. Some previous analysis on a surface brightness profile involved
the correction of a single exposure map.
However, the \chandra\ exposure map is energy and position dependent.
It is not accurate to use a single exposure map (even one convolved with the group
spectrum) as there is spectral variation in a group. The most accurate approach is
to generate response files for each bin of the surface brightness and convert the
raw count rate (without corrections on vignetting and other response) to density, from
the derived 3D temperature and abundance profiles and response files.
To achieve that, we use XSPEC to generate an MEKAL emissivity matrix that depends on temperature,
abundance and position of the radial bin for surface brightness (or response files there),
once the absorption is determined. The ranges for
temperatures and abundances cover the observed ranges for any particular group.
This emissivity matrix provides the conversion factor needed to transfer the
observed count rate to the emission measure.
We assume the following density model (with 11 free parameters):

\begin{eqnarray}
n_{e}^{2} = n_{0}^{2} \frac{(r/r_{c})^{-\alpha}}{[1+(r/r_{c})^{2}]^{3\beta-\alpha/2}}
\frac{1}{[1+(r/r_{s})^{\gamma}]^{\delta/\gamma}} \nonumber \\
+ n_{02}^{2} [1-(r/r_{c2})^{\alpha_{2}}]^{3\beta_{2}/\alpha_{2}} (r < r_{c2})
\end{eqnarray}

This model combines the profile used in V06 and the profile proposed by Ettori (2000) for
cool cores. We find that this model provides very good fits for all groups in our sample.
The density profile is then converted to the emissivity profile from the emissivity matrix.
The emissivity profile is then projected
along line of sight with the formula of geometrical deprojection (e.g.,
McLaughlin 1999). The local CXB can be added later with the known CXB spectra and the radial
set of the response files. The resulting surface brightness profile can be compared with
the observed one. 
To avoid the binsize effect that may especially affect outskirts as wider bins are required
there, we further divide bins within 100 kpc to 3 sub bins and bins beyond the central
100 kpc to 6 sub bins. The conversion and deprojection is done in these sub bins.
They are later merged to compare with the observed profile.
The density errors are estimated from 1000 Monte Carlo simulations, with 1000 corresponding
simulated abundance and temperature profiles as inputs.

This method is similar to that used in V06, but we use the 3D temperature and abundance
profiles to convert count rates to density. Like V06,
we also analyzed the \rosat\ PSPC pointed observations for the purpose of constraining
gas density profiles at large radii. The inclusion of the PSPC data is especially
useful to $z<0.04$ groups. We used the software developed by Snowden et al.
(1994) to produce flat-fielded PSPC images in the 0.7 - 2 keV band (more exactly, R567
bands, or PI channels of 70 - 201). The images are further analyzed as described in
Vikhlinin et al. (1999).
Seventeen groups have sufficiently long PSPC data (listed in Table 1 with effective exposure)
and we included the PSPC surface brightness profile in the modeling of the density profile.
The probed outermost radius of the PSPC data for each group is listed in Table 2.
We only used the PSPC surface brightness profiles outside the central
3$'$ to avoid the PSF correction in the core. In all cases, there is good
agreement between the \chandra\ and PSPC surface brightness data.
One example is shown in Fig. 7.

\section{Characteristic radii and definition of system temperatures}

As stated above, we generated 1000 simulated profiles of the 3D temperature,
abundance and density profiles, which cover the ranges of the measurement errors.
Assuming hydrostatic equilibrium,
each set of the 3D temperature and density profiles determine a set of characteristic
radii of the group ($r_{500}$,
$r_{1000}$, $r_{1500}$ and $r_{2500}$). The profiles of interesting quantities, e.g., total
mass, gas mass and entropy, and their values at the characteristic radii are also
determined. For each quantity, the peak in the distribution from 1000 simulations
defines the most probable value (or ``best fit''). The 1$\sigma$ errors at two sides
are estimated by determining the regions that contain 68\% of realizations at each side.
The results are listed in Table 3 and 4.
$r_{500}$ is the basic characteristic radius that is used in most scaling relations in
this work. However, it cannot be robustly determined for all groups in our sample so
the $r_{500} - T$ scaling relation is required.
As the radial coverage of data in this sample differs from one group to
another, we define four tiers of groups for which group properties (entropy, mass,
$Y_{\rm X}$, and gas fraction) are derived to different characteristic radius.

\begin{itemize}
\item Tier 1: groups with surface brightness (including the PSPC profile) derived at
$> 2\sigma$ levels to $> r_{500}$ (note we are conservative to estimate the errors of
the local CXB, see notes of Table 2) and a temperature profile derived to
$>$ 80\% of $r_{500}$. Eleven groups in our sample are in this tier and their
temperature profiles are derived to 81\% - 117\% of $r_{500}$ with a median of 97\%.
Groups in this tier have density profiles derived to $> r_{500}$ so the density
gradient at $\sim r_{500}$ is constrained. Temperature profiles are derived
beyond $r_{500}$ or sufficiently close to $r_{500}$ for reasonable extrapolation.
We note that in V06, there are three clusters and groups (A383, MKW4 and A1991)
with temperature profiles derived only to 73\% - 89\% of $r_{500}$, but properties
at $r_{500}$ are still derived from extrapolation. Thus, our criteria are similar
to those in V06 and V08.
\item Tier 2: groups with surface brightness and temperature profiles all derived
to at least $r_{1000}$ but not in tier 1. Twelve groups are in this tier and their temperature
profiles are derived to 68\% - 87\% of $r_{500}$ with a median of 77\%.
We consider that groups in this tier have properties well constrained close to
$r_{500}$ so $r_{500}$ is determined in these groups. Group properties at $r_{500}$
are also derived from extrapolation. However, in figures of scaling relations at
$r_{500}$, tier 2 groups are always marked differently from tier 1 groups. The fits
with or without them are both listed.
\item Tier 3: groups with surface brightness and temperature profiles derived
to at least $r_{1500}$ but not in tiers 1 and 2. Eleven groups are in this tier.
Nine of them have temperature profiles derived to 52\% - 72\% of $r_{500}$, which is
close to $r_{1000}$ ($\sim 0.73 r_{500}$ in this sample). Thus, group properties
at $r_{1000}$ are also derived from extrapolation. A1238 and RXJ~1206-0744 are
put in this tier as the temperature and density errors are large although $r_{1000}$
is reached in both cases.
\item Tier 4: nine other groups with surface brightness and temperature profiles
derived to at least $r_{2500}$. Group properties at $r_{1500}$ are also derived from
extrapolation in this tier.
\end{itemize}

Previous X-ray work on clusters and groups often defined $<T>$, which is the
emission-weighted temperature within a certain aperture. As definitions
are generally different, it is necessary to use an unified definition that
is easily accessible from observations. In this work, we define the system temperature as:

\begin{itemize}
\item  $T_{500}$: the spectroscopic temperature measured from the integrated spectrum
in the projected 0.15 $r_{500}$ - $r_{500}$ annulus.
\end{itemize}

We derive $T_{500}$ by integrating the 3D temperature profile from 0.15 $r_{500}$ to
1.6 $r_{500}$ (or $\sim r_{180}$), in an annular cylinder with projected radii
of 0.15 $r_{500}$ - $r_{500}$ along the line of sight, with the algorithm by
Vikhlinin (2006). The choice of the outer radius in the 3D integration little
affects $T_{500}$ as it is emission-weighted. The same definition of the system
temperature was also used in Nagai, Kravtsov \& Vikhlinin (2007b, NKV07 hereafter),
Maughan (2007) and V08.
This definition excludes the central region, where a cool core or a locally heated
region may exist. Indeed the group temperature profiles are much more similar
at $r > 0.15 r_{500}$ ($\S$5 and Fig. 8).
This temperature can also be directly derived from data, provided that the full coverage of
$r_{500}$ is achieved (generally not the case for groups in our sample).
Similarly, we can define $T_{1000}$, $T_{1500}$ and $T_{2500}$, with the projected
inner boundary always at 0.15 $r_{500}$ and the projected outer boundary at $r_{1000}$,
$r_{1500}$ and $r_{2500}$ respectively (3D temperature profile still integrated to
1.6 $r_{500}$). Their empirical relations can also be determined.

As we cannot derive $r_{500}$ for groups in tiers 3 and 4, the $r_{500} - T_{500}$
relation needs to be determined. The $r - T$ relation is just a manifestation
of the $M - T$ relation so it is presented in $\S$7.1.
We also derived the average ratios of the characteristic radii from 23 tier 1 and 2
groups: $r_{1000}$ / $r_{500}$ = 0.741$\pm$0.013, $r_{1500}$ / $r_{500}$ = 0.617$\pm$0.011
and $r_{2500}$ / $r_{500}$ = 0.471$\pm$0.009, which are about what are expected for
the average $c_{500}$ of this sample ($\sim 4.2, \S$7.3), 0.727, 0.599 and 0.465
respectively (assuming an NFW profile).

We also need empirical relations between $T_{500}$ and $T_{1500}$ (or $T_{2500}$)
to estimate $T_{500}$ for groups in tiers 3 and 4.
For 23 groups in the first and second tiers, we found:

\begin{eqnarray}
kT_{500} / kT_{1500} = 0.93 \pm 0.02
\end{eqnarray}

This ratio is not temperature dependent in our sample and the fit is very
good ($\chi^{2}$/dof = 6.6/22). If we only fit 11 groups in the first tier,
the ratio is the same. We notice that V08 derived a similar empirical
relation between $T_{500}$ and temperature measured at 0.15 - 0.5 $r_{500}$ (close
to our $T_{2500}$). V08 also included a linear term as they explored a wider
temperature range. We also derived $kT_{500} / kT_{2500} = 0.89 \pm 0.02$,
but the scatter is larger as shown by the poorer fit ($\chi^{2}$/dof = 18.6/22).
The empirical relation between $kT_{500}$ and $kT_{1500}$, combined with the
$r_{500} - T_{500}$ scaling, allows us to estimate $r_{500}$ for groups in tiers 3
and 4, in a few iterations. We specifically used the $r_{500} - T_{500}$ relation
from 23 tier 1 and 2 groups and 14 clusters in V06 and V08 (The fourth row of Table 6). 
For reference, we also give the best fit of the $r_{2500} - T_{500}$ relation
for all 43 groups in this sample:
$(E(z) r_{2500} / {\rm 155\pm4 h^{-1} kpc}) = (T_{500} / \rm 1 keV)^{0.520\pm0.040}$,
which is similar to the V06 result,
$(E(z) r_{2500} / {\rm 146\pm3 h^{-1} kpc}) = (T_{500} / \rm 1 keV)^{0.547\pm0.020}$.

\section{Temperature profiles}

Scaled temperature profiles for these groups are shown in Fig. 8, in logarithmic
and linear scales. We scale temperatures with $T_{2500}$, which is robustly
determined for each group. While the scatter within the central 0.15 $r_{500}$
is large, the group temperature profiles are more similar beyond 0.15 $r_{500}$,
with a declining slope similar to that predicted from simulations (e.g., Loken et al. 2002).
From 0.15 $r_{500}$ to $\sim r_{500}$, we can fit the projected temperature profiles
with this simple form:

\begin{eqnarray}
T/T_{2500} = (1.22\pm0.02) - (0.79\pm0.04) r/r_{500} 
\end{eqnarray}

Interestingly, V05 derived an average form of $T/<T> = 1.22 - 1.2 r/r_{180}$
(0.125 $< r/r_{180} <$ 0.6) for 13 systems in their sample, which is very similar
to ours as $r_{500} \sim 0.62 r_{180}$.
If we use a similar form as used in Loken et al. (2002), we can also fit the
projected temperature profiles at 0.15 $r_{500}$ - $r_{500}$ with this form:

\begin{eqnarray}
T/T_{2500} = (1.37\pm0.03) (1 + r/r_{500})^{-(1.34\pm0.21)} 
\end{eqnarray}

Based on \asca\ data, Markevitch et al. (1998) first suggested that temperature
profiles of clusters are self-similarly declining with radius. This result
was later confirmed by De Grandi \& Molendi (2002) with the \sax\ data, and
by V05 with the \chandra\ data, and by Piffaretti et al. (2005),
Pratt et al. (2007) and Leccardi \& Molendi (2008, LM08 hereafter) with the \xmm\
data. Self-similarly declining temperature profiles
are also generally observed in simulations (e.g., Loken et al. 2002; Borgani et al.
2004; Kay et al. 2004). The self-similar decline in groups was also suggested
by Sun et al. (2003), G07 and Rasmussen \& Ponman (2007).
Our group sample is the largest one so far with detailed studies.
As shown in Fig. 8, group temperature profiles are generally self-similar with
a slope consistent with that in simulations, although there is some scatter.
We also combine all data points to make a mean temperature profile (Fig. 9).
Fig. 9 also plots the mean temperature profile of hot clusters at $z$=0.1-0.3
from LM08, as well as the mean temperature profile
of 1-3 keV systems from the Borgani et al. (2004) simulations.
We do not plot the mean temperature profiles from De Grandi \& Molendi (2002),
V05 and Pratt et al. (2007), but they are close to LM08's (Fig. 21 in LM08).
Clearly, the group temperature profiles are more peaky than those of clusters,
starting to decline at $\sim 0.2 r_{500}$, where the mean cluster temperature
profile is still flat. This difference was also noticed by V06.
The group temperature profiles actually agree more with simulations (e.g., Loken et al. 2002;
Borgani et al. 2004), which generally have problems explaining the flat
temperature profiles of clusters at $0.15 r_{500} - 0.3 r_{500}$ (e.g., Borgani et al. 2004).

\section{Entropy}

The ICM entropy, defined as $K = T / n_{e}^{2/3}$, is a fundamentally important
quality of the ICM as summarized in Voit (2005). Entropy
records the thermodynamic history of the ICM, as non-gravitational processes (e.g.,
AGN heating, cooling and star formation) deviate the entropy relations from the
self-similar relations determined with only gravity and accretion shocks
(e.g., Ponman et al. 1999; Voit \& Bryan 2001).
Voit et al. (2005, VKB05 hereafter) derived the baseline ICM
entropy profile from simulations in the absence of non-gravitational processes,
$K(r) / K_{\rm 200, adi} = 1.32 (r/r_{200})^{1.1}$.
$K_{\rm 200, adi}$ is an entropy scale of a non-radiative cluster at $r_{200}$ (VKB05).
This baseline entropy profile can be compared with our observational results
to measure the impact of non-gravitational processes on the ICM. 
With a weighted average of $\sim$ 4.2 for $c_{500}$ in this sample ($\S$7.3),
$r_{500}/r_{200}$ = 0.669 for an NFW mass profile and the baseline entropy relation is converted to:
$K(r) / K_{\rm 500, adi} = 1.40 (r/r_{500})^{1.1}$ (note that we use $K_{\rm 500, adi}$
here for the adiabatic entropy scale at $r_{500}$, see the definition of $K_{\rm 200, adi}$
in VKB05 and Voit 2005).
For the cosmology assumed in this paper and a baryon fraction ($f_{\rm b}$) of 0.165
(Komatsu et al. 2008),

\begin{equation}  \label{K500}
K_{\rm 500, adi} = 342 \hspace{0.16cm} {\rm keV cm^{2}} (\frac{M_{500}}{10^{14} M_{\odot}})^{2/3} E(z)^{-2/3} h_{73}^{-4/3}
\end{equation}

Note that $K_{\rm 500, adi}$ has a different $h$ dependence from the observed entropy
($K \propto h_{73}^{-1/3}$).
From the derived temperature and density profiles, we obtained entropy profiles
for each group. The best-fit scaled entropy profiles are shown in Fig. 10.
Substantial scatter is present within 0.3 - 0.4 $r_{500}$.
As Fig. 10 is basically a plot showing the large scatter of the density profiles,
it is clear that there is a wide range of X-ray luminosities in this sample,
which is also implied by the $K - T$ relations shown next.
We discuss the ICM entropy of these groups in the following three sections.

\subsection{$K - T$ relations}

We first examined the $K - T$ relations at 30 kpc, 0.15 $r_{500}$, $r_{2500}$,
$r_{1500}$, $r_{1000}$ and $r_{500}$ (Fig. 11). Entropy values at these radii are:
$K_{\rm 30 kpc}, K_{\rm 0.15 r500}, K_{2500}, K_{1500}, K_{1000}, K_{500}$
respectively (note the difference between $K_{500}$ and $K_{\rm 500, adi}$
defined earlier). We chose 30 kpc to represent the entropy level around the BCG.
Previous studies often discussed entropy at 0.1 $r_{200}$ (e.g., Ponman et al. 2003),
which is about 0.15 $r_{500}$. We also performed the BCES (Y$|$X) regression
(Akritas \& Bershady 1996) to these relations and the results are listed in Table 5. 
The entropy values at 30 kpc radius hardly show any correlation with the system
temperature. The $K - T$ relation is stronger at 0.15 $r_{500}$, but the
intrinsic scatter is still substantial. The large scatter of gas entropy
around the center has been known for a while (e.g., Ponman et al. 2003).
We also examined the connection between the luminosity of the
central radio source (see Table 1) and the
entropy scatter at 0.15 $r_{500}$. No correlation is found, before or after the
temperature dependence of entropy is removed. However in $\S$7.2, we show that
the entropy scatter at 0.15 $r_{500}$ is tightly correlated with the scatter of
the gas fraction within $r_{2500}$.

Despite large entropy scatter from the core to at least 0.15 $r_{500}$,
the $K - T$ relations are tighter at $r_{2500}$ and beyond.
We derived the intrinsic scatter in these relations,
using the method described in Pratt et al. (2006) and Maughan (2007). As shown
in Fig. 11, the intrinsic scatter reduces to 10\% at $r_{2500}$ and stays at that
level to $r_{500}$. As the derived intrinsic scatter decreases with 
increasing measurement errors, could this trend be due to the increasing
measurement errors with radius? To answer this question,
we compared $K_{2500} - T_{2500}$ and $K_{0.15 r500} - T_{2500}$ relations.
If we use the relative measurement errors of $K_{0.15 r500}$ for the corresponding
$K_{2500}$, the intrinsic scatter only increases to 12\%. Similarly, if we use the
relative measurement errors of $K_{2500}$ for the corresponding $K_{0.15 r500}$, the
intrinsic scatter only decreases to 28\%. Therefore, the significant tightening of the
$K - T$ relation from 0.15 $r_{500}$ to $r_{2500}$ is not caused by different measurement
errors. Similarly, we find that the intrinsic scatter in entropy at radii from
$r_{2500}$ to $r_{1000}$ is affected little by switching measurement errors.
The $K_{500}$ values have large measurement errors and half of systems
are extrapolated from $r_{1000}$. Clearly more groups with $K_{500}$ robustly determined
are required, but the intrinsic scatter in the current $K_{500} - T_{500}$ relation is
consistent with the level from $r_{2500}$ to $r_{1000}$.
Thus, groups behave more regularly from $r_{2500}$ outward than inside 0.15 $r_{500}$.

We also include 14 clusters from V06 and V08 for comparison and to constrain the
$K - T$ relations in a wider temperature range (Fig. 12 and Table 5). The measured
slopes increase from $\sim$ 0.50 at 0.15 $r_{500}$ to $\sim$ 1 at $r_{500}$, mainly
caused by the excess entropy of groups at their centers.
The slope we find at 0.15 $r_{500}$, 0.494$\pm$0.047, is consistent with what
Pratt et al. (2006) found at 0.1 $r_{200}$ for 10 groups and clusters with the \xmm\
data, 0.49$\pm$0.15, but smaller than what Ponman et al. (2003) found at 0.1 $r_{200}$
with \rosat\ and \asca\ data, $\sim$ 0.65. We however caution that the V08 cluster sample
lacks non-cool-core clusters. Pratt et al. (2006) also gave the $K - T$ relation at
0.3 $r_{200}$, which is about $r_{2500}$. The slope they found, 0.64$\pm$0.11, is
also consistent with our result, 0.740$\pm$0.027.
At $r_{500}$, the slope we found is consistent with the expected value from the
self-similar relation (1.0).
Interestingly, the derived $K_{500} - T_{500}$ relation agrees very well with the
NKV07 simulations with cooling + star-formation (Fig. 12), although the agreement
is progressively worse with decreasing radius. The tightening of the $K - T$ relation
at $r_{2500}$ and beyond was also reported in the NKV07 simulations, but the predicted
$K_{2500} - T$ relation by NKV07 lies below all groups in our sample (Fig. 11 and 12).
The NKV07 simulations with cooling + star-formation achieve entropy amplification
with strong condensation to drop dense materials out of the X-ray phase. The
resulting stellar fraction is about twice the observed value as too much
material has cooled. If the star formation is suppressed with more efficient
feedback, entropy is lower in the NKV07 simulations, between the results from the
simulations with cooling + star-formation
and the non-radiative simulations, which further disagrees with observations.
It seems that the most challenging task is to explain the excess entropy of groups at $r_{2500}$.
As shown in Borgani \& Viel (2008), simply increasing entropy floor from pre-heating
produces too large voids in the Lyman-$\alpha$ forest.
Thus, it is still an open question on how to generate enough entropy in groups
from the center to $r_{500}$ and still preserve other relations like the condensed
baryon fraction and properties of the Lyman-$\alpha$ forest.

\subsection{Entropy ratios}

We derived the ratios of the observed entropy to the baseline entropy
from VKB05 at $r_{500}$, $r_{1000}$ and $r_{2500}$ (Fig. 13).
For comparison, we also include the entropy ratios
for 14 clusters from V06 and V08. The entropy ratios are always larger than or comparable
to unity at all radii. The average ratio decreases with radius, for both groups and
clusters, although the decrease is more rapid in groups. For groups, the weighted 
mean decreases from 2.2 at $r_{2500}$ to 1.57 - 1.60 at $r_{500}$, while the mean
decreases only $\sim$ 9\% for clusters. Thus, there is still a significant entropy excess
over the VKB05 entropy baseline at $r_{500}$, in both groups and clusters. The weighted mean
entropy of groups at $r_{500}$ is $\sim$ 18\% larger than that of clusters. If we consider
only the $T_{500} \lsim 1.4$ keV groups, the weighted mean for 7 groups is 1.68.
Similar studies have been done with \asca\ and \rosat\ data (Finoguenov et al. 2002;
Ponman et al. 2003). When the observed entropy values at $r_{500}$ are scaled
with $M_{500}^{-2/3}$ (which is proportional to $K_{\rm 500, adi}^{-1}$), both works
found that groups have on average twice the scaled entropy of clusters. Ponman et al. (2003)
further concluded that the excess entropy is observed in the full mass range.
Thus, the results from this work, V06 and V08 confirm the previous finding of the excess
entropy at $r_{500}$ over the full mass range, but the magnitude of the excess is smaller
and the difference between groups and clusters is also smaller than previous results,
$\sim$ 15\% - 20\% vs. $\sim$ 100\% in Finoguenov et al. (2002) and Ponman et al. (2003). 

As pointed out by Pratt et al. (2006), the magnitude of the excess may be
affected by the systematics of the $M - T$ relation, or how robust the hydrostatic
equilibrium (HSE) mass is. The NKV07 simulations show that the HSE mass is systematically
lower than the real mass and the difference is biggest in low-mass systems. The real mass
is 45\% higher than the HSE mass for 1 keV groups, while the difference reduces to
13\% for 10 keV clusters. The best-fit $M_{500} - T_{500}$ relation from this work and
V06 ($\S$7.1) is close to the relation for the HSE mass in NKV07. Thus, if this
bias is real, the actual entropy ratios at $r_{500}$ are smaller. For 1 keV groups,
a 45\% higher total mass means a 13\% larger $r_{500}$. For a typical entropy slope of
0.7 in this sample ($\S$6.3), the entropy ratio at $r_{500}$ is reduced by 17\%. Similarly
for 5 keV clusters, the entropy ratio at $r_{500}$ is reduced by 8\% for an average entropy
slope of 0.9 at large radii. Thus, this bias can explain only part of the entropy excess
observed. The entropy baseline we adopted in this work is from the SPH simulations,
while the AMR simulations produce a baseline with ~ 7\% higher normalization (VKB05),
likely because of its better capability to catch shocks. However, the entropy excess
for groups at $r_{500}$ remains significant.

There are several mechanisms to achieve entropy amplification at $r_{500}$.
The first idea relies on modification of accretion, without extra non-gravitational
processes (Ponman et al. 2003; Voit et al. 2003; Voit \& Ponman 2003).
If preheating or feedback in small subhalos that are being accreted can eject gas out
of the halo and thicken the filaments significantly, the accretion may be smoother
than the lumpy accretion in hierarchical mergers. Voit (2005) showed that the smooth
accretion can generate $\sim$ 50\% more entropy throughout the cluster than would
lumpy hierarchical accretion. However, Borgani et al. (2005) argued that this
entropy amplification effect is substantially reduced by cooling.
The other ideas resort to non-gravitational processes.
As discussed in the last section, although the NKV07 simulations predict the $K_{500} - T_{500}$
relation very well, they produce too many stars. Thus, models with only cooling may
not be enough. Borgani et al. (2005) showed that galactic winds from SN explosions
are rather localized and cannot boost entropy enough at large radii.
Thus, a feedback mechanism that can distribute heat in a very diffuse way is required.
As groups are smaller than hot clusters, AGN outflows from the central galaxy
can reach a larger scaled radius. One good example in this sample is 3C449 ($\S$8 and
Appendix), with radio lobes extending to at least 3.7 $r_{500}$.
Thus, it is still an open question to explain the entropy excess at $r_{500}$, especially
in groups. But the much reduced entropy ratios from this work largely alleviate the
problem.
It is also clear that more systems with the entropy ratio constrained at $r_{500}$
are required for better comparison, especially $T_{500} \lsim 1.4$ keV groups.

\subsection{Entropy slopes}

We also derived the entropy slopes at 30 kpc$ - 0.15 r_{500}$, 0.15 $r_{500} - r_{2500}$,
$r_{2500} - r_{1500}$ (or $r_{\rm det, spe}$) and $r_{1500} - r_{\rm det, spe}$ (Fig. 14).
The scatter is large but the average slopes are about the same ($\sim$ 0.7) beyond
0.15 $r_{500}$. The slope is always shallower than that from pure gravitational processes
($\sim$ 1.1).
Mahdavi et al. (2005) analyzed the \xmm\ data of 8 nearby groups and found their
entropy profiles are best fitted by a broken power law with the break radius of
$\sim 0.1 r_{500}$. Across the break radius, the entropy slope decreases from 0.92
to 0.42. As the break radius is very near the core, this kind of entropy profiles
may be caused by a cool core within 0.1 $r_{500}$. From 0.15 $r_{500}$, there are
systems in our sample with an entropy slope of around 0.42, but the weighted mean
is significantly larger. The measured average slopes in this work are consistent with
the result by Finoguenov et al. (2007) (0.6 - 0.7) for groups.

\section{Mass and gas fraction}

\subsection{$M - T$ and $M - Y_{\rm X}$ relations}

One of the most important aspects of cluster science is to use clusters to study cosmology,
which often involves derivation of a cluster mass function. Assuming hydrostatic equilibrium,
X-ray observations can be used to derive the cluster mass (at least the HSE value),
provided the radial coverage of the data is good. Often we estimate cluster mass from 
another X-ray observable used as a mass proxy. A frequently used
mass proxy is gas temperature so it is important
to understand the cluster $M - T$ relation. There had been a lot of work on the
cluster $M - T$ relation before the \chandra\ and \xmm\ era (e.g., Finoguenov et al. 2001;
Sanderson et al. 2003). However, as emphasized in V06, it is crucial to
constrain the gas properties (both the temperature gradient and the density gradient)
at large radii (e.g., around $r_{500}$). Systematics like the assumption of a polytropic
equation of state and the inadequate fit of the density profile at large radii 
can bias the HSE mass to lower values (Borgani et al. 2004; V06).
The $M - T$ relations have also been constrained with \chandra\ and \xmm\ (Arnaud et al.
2005 with the \xmm\ data; V06 and V08 with the \chandra\ data).
However, the number of clusters used to
constrain the $M - T$ relation is still small (10 in Arnaud et al. 2005 and 17 in V08).
There are only 4 systems with temperatures of 2.1 - 3.0 keV
in Arnaud et al. (2005) and only three systems with temperatures of
1.6 - 3.0 keV in V08 (two overlapping with the Arnaud et al. sample).
Our paper adds more $<$ 2.7 keV systems (23 tier 1/2 groups).

The total mass values and uncertainties at interesting radii are derived from
the 1000 simulated density and temperature profiles. The determination of the
best-fit value and the 1$\sigma$ errors (at two sides) is mentioned in $\S$4.
One difference from the determination of entropy is that we include only
simulations that produce physically meaningful mass density profiles (mass
density always larger than zero). The derived $M_{500} - T_{500}$ relation
with the tier 1+2 groups is shown in Fig. 15. The BCES fits are listed in
Table 6. We also included 14 clusters from V08 to constrain the $M - T$
relation in a wider mass range. Our results show that the $M_{500} - T_{500}$
relation can be described by a single power law down to at least $M_{500} =
2 \times 10^{13}$ h$^{-1}$ M$_{\odot}$. At $T_{500} < 1$ keV, more systems
are needed to examine whether the relation steepens or not. Our $M_{500} - T_{500}$
relation is steeper than but still consistent with V08's (1.65$\pm$0.04 vs. 1.53$\pm$0.08).
Our slope is consistent with the Borgani et al. (2004) simulations with non-gravitational
processes included (1.59$\pm$0.05), especially if two groups at $T_{500} < 1$ keV
are excluded. The derived $M_{500} - T_{500}$
relation can also be compared with that by Arnaud et al. (2005), which has a slope
of 1.71$\pm$0.09. Arnaud et al. (2005) defined the system temperature as the overall
spectroscopic temperature of the 0.1 $r_{200} - 0.5 r_{200}$ region, which should
be close to $T_{1000}$ defined in this work (note $r_{500} \sim 0.66 r_{200}$). The
data of the tier 1 and 2 groups give $T_{500} / T_{1000}$ = 0.96 on average. Then
we find that the \chandra\ $M_{500} - T$ relation constrained from 23 groups + 14 V08
clusters is 18\% - 3\% higher than that by Arnaud et al. (2005) at 1 - 10 keV.
We also notice that at 1 keV, the normalization of the \chandra\ $M_{500} - T_{500}$
relation is 54\% higher than that by Finoguenov et al. (2001). This is
expected from the generally higher density gradient around $r_{500}$ derived
in this work (Fig. 16) than the typical values in Finoguenov et al. (2001)
(also see the Appendix of V06).

As shown in Fig. 15, our $M_{500} - T_{500}$ relation is offset
from the true $M_{500} - T_{500}$ relation in the NKV07 simulations, from
33\% lower at 1 keV to 9\% lower at 10 keV. The agreement with the Borgani et al.
(2004) simulations is better, from 18\% lower at 1 keV to 6\% lower at 10 keV.
Interestingly, the $M_{\rm 500, HSE} - T_{500}$ relation from the NKV07
simulations is almost the same as the \chandra\ $M_{500} - T_{500}$ relation
from 23 groups + 14 V08 clusters. NKV07 attribute turbulence as the extra pressure
to deviate the HSE mass from the true mass (also see Rasia et al. 2004; Kay et al.
2004). Indeed for the ICM without a magnetic field, the dynamic
viscosity is roughly proportional to $T_{\rm ICM}^{2.5}$. Thus, it may not
be surprising that cool groups can develop stronger turbulence. However,
the ICM is magnetized and the real magnitude of the ICM turbulence is unknown.
The NKV07 simulations only have numerical viscosity that is small. 
Simulations with viscosity at different strengths are required to better
determine this bias term. High resolution X-ray spectra of the ICM may be
ultimately required to constrain the turbulence pressure in the ICM.

Kravtsov et al. (2006) suggested a new mass proxy, the $Y_{\rm X}$
parameter (product of the gas temperature and the gas mass derived from the
X-ray image, or $M_{\rm gas, 500} T_{500}$ in this work), which in simulated
clusters has a remarkably low scatter of only 5\%-7\%, regardless of whether
the clusters are relaxed or not. The agreement between simulations and observations
is also better for the $M - Y_{\rm X}$ relation than the $M - T$ relation (NKV07).
We examined the $M_{500} - Y_{\rm X, 500}$ relation for 23 groups + 14 V08 clusters
(Fig. 15 and Table 7). Our results indicate that a single power law
relation can fit the data very well. Our best-fit (0.57$\pm$0.01) is the same as
the V08 best-fit at $Y_{\rm X} > 2\times10^{13} M_{\odot}$ keV (0.57$\pm$0.05),
implying the groups aligned well with clusters.
Our best-fit is also consistent with the \xmm\ result by Arnaud et al. (2007)
(10 clusters at $Y_{\rm X} > 10^{13} M_{\odot}$ keV), on both the slope
(0.548$\pm$0.027) and the normalization (within $\sim$ 3\%).
Maughan et al. (2007) assembled 12 clusters at $z=0.14-0.6$ from their work and
literature ($Y_{\rm X} > 8\times10^{13} M_{\odot}$ keV) and found that the slope of the
$M_{500} - Y_{\rm X, 500}$
relation is consistent with the fit to the V06 clusters (0.564$\pm$0.009).
Intrinsic scatter in both the $M-T$ and $M - Y_{\rm X}$
relations are consistent with zero as the measurement errors for groups are large.
If we simply move the best-fit lines up or down to estimate the range of the scatter
from the best-fit mass values, the scatter in the $M - Y_{\rm X}$ relation is about the
half that in the $M - T$ relation. The slope is very close to the self-similar value of
0.6, especially if only tier 1 groups are included (0.588$\pm$0.012).
Our best fits lie between the true mass and the HSE mass from the NKV07 simulations,
but the offset is much smaller than that in the $M_{500} - T_{500}$ relation.
Thus, the $Y_{\rm X}$ parameter appears to be a robust mass proxy down to at least
2 $\times 10^{13}$ h$^{-1}$ M$_{\odot}$.
With the derived $M-T$ and $M - Y_{\rm X}$ relations in this work, the slope of
the $M_{\rm gas, 500}-T_{500}$ relation is 1.89$\pm$0.05 (for tier 1+2 groups +
V08 clusters), or 1.86$\pm$0.06 (for tier 1 groups + V08 clusters).
This value of the slope is consistent with the result by Mohr et al. (1999),
1.98$\pm$0.18 (90\% confidence).

\subsection{Gas fraction}

We derived the enclosed gas fraction profile for each group. The enclosed gas fraction
generally increases with radius and this trend continues to the outermost radius
in our analysis, as generally found in V06. The enclosed gas fractions for groups at
$r_{2500}$ have a large scatter (Fig. 17). For groups with similar $T_{500}$,
$f_{\rm gas, 2500}$ can be different by a factor as large as 2.5.
Both the weighted average and the median of $f_{\rm gas, 2500}$ in our sample is
0.043, much smaller than the typical value of $\sim$ 0.09 for V08 clusters.
This mean can be compared with the average $f_{\rm gas, 2500}$ by G07 (0.050$\pm$0.011),
and we note that the 16 groups in G07 are on average brighter than our
systems in a similar redshift range.
The intrinsic scatter of $f_{\rm gas, 2500}$ is tightly correlated with the intrinsic
scatter of $K_{\rm 0.15 r500}$, after the temperature dependence of both variables are removed
from their relations with temperature (Fig. 18).
Groups with low $f_{\rm gas, 2500}$ have high $K_{\rm 0.15 r500}$, relative to
the average relations. Thus, the large scatter of gas fraction within $r_{2500}$ for
groups is tightly correlated with the large scatter of entropy at 0.15 $r_{500}$,
and likely also the large scatter of X-ray luminosities for groups. Group properties (e.g.,
luminosity and central entropy) have large scatter because of the large scatter of
the gas fraction around the center (e.g., $r_{2500}$). Groups are on average
fainter than what is expected from the self-similar $L - T$ relation because groups
are generally ``gas-poor'' within $r_{2500}$, compared with clusters.

The enclosed gas fraction within $r_{500}$ is also derived for tier 1 + 2 groups
(Fig. 19). We added 14 clusters from V06 and V08 to constrain the $f_{\rm gas} - T_{500}$
relations. The $f_{\rm gas, 500} - T_{500}$ relation has a slope of $\sim$ 0.16 - 0.22,
depending on whether both tier 1 and 2 are included (Fig. 19). 
We also give the $f_{\rm gas, 500} - M_{500}$ relation from the BCES orthogonal fit.
For tier 1 + 2 groups + clusters:

\begin{equation}
f_{\rm gas, 500} = (0.0616\pm0.0060) \hspace{0.1cm} h_{73}^{-1.5} (\frac{M_{500}}{10^{13} h_{73}^{-1} M_{\odot}})^{0.135\pm0.030}
\end{equation}

For tier 1 groups + clusters:

\begin{equation}
f_{\rm gas, 500} = (0.0724\pm0.0078) \hspace{0.1cm} h_{73}^{-1.5} (\frac{M_{500}}{10^{13} h_{73}^{-1} M_{\odot}})^{0.093\pm0.031}
\end{equation}

We notice that the
NKV07 simulations predict $f_{\rm gas, 500} \propto T_{500}^{0.152}$ (or $M_{500}^{0.10}$),
from their
best-fit $M_{500} - T_{500}$ and $M_{500} - Y_{\rm X, 500}$ relations. The gas fraction
predicted in simulations depends on the modeling of cooling (e.g., Kravtsov et al. 2005)
and is often tangled with the problem of predicting the right stellar mass fraction
in clusters. We also derived the enclosed gas fraction between $r_{2500}$ and
$r_{500}$ for 23 groups in tiers 1 and 2 (Fig. 19). Combined with the V06 and V08 results
for clusters, the average $f_{\rm gas} - T_{500}$ has little or no temperature
dependence with an average value of $\sim$ 0.12, although the measurement errors are not small.
$f_{\rm gas, 2500-500}$ can also be derived as:

\begin{eqnarray}
f_{\rm gas, 2500-500} = f_{\rm gas, 500} (\frac{1 - a \frac{f_{\rm gas, 2500}}{f_{\rm gas, 500}}}{1 - a}) \\
  ( a = M_{2500} / M_{500} = 5 (r_{2500}/r_{500})^3 ) \nonumber
\end{eqnarray}

We use: $f_{\rm gas, 2500} = 0.0347 T_{500}^{0.509}$ (from the BCES fit to 43 groups
and 14 clusters, Fig. 17) and $f_{\rm gas, 500} = 0.0708 T_{500}^{0.220}$ (from the
BCES fit to 23 groups and 14 clusters, Fig. 19). Combining our results on
$c_{500}$ ($\S$7.3) with V06's for clusters, roughly we have
$c_{500} = 5.0 (M_{500} / 10^{13} M_{\odot})^{-0.09}$.
Assuming an NFW profile, the $r_{2500}/r_{500}$ ratio can be well approximated as:
$r_{2500}/r_{500} \approx$ 0.322 + 0.178 $lg(1.523 c_{500})$ at $c_{500} = 1.3 - 5.0$.
Thus, combined with the $M_{500} - T_{500}$ relation derived in this work, we can
estimate the $f_{\rm gas, 2500-500} - T_{500}$ relation. Indeed as shown in Fig. 19,
$f_{\rm gas, 2500-500}$ is nearly constant at 1 - 10 keV.
The average $f_{\rm gas, 2500-500}$ is still $\sim$ 27\% lower than the universal
baryon fraction (0.1669$\pm$0.0063 from Komatsu et al. 2008). However, one should
be aware that the enclosed gas fraction still rises beyond $r_{500}$, as generally
found in V06 and this work. We also notice that the observed $f_{\rm gas, 2500-500}$
is consistent with what was found in the simulations of Kravtsov et al. (2005) with
cooling and star formation. We conclude that
the low gas fraction generally observed in groups is mainly due to the low
gas fraction of groups within $r_{2500}$, or the generally weak ability of the
group gas to stay within $r_{2500}$. Beyond $r_{2500}$, the groups
are more regular and more similar to hot clusters, as also shown by the smaller
scatter of their entropy values at $r \gsim r_{2500}$.

A natural question motivated by these results on entropy and gas fraction is:
what is the fraction of the group luminosities within 0.15 $r_{500}$ or $r_{2500}$?
Detailed work on the group $L_{\rm X} - T$ relation is beyond the scope of this paper
and will be presented in a subsequent paper with an extended sample (including
non-relaxed groups). Here we give the curves of the enclosed count fluxes for
17 groups that $r_{500}$ is reached by \chandra\ or (and) PSPC (Fig. 20).
As the previous $L_{\rm X} - T$ relations only used the global spectrum to convert
the count rate to the group flux, Fig. 20 can be regarded as the growth curves of the enclosed
group luminosities. At 0.15 $r_{500}$, the fraction ranges from 13\% to 69\%.
At $r_{2500}$, the fraction ranges from 51\% to 94\%. Two extreme cases defining the
boundaries are A1238 and AS1101, also two systems with similar $T_{500}$ but very
different $f_{\rm gas, 2500}$ (Table 3). A system with a bright cool core (like
AS1101) has a large fraction of the X-ray luminosity within 0.15 $r_{500}$.
Its system temperature without excluding the central core (e.g., 0.15 $r_{500}$)
therefore is biased to a lower value.
A system without a bright cool core (like A1238) has a small fraction of the X-ray
luminosity within 0.15 $r_{500}$, and its system temperature without excluding
the central core (e.g., 0.15 $r_{500}$) is usually biased to a higher value
(Fig. 3-5). All these factors contribute to the large scatter of the group
$L_{\rm X} - T$ relation (e.g., Osmond et al. 2004). Thus,
the group $L_{\rm X} - T$ relation can be significantly contaminated by the large
difference in the cores (e.g., within 0.15 $r_{500}$). 

\subsection{$c_{500}$}

We also fitted the total mass density profile with the NFW profile and derived the concentration
parameter, $c_{500} = r_{500} / r_{\rm s}$, where $r_{\rm s}$ is the characteristic
radius of the NFW profile. V06 used an inner radius of 0.05 $r_{500}$, since the stellar
mass of the cD is dominant in the center. The V06 sample is mainly composed of clusters.
The groups in our sample have $r_{500}$ of 440 - 800 kpc, so the contribution
of the stellar mass at 0.05 $r_{500}$ is still significant for low temperature systems
(see e.g., G07). Thus, we use a fixed inner radius of 40 kpc. The outer radius is
the outermost radius for the spectral analysis ($r_{\rm det, spe}$ in Table 2). 
We derived the total mass with 1 $\sigma$ uncertainties at radii corresponding to
the boundaries of radial bins for spectral analysis between 40 kpc and $r_{\rm det, spe}$
(Fig. 3-5). The resulting mass density profile (at 4 - 10 radial points in this work) is fitted
with an NFW profile and the uncertainty is estimated from 1000 Monte Carlo simulations.
The results for 33 groups are present in Table 3 and are plotted with the system mass in Fig. 21.
For the other 10 groups, $c_{500}$ is very poorly constrained.
Our errors on $c_{500}$ are larger than those in V06 and G07 as our errors on mass
are larger. 

We compare our results with the best-fit $c - M$ relation from G07.
G07 gave the best-fit $c_{\rm vir} - M_{\rm vir}$ relation. For their range of
$c_{\rm vir}$, $c_{500} \sim 0.51 c_{\rm vir}$ (for $c_{\rm vir}$ = 10.35). For
$\Delta=101$, we convert G07's best-fit relation to: $c_{500} (1+z) = 
3.96 (M_{500}/10^{14} M_{\odot})^{-0.226}$ (adjusted to our cosmology).
As shown in Fig. 21, at $M_{500} > 4.5\times10^{13} M_{\odot}$, the G07 fit
describes our results very well. But our results do not show significant mass
dependence, so at $M_{500} < 4.5\times10^{13} M_{\odot}$, our results are
systematically below the G07 fit, although the errors are not small. 
However, the difference mainly comes from three groups (NGC~1550, NGC~533 and NGC~5129)
for which G07 found $r_{\rm s}$ = 41 - 46 kpc (adjusted to our cosmology), and our
inner radius cut at 40 kpc prevents us from measuring such a small value of $r_{\rm s}$.
In fact, the overdensity radii
of these three groups agree better between G07 and this work:
$r_{2500} = 206\pm2$ kpc (G07) vs. 222$\pm$6 from this work for NGC~1550,
$r_{1250} = 251\pm2$ kpc (G07) vs. 275$\pm$30 for NGC~533,
$r_{1250} = 217\pm7$ kpc (G07) vs. 236$\pm$13 for NGC~5129.
Excluding these three groups, our results agree well with G07's. Thus, the difference
mainly hinges on the determination of $r_{\rm s}$ that is related to the subtraction
of stellar mass, while the results at large radii agree better.
We also compare our results with the simulations of Bullock et al. (2001) and further work
\footnote{http://www.physics.uci.edu/\~{{}}bullock/CVIR/}. As $c_{500}$ is sensitive
to the halo formation time, smaller $\sigma_{8}$, $\Omega$$_{\rm M}$ and tilt
drive $c_{500}$ smaller. We used the parameters derived in Komatsu et al. (2008)
(see the caption of Fig. 21). As shown in Fig. 21, our results are generally consistent with
the Bullock et al. simulations. Detailed discussions on the difference of the observed
concentration parameter from the prediction can be found in Buote et al. (2007).
We should point out that both our analysis and the V06 analysis do not subtract the
stellar mass and the X-ray gas mass, while G07 subtracted both components.
G07 also included the \xmm\ data for most of their groups.
Future work on $c_{500}$ may need very good measurement of the gas properties to $> r_{500}$
and careful modeling of the stellar and gas components (e.g., G07). The group dark matter
mass profile may also need to be examined first to see whether a single NFW profile
is the best fit.

\subsection{The baryon fraction and fossil groups in this sample}

We can estimate the enclosed baryon fraction from the cluster stellar fraction
estimated before. Lin et al. (2003) gave: $M_{*, 500} = 7.30\times10^{11} M_{\odot}
(T_{\rm X} / 1 {\rm keV})^{1.169}$ (adjusted to our cosmology) from the 2MASS
data of nearby groups and clusters, which includes the stellar mass in cluster galaxies.
With the $M - T$ relation derived in this work from 23 groups and 14 V08 clusters,
$f_{\rm *, 500, Lin} = 0.0263 (T_{\rm X} / 1 {\rm keV})^{-0.481}$.
Gonzalez et al. (2007) included the intracluster stellar mass and found:
$f_{\rm *, 500, Gonzalez} = 0.0380 (M_{500} / 10^{14} M_{\odot})^{-0.64}$ (adjusted to our cosmology).
With our $M_{500} - T_{500}$ relation,
$f_{\rm *, 500, Gonzalez} = 0.0864 (T_{500} / 1 {\rm keV})^{-1.056}$.
Adding the relation for $f_{\rm gas, 500}$, the total baryon fraction from groups
to clusters can be estimated. As shown in Fig. 19, there is substantial difference
for groups. 
We also examined the stellar mass of the cD and its relation with
other group properties. As shown in Fig. 22, the stellar mass of the group cD (which is
proportional to its $K_{s}$ band luminosity) is weakly correlated with the system mass.
Low-mass cDs generally reside in low-mass groups. We also examined the relation between
$L_{Ks}$ of the cD and $f_{\rm gas, 2500}$ but found no correlation. However, in low-mass
systems with low $f_{\rm gas, 2500}$, the stellar mass of the cD can be comparable to
the gas mass within $r_{2500}$.

We also searched for fossil groups in this sample. From Jones et al. (2003), fossil
groups are defined as a bound system of galaxies with the $R$ band magnitude difference
of the two brightest galaxies within half the virial radius larger than 2 mag.
In this work, because we do not have homogeneous $R$ band magnitudes for group galaxies
in this sample, we used the 2MASS $K_{\rm s}$ band magnitude, which is
a good indicator of the stellar mass. To ensure large and blue spirals are not left
out, we also checked NED and HyperLeda to examine the $B$ band magnitude difference.
Jones et al. (2003) used $r_{200}$ as the virial radius, while we use the exact
definition of the virial radius for our cosmology, $r_{\rm vir}$ = $r_{\Delta}$ ($\Delta$
$\sim$ 101). For the typical mass concentration of groups in this sample,
0.5 $r_{\rm vir} \sim r_{500}$. Thus, we examined the $K_{\rm s}$ and $B$ band magnitude
difference for group galaxies within $r_{500}$.
Six fossil groups are selected: NGC~741, ESO~306-017, RXJ~1159+5531, NGC~3402, ESO~552-020
and ESO~351-021. RXJ~1159+5531 and ESO~306-017 are known fossil groups (Vikhlinin et al. 1999;
Sun et al. 2004). We note that NGC~1132 was considered a fossil group (Mulchaey \& Zabludoff,
1999). However, NGC~1126 is 8.4$'$ from NGC~1132 with a velocity difference of 438 km/s
($r_{500}=16.2'$ or 440 kpc for the NGC~1132 group). The 2MASS $K_{\rm s}$ magnitude
difference is 1.49 mag and the $B$ magnitude difference is 2.07 mag. Thus, we do not
consider NGC~1132 a fossil group based on the Jones et al. (2003) definition.

Although the sample is small and not representative, we examined whether these
fossil groups populate a different position of the scaling relations than
non-fossil groups. No significant difference in $K - T$, $f_{\rm gas} - T$ and $M - T$
relations is found. As shown in Fig. 22, these fossil groups indeed have massive cDs.
The $c_{500}$ of these fossil groups (four listed in Table 3, others poorly constrained)
are $\sim$ 2.4 - 4.3, on average smaller than the average of this sample, which differs
from the claim by Khosroshahi et al. (2007) that fossil groups have higher mass
concentration than non-fossil systems.

\section{AGN heating and Radio galaxies}

\subsection{Sign of heating in entropy profiles}

Besides pre-heating (from SNe and AGN) and cooling, impulsive heating from
the central AGN is often required to explain the observed scaling relations
like $L - T$ and $K - T$ (e.g., Lapi et al. 2005). Strong shocks driven by the
central AGN may boost the ICM entropy and create an entropy bump. This transient
anomaly in the entropy profile may be detected in a large sample. We have searched
for such entropy features in our sample and find two promising cases: UGC~2755 and
3C~449 (Fig. 23). There are also three groups with a significant break observed in their
surface brightness profiles (3C442A, IC1262 and A2462) that may be related to AGN
heating. We briefly discuss them in the Appendix.
Both UGC~2755 and 3C~449 host a strong FR I radio source with two-sided radio lobes.
UGC~2755 has a central corona with a radius of $\sim$ 3 kpc (see Sun et al. 2007 for
the connection of thermal coronae with strong radio sources).
From 10 kpc to 80 kpc in radius, the surface brightness profile
is very flat. Then there is a sharp break at $\sim$ 90 kpc.
UGC~2755's radio lobes extend to $\sim$ 100 kpc from the nucleus in the NVSS image,
which may naturally explain the entropy bump within $\sim$ 90 kpc.
3C449 also has a central corona with a radius of $\sim$ 3 kpc.
Its entropy bump is at $\sim$ 40 kpc - 100 kpc and less significant as
that in UGC~2755. 3C449's radio outflow is spectacular and can
be traced to at least 1.6 Mpc in radius from the NVSS image (compared to its
$r_{500}$ of 433 kpc estimated from $r - T$ relation).
Nevertheless, the brighter inner part of the radio jets/lobes
ends at $\sim$ 100 kpc from the nucleus. Thus, the entropy bump we observed
in 3C~449 may represent the most recent heating event.

Strong shocks are required to effectively boost entropy. Using the
standard Rankine-Hugoniot conditions, the entropy increase after shocks is a
sensitive function of the shock Mach number, and weak shocks have little effect
on amplifying entropy. A shock with
a Mach number of 1.2 only increases entropy by 0.44\%. A Mach 2 shock
increases entropy by 20\%, while Mach 3.3 and 4.4 shocks increase
entropy by 100\% and 200\% respectively.
The entropy bump in 3C449 is only at the level of 20\%, which can be produced
by a single Mach 2 shock. The entropy boost in UGC~2755 is 1.5 - 2 times,
which can be produced by a single Mach 3 shock. The adiabatic sound speed
in groups is not high, 540 - 630 km/s in 1.1 - 1.5 keV ICM for these two groups.
In these low density groups, the ambient pressure is much lower than that in
the dense cores of hot clusters. Shock deceleration may also be slower.
The velocity of the outflow-driven shock
is $\sim f_{P} (P_{\rm kin}/\rho r^{2})^{1/3}$, where $\rho$ is the ICM density and
$f_{P}$ is a structure factor of order unity that depends on the outflow geometry
and the preshock density profile (e.g., Ostriker \& McKee 1988). $n_{e} \sim
10^{-3}$ cm$^{-3}$ around the entropy bumps and $P_{\rm kin} \sim 10^{44}$
ergs s$^{-1}$ from the radio luminosities of two radio AGN and the relation
derived by B\^{i}rzan et al. (2004). The estimated velocity is then
$\sim 1300 f_{P}$ km/s, which is comparable to the requirement of
the entropy boost. Therefore, radio outflows in these two groups are capable of
driving Mach 2 - 3 shocks to produce the observed entropy bumps.

\subsection{Strong central radio sources in this sample}

There are ten groups hosting a central radio source that is more luminous
than $L_{\rm 1.4 GHz} = 10^{24}$ W Hz$^{-1}$:
3C~31, 3C~449, UGC~2755, 3C~442A, A160, A2717, AS1101, A3880, A1238 and A2462.
What is the typical X-ray gas environment around these radio sources?
Six of them lack large cool cores (e.g., $\gsim$ 30 kpc radius), 3C31, 3C~449, UGC~2755, A160,
A1238 and A2462. However, all of them host a central corona with a radius of $\sim$
3 - 8 kpc, typical for massive cluster and group galaxies as discussed in
Sun et al. (2007).
This component is reflected in their temperature profile, except for
A1238 as it is faint. The spectrum of A1238's central source can be
described by a $\sim$ 0.8 keV thermal component. Its X-ray luminosity and
$K_{\rm s}$ band luminosity fall on the typical region for coronae and
its properties are similar to ESO~137-006 in A3627 (a nearby bright corona, Sun et al. 2007).
Four other groups (3C442A, A2717, A3880 and AS1101) host larger cool cores
with a radius of $\sim$ 30 kpc or larger. The cool cores in 3C442A and
A2717 are clearly disrupted, likely by the radio sources.
Thus, all these strong radio sources have low-entropy ICM ($< 30$ keV cm$^{2}$) at
the center.

\section{Systematic errors}

We follow Humphrey et al. (2006), G07 and V08 to discuss the systematic error
budgets in our results. The uncertainties of the local X-ray background are the
main error budget at large radii. As shown in $\S$3.2, we have included a
conservative estimate of the background uncertainties into the errors of
temperatures and densities. This is the primary reason that our results
at large radii have larger errors than those of V06 for the same systems.

We used the LAB survey data (Kalberla et al. 2005) for the galactic hydrogen
column density (Table 2) and examined the \chandra\ spectra for excess absorption.
The LAB column density is on average $\sim$ 6\% lower than the column density
from Dickey \& Lockman (1990). Out of 43 groups we studied, 19 show significant
excess absorption. We can compare this fraction to that of V05, 6 out of 13 with
excess absorption. For the five groups that were studied in V05 and this
work, both works find excess absorption for the same three groups.
On the other hand, G07 used the galactic hydrogen column density from Dickey \& Lockman
(1990) for all groups.
The incidence of excess absorption in our work increases with the galactic hydrogen
column density. At $N_{\rm 21 cm} > 4\times10^{20}$ cm$^{-2}$, eight of eleven
groups (most of them at $z < 0.03$) show excess absorption. This trend is
qualitatively consistent with the result by Arabadjis \& Bregman (1999).
We also examined the effects of a conservative $N_{\rm H}$ uncertainty of $\pm2\times10^{20}$
cm$^{-2}$ (see e.g., V08) on our results.
For $T <$ 1.6 keV gas, the determination of temperature is little affected by  
absorption. An $N_{\rm H}$ change of $\pm2\times10^{20}$ affects the temperature
by $\sim \mp$ 1.4\% and the density by $\sim \pm$ 6.1\%. The subsequent changes
on $K, M, Y_{\rm X}$ and $f_{\rm gas}$ at an overdensity radius are $\sim \mp$6.3\%,
$\mp$2.1\%, $\pm$3.9\% and $\pm$7.6\% respectively, assuming an entropy slope
of 0.7 (note that the overdensity radius also depends on mass).
For $T >$ 1.6 keV gas, an $N_{\rm H}$ change of $\pm2\times10^{20}$ affects the
temperature by $\sim \mp$ 5.3\% and the density by $\sim \pm$ 3.7\%.
The subsequent changes
on $K, M, Y_{\rm X}$ and $f_{\rm gas}$ at an overdensity radius are $\sim \mp$5.9\%,
$\mp$8.0\%, $\mp$4.4\% and $\pm$9.7\% respectively.
At small radii, uncertainties at these levels are not important as they are
smaller than the intrinsic scatter (e.g., $K_{\rm 0.15 r500}$ and $f_{\rm gas, 2500}$).
At large radii, the statistical errors (including uncertainties from the local X-ray
background) overwhelm. Nevertheless, the systematic errors from $N_{\rm H}$ should
be kept in mind.

We use the deprojection algorithm derived by Vikhlinin (2006). The form of the
3D temperature profile is the same as that used in V06.
The robustness of this deprojection algorithm has been presented in Vikhlinin (2006),
with uncertainties of $\sim$ 0.05 keV. Nagai et al. (2007a) presented mock \chandra\
analysis of cluster simulations, using the deprojection algorithm proposed by
Vikhlinin (2006) and the form of the temperature profile used in V06. The best-fit
3D temperature profile is well consistent with the true temperature profile in simulations,
with residuals in a similar level as shown in Vikhlinin (2006). This test further
validates the approach used in the present work. We have also performed a test
on a sub-sample of 6 groups with best-quality data. The traditional onion-peeling
method (e.g., McLaughlin 1999; G07) was used. The resulting deprojected temperature
profiles often show small magnitude of oscillating. If we use the form of equ (2)
to fit them, the best-fits are consistent with those shown in Fig. 3 - 5 within
$\sim$ 0.05 keV on average. The uncertainties on temperatures range from 2\% - 6\%,
while the uncertainties on densities are 0.5\% - 3\% (both directions).
Thus, this systematic error (larger for $<$ 1 keV groups) are smaller than the
systematic error related to the uncertainty of $N_{\rm H}$.

There are other systematic errors, like the choice of the plasma spectral codes,
and departures from spherical symmetry in the group gas. Those are either small
in magnitude or have little impact on the scaling relations, as discussed in
Humphrey et al. (2006), G07 and V08. As mock data from simulations are starting
to be analyzed in the same way as the observational data (Nagai et al. 2007a;
Rasia et al. 2008), these factors are becoming better controlled.

\section{Summary and conclusions}

We present an analysis of 43 galaxy groups with \chandra\ observations. With inclusion of
many faint systems (e.g., the ones hosting strong radio sources), our sample is not
much biased to X-ray luminous groups, as shown by the wide range of ICM entropy
values around the group center ($\S$6). We used the ACIS stowed background and
modeled the local CXB for each group. Uncertainties of local background are folded
into the derived temperature and density profiles. The projected temperature profile and
the surface brightness profile are modeled with sophisticated models, which have
enough freedom to describe the data from the core to the outskirts. The 3D abundance
profile is also derived. The 3D temperature and density profiles are constrained
through iterative fitting. The uncertainties are estimated from 1000 Monte Carlo
simulations. Gas properties are derived to at least $r_{2500}$ for all 43 groups.
For 11 groups, we can derive gas properties robustly to $r_{500}$. For another
12 groups, gas properties can be robustly derived to $\gsim r_{1000}$, so we
extrapolate the results to $r_{500}$. The main results of this paper are:

1) We present the $M_{500} - T_{500}$ and $M_{500} - Y_{\rm X, 500}$ relations in
$M_{500} = 10^{13} h^{-1} M_{\odot}$ - 10$^{15} h^{-1} M_{\odot}$, combined
with the V08 results on 14 $T_{500} >$ 3.7 keV clusters. Both relations
are well behaved at the low-mass end and can be well fitted with a single power
law ($\S$7.1 and Fig. 15). The $M_{500} - Y_{\rm X, 500}$ relation indeed
has a smaller scatter than the $M_{500} - T_{500}$ relation (about half).
The $M_{500} - T_{500}$ relation from observations is still offset
from simulations (e.g., NKV07). Interestingly, the \chandra\ $M_{500} - T_{500}$ relation
is very close to the $M_{\rm 500, HSE} - T_{500}$ relation in the NKV07 simulations.
Although it is tempting to attribute the difference to a mass
bias, better understanding of the ICM viscosity is required.

2) The group gas fraction within $r_{2500}$ is on average much smaller than
that of clusters (e.g.,  $\sim$ 0.043 for $\sim$ 1.5 keV groups vs. $\sim$ 0.09
for $\sim$ 6 keV clusters, $\S$7.2 and Fig. 17), which is consistent with G07's
results. The group gas fraction within
$r_{2500}$ also has a large scatter, spanning a factor of $\sim$ 2 at any fixed
temperature. On the other hand, the gas fraction measured between $r_{2500}$ and
$r_{500}$ has no temperature dependence with an average value of
$\sim$ 0.12 (Fig. 19). Thus, the generally low gas
fraction in groups is due to the general low gas fraction within $r_{2500}$.

3) We derived the $K - T$ relations at 30 kpc, 0.15 $r_{500}$, $r_{2500}$,
$r_{1500}$, $r_{1000}$ and $r_{500}$ ($\S$6.1, Fig. 11). The large scatter of
the entropy values at 30 kpc
and 0.15 $r_{500}$ reflect the wide luminosity range of groups in this sample.
The $K - T$ relation is significantly tighter beyond $r_{2500}$ and the intrinsic
scatter of entropy is the same at 10\% - 11\% from $r_{2500}$ to $r_{500}$.
Thus, the group properties are more regular from $r_{2500}$ outward, in line with the
gas fraction results.
With 14 clusters from V06 and V08 included, we also present $K - T$ relations
in the full temperature range (Fig. 12 and Table 5). At $r_{500}$, the slope
of the $K - T$ relation is consistent with the value from self-similar
relation (1.0).

4) The ratios of the observed entropy values to the baseline values (from
adiabatic simulations) decrease with radius ($\S$6.2 and Fig. 13). At $r_{2500}$,
the ratio ranges
from $\sim$ 1.8 - 3.5, with a weighted mean of 2.2. The weighted mean decreases
to $\sim$ 1.8 at $r_{1000}$ and $\sim$ 1.6 at $r_{500}$. The still significant
entropy excess at $r_{500}$ in groups may require a diffuse way to distribute
heat (e.g., AGN heating, see $\S$8.1), but it may also be understood with smoother accretion
and the mass bias that may be especially large in groups ($\S$6.2).
The entropy excess at $r_{500}$ is also detected for 14 clusters from V06 and
V08 ($\sim$ 35\%, $\S$6.2). The difference in the entropy excess at $r_{500}$ between groups
and clusters ($\sim$ 17\%) is not as large as previously claimed from the \rosat\
and \asca\ data.

5) The entropy slopes are determined at 30 kpc - 0.15 $r_{500}$, 0.15 $r_{500}$ -
$r_{2500}$, $r_{2500}$ - $r_{1500}$ and $r_{1500}$ - $r_{\rm det, spe}$ ($\S$6.3
and Fig. 14) . The slopes are all shallower than 1.1 beyond 0.15 $r_{500}$.
Scatter is large but the average slope is $\sim$ 0.7 beyond 0.15 $r_{500}$.

6) The group temperature profiles are similar at $> 0.15 r_{500}$, despite large
scatter within 0.15 $r_{500}$ ($\S$5 and Fig. 8). The average slope is consistent
with the ``Universal temperature profile'' (Markevitch et al. 1998; De Grandi \& Molendi
2002; Loken et al. 2002; V05; Pratt et al. 2007; G07; Rasmussen \& Ponman 2007; LM08)
but there is still scatter. The group temperature profiles also appear more peaky
than those of clusters (Fig. 9).

7) We also derived the concentration parameter ($c_{500}$) for 33 groups ($\S$7.3).
Our results are generally consistent with the Bullock et al. (2001) simulations under
the current WMAP5+SN+BAO cosmology.

8) We selected six fossil groups in this sample and four are new. The X-ray gas properties
of these fossil groups have no significant difference from non-fossil groups
($\S$7.4) in scaling relations.

9) We found two groups with substantial entropy bumps ($\S$8.1), which may indicate
a recent strong heating episode. Both host strong radio galaxies at the center and we
estimate that the radio AGN is capable of driving shocks to boost entropy to the
observed level.

The emerging picture of groups from this work is that the main difference
between groups and hotter clusters is the general weak ability of the group gas to stay
within $r_{2500}$, which explains most of deviation of the group properties
from the self-similar relations (e.g., entropy and luminosity).
The group properties within $r_{2500}$ have large scatter, but may all be
related to variations in the level of the enclosed gas fraction within $r_{2500}$.
Beyond $r_{2500}$, groups are more regular and more like clusters, making
them promising tools for cosmology, as shown by the well-behaved $M - T$
and $M - Y_{\rm X}$ relations derived in this work.

\acknowledgments

This work would be impossible without the rich data sets in the \chandra\ archive.
We would like to thank all PIs of the \chandra\ observations for their
original efforts. We thank M. Markevitch for his help on the ACIS stowed background.
We thank the referee, D. Buote, for helpful and prompt comments.
We thank helpful discussions with M. Balogh, S. Borgani, D. Nagai and G. Pratt.
The financial support for this work was provided by the NASA LTSA grant NNG-05GD82G.
We made use of the NASA/IPAC Extragalactic Database and the HyperLeda database.

\begin{appendix}

\subsubsection{Components of the \chandra\ background}

The \chandra\ background has been extensively discussed before (e.g., Markevitch et al.
2003; Wargelin et al. 2004; V05; HM06; Humphrey \& Buote 2006). We present here
a brief summary, focusing in particular on the cosmic X-ray background (CXB). There are
two basic components in the quiescent \chandra\ background, particles and photons
(or CXB). The dominant background component is the charged particle background
(PB), which is not vignetted. The spectrum of the \chandra\ PB has been remarkably
stable since 2000 (Vikhlinin et al. 2005, V05 hereafter; Hickox \& Markevitch 2006,
HM06 hereafter), although the absolute flux changes with time and is related to
the solar cycle. The only exception so far is for BI data
after the middle of 2005, which is discussed in the next section.

The cosmic hard X-ray background is considered to be composed of unresolved
X-ray point sources, mostly AGN.
This component can be described by an absorbed power-law with a photon index of $\sim$ 1.5
(HM06). Its flux depends on the level of point source excision or the limiting flux
for point sources. HM06 analyzed the two deepest \chandra\ fields at that time,
\chandra\ deep field north (a combined
clean exposure of 1.01 Ms) and south (a combined clean exposure of 0.57 Ms).
The unresolved hard X-ray background flux density is (3.4$\pm$1.7)$\times10^{-12}$ ergs
s$^{-1}$ cm$^{-2}$ deg$^{-2}$ in the 2 - 8 keV band, which represents the lower
limit of the hard CXB flux in shorter \chandra\ observations.
The X-ray $logN - logS$ relation has been well studied allowing us to predict
the average unresolved cosmic hard X-ray background below the point source limiting flux.
Kim et al. (2007; K07 hereafter) presented the average relation between point
source limiting flux and expected hard CXB flux density (Fig. 19 of K07).
For the limiting flux for point sources in the outermost bins of groups in our sample,
we expect a resolved fraction of 25\% - 65\% in the 2 - 8 keV band,
which corresponds to a flux density of 6.1 - 13.1 $\times10^{-12}$ ergs
cm$^{-2}$ s$^{-1}$ deg$^{-2}$ for the unresolved hard CXB background.
However, one should be aware that the K07 relation is just an average. About 30\%
variation can be expected over the angular scale we study (HM06).

The soft X-ray background is composed of several components, Galactic, local bubble,
geocoronal and heliospheric emission (e.g., Wargelin et al. 2004; Snowden et al. 2004;
Koutroumpa et al. 2007). The latter two components are primarily from the solar wind
charge exchange (SWCX) emission and are time variable, with a contribution to
the O VII and O VIII lines as much as the Galactic component (Koutroumpa et al. 2007).
The strength of the soft X-ray background varies
with the sky position (as shown in the R45, or 3/4 keV \rosat\ all sky survey map).
There has been lots of work done to quantify its spectral properties. With the
\chandra\ data, Markevitch et al. (2003) and HM06 have shown that, the soft
X-ray background beyond the regions with strong RASS R45 flux (e.g., the
North polar spur), can be well described by a single unabsorbed thermal component with
a temperature of $\sim$ 0.2 keV. Its spectrum typically shows a broad line hump
around 0.6 keV, mainly from the 0.57 keV O VII and 0.65 keV O VIII lines.
The soft X-ray background has also been studied with the \suzaku\ data, which
have the higher spectral resolution.
Fujimoto et al. (2007) analyzed the \suzaku\ data of the North Ecliptic Pole
region (R45 = 140 $\times10^{-6}$ counts s$^{-1}$ arcmin$^{-2}$) and found
that the soft X-ray component at the non-flare period has a temperature of 0.18 keV
with over-solar abundances.
Miller et al. (2008) analyzed the \suzaku\ data of the brightest
region of the North Polar Spur (NPS, R45 = 748 $\times10^{-6}$ counts s$^{-1}$
arcmin$^{-2}$) and found that the NPS thermal component has a temperature of
$\sim$ 0.28 keV with generally sub-solar abundances, besides the assumed 0.1 keV
local bubble and galactic halo components with solar abundance. Snowden et al. (2008)
adopted a soft CXB model composed of an unabsorbed $\sim$ 0.1 keV component for
the local hot bubble, an absorbed $\sim$ 0.1 keV component for the cooler Galactic
halo emission and an absorbed $\sim$ 0.25 keV component for the hotter Galactic
halo emission. All abundances are fixed at solar.
Thus, we have enough knowledge to model the soft X-ray background.

\subsubsection{``Blank sky background'' and stowed background}

Much previous \chandra\ work on clusters used the blank-sky
background data\footnote{http://cxc.harvard.edu/contrib/maxim/acisbg/}, which are
good enough in high surface brightness regions.
However, the averaged soft X-ray component in the standard blank-sky background
data is very likely different from the actual soft X-ray foreground
in any particular field, as pointed out previously by e.g., Markevitch et al. (2003),
V05, Humphrey \& Buote (2006) and G07. Taking the example of the blank sky background
data in period D,
the ACIS-I file combines 29 pointings from Jan., 2001 to Nov. 2004, at regions where
R45=90 - 150. The exposures range from 27 ks to 165 ks with a median of $\sim$ 66 ks.
The ACIS-S file combines 12 pointings from Nov. 2001 to Oct. 2003, at regions where
R45=90 - 150. The exposures range from 20 ks to 114 ks with a median of $\sim$ 30 ks.
In 5 of 12 pointings, there are no S1 data. The ACIS response has been changed
significantly from 2001 to 2004 (especially because of the contamination on the optical
blocking filter). Thus, it is often inadequate to only use the blank-sky background
to constrain the ICM properties at low surface brightness regions. Moreover, the PB
and the CXB are not separated in the blank-sky background data. When the blank-sky
background is scaled to account for the flux change of the PB, the CXB in the
blank-sky background data is unphysically scaled. While the PB rate did not change
much between the spring of 2000 and the spring of 2004, it has been significantly
increasing ever since. In 2006, the PB rate was on average 50\% higher that the average
value between the spring of 2000 and the spring of 2004 (Fig. 6.24 from \chandra's
Proposers' Observatory Guide v.9\footnote{http://asc.harvard.edu/proposer/POG/}).
Thus, any analysis for data taken after the middle of 2004 involves a large scaling
of the PB, often resulting significant over-subtraction of the CXB.
Therefore, a second correction besides subtracting the scaled blank-sky background
is required (e.g., Vikhlinin et al. 2005). This ``double subtraction''
is often efficient but requires the presence of source-free regions in the
\chandra\ field, which is not true for many nearby groups in our sample.
In this work, we utilize the newly available ACIS stowed background data
to subtract the PB component. The CXB
is modeled and the uncertainties are folded into the final error budgets.

Since Sep., 2002, ACIS observations have been carried out twice a year in the
stowed position, shielded from the sky by the science instrument module
structure and away from the on-board calibration source. By the end of May, 2007,
415 ksec data had been collected.
Background flares have never been observed in the stowed data.
The comparison with the dark moon observations indicates that the stowed background
is the same as the quiescent PB collected by the CCDs in the normal
focal position (Markevitch et al. 2003; the CXC calibration website  $^{8}$.).
HM06 also used the stowed background (236 ks at the time of their work)
to carry out absolute measurement of the unresolved CXB.
They show that between Jan. 30, 2000 and Sep., 2002 when the stowed background
data are not available, the spectral slope of the PB is the same.
Thus, we can apply the stowed background to early data.
The stowed background allows us to separate the non-vignetted PB from the vignetted CXB.
The main reason for our preference for the stowed background over the blank-sky
background is that we have better control of the local background for nearby
groups where sources fill the whole \chandra\ field. The ``double subtraction''
method with the blank-sky background cannot be applied for these nearby groups, as
there is no region that is free of group emission. The derived local X-ray background
based on stowed background also has a clear physical meaning and can be compared
between \chandra\ observations with very different PB fluxes or observations with
other telescopes like \xmm. Another subtle advantage of the stowed background
is related to the telemetry limit of \chandra, especially for the VFAINT mode data.
The blank-sky background data of each specific ACIS CCD may come from different
combination of observations, especially for ACIS-S (e.g., S1 vs. S2+I3). Therefore,
the residual or decremental background on the S1 CCD is in principle different
from that on FI chips, which complicate the analysis. Similarly, even if S3 is turned
on for ACIS-I observations, it cannot be used for local background study as the
blank-sky background of the S3 chip (when the aimpoint is on ACIS-I) is only a subset
of the ACIS-I blank-sky background data. On the contrary, with the stowed background,
we are analyzing the absolute CXB in the interested field so data on different CCDs
can be fitted jointly to make better constraints.
Therefore, we used the stowed background to subtract the PB in our observations.

The spectra of the \chandra\ PB have been very stable (e.g., HM06).
However, a small change of the spectral shape on the BI CCDs (S3 and S1)
has been identified from around the middle of 2005, while the
spectra of the PB on the FI chips still keep the same $^{8}$.
The change appears abrupt around the middle of 2005, while the spectral shape of
BI data remain the same from that time to at least the middle of 2007 (private
communication with Maxim Markevitch). The
stowed background data have been broken into 2 periods, one with 235 ksec total
exposures from five observations of Sep. 3, 2002 to June, 10, 2005 (period D), the other
with 180 ksec total exposure from four observations of Nov., 13, 2005 to May, 28, 2007
(period E). We emphasize that the notation
adopted here is only for the purpose of this work.
We examined the spectral difference between these two periods, after matching
their fluxes in the 9.5 - 12 keV band. The flux of the period D background is
always a little lower than that of the period E background, after re-scaling.
The biggest difference is seen on the S3 chip, with a 6.0\% difference in the
0.35-7.0 keV band (Fig. 24). The residual emission is very
flat (note it is non-vignetted) and can be removed by increasing the D period
background by 5.7\%. The difference is smaller on the S1 chip, 2.4\% in the
0.35-5.5 keV band (note that the S1 PB increases rapidly at $>$ 5.5
keV, Fig. 6.21 of \chandra's Proposers' Observatory Guide v.9). The difference
is consistent with zero in FI chips, $\sim$ 1\% in the 0.5-7.0 keV band,
excluding the Au line in the 2.0-2.3 keV band. Thus, we can apply the total
stowed background (415 ks exposures) to the FI data. For BI data, we use the
stowed background in their corresponding periods. There are only two groups in our sample
with BI data taken between June, 10, 2005 and Nov., 13, 2005 (NGC~1550 and NGC~5098).
Both were observed after Oct. 22, 2005 and both have earlier FI data. We used the
period E stowed background for the BI data of both groups.
In this work, we also take a larger uncertainty on the normalization of
period E PB for BI chips (5\% compared to 3\% for FI data and the period D BI data).

\section{Notes on some groups}

In this section, we present notes on some groups, mainly on the comparison
with previous work on the gas properties at large radii. Thus, the cited references
are usually not complete for each group as the detailed dynamical and thermal
structures of the group cores are beyond the scope of this work.

\paragraph{NGC 1550} was examined by Sun et al. (2003), with two ACIS-I observations.
Now with two additional longer ACIS-S exposures in the offset positions, the gas
properties in this system can be
constrained much better. NGC~1550's temperature profile is among the best determined
for 1 keV groups, with the good \chandra\ coverage. We can compare our results with
those from G07 who analyzed two short ACIS-I observations and an \xmm\ observation.
The temperature profiles agree well although we constrain the temperature to
larger radii from the ACIS-S observations. Our $r_{500}$ and $M_{500}$ are 10\% and
50\% higher than those derived
by G07. Our $c_{500}$ (4.93$^{+0.50}_{-0.46}$) is smaller than G07's (9.0$\pm$0.6).

\paragraph{NGC~3402} has been studied by V05 and V06. Although our temperature profile
is consistent with V05, the slope of the decline is smaller.
The \iras\ $100\mu$m map shows the presence of Galactic cirrus around the
group. We indeed derived a higher absorption column, 1.1$\pm0.1 \times 10^{21}$
cm$^{-2}$, than the Galactic value (4.0$\times10^{20}$ cm$^{-2}$ from LAB).
This value is smaller than that derived in V05, 1.55$\pm0.1 \times 10^{21}$
cm$^{-2}$. However, the absorption difference has little effect on the derived gas
temperature. As the gas temperature is mainly determined by blended line centroid,
gas temperature remains almost the same, with the higher $N_{\rm H}$ in V05.
Because of higher temperatures derived at large radii, our derived $M_{2500}$ is
larger than that from V06 and $f_{\rm gas, 2500}$ is smaller than in V06.

\paragraph{Abell~262} is a nearby luminous system in which the X-ray emission can
be traced to over 800 kpc in the 7.6 ks PSPC data. A262 was included in V06 and G07
samples. We included a new deep \chandra\ exposure (110 ks) taken in 2006 in our
analysis, while G07 also analyzed an \xmm\ observation. Our results of
$r_{2500}, M_{2500}$ and $f_{gas, 2500}$ are consistent with those in V06 and G07.
Our $c_{500}$ (3.48$^{+0.49}_{-0.45}$) is well consistent with V06's result
and is close to G07's result (4.5$\pm$0.4), but smaller than the result by
Piffaretti et al. (2005) (5.8$\pm$1.2).

\paragraph{NGC~383} hosts a bright FRI radio source 3C31.
There is a background cluster centered on 2MASX~J01065891+3209285 at
$z$=0.1116. We derived the surface brightness profile centered on the
background cluster and also analyzed a short exposure (ObsID 3555, 5.1 ks) 
centered on the background cluster. The cluster emission is detected to
$\sim 4'$ radius. In the analysis for NGC~383, we
excluded the region within 6.5$'$ of 2MASX~J01065891+3209285, which is a bit
larger than its $r_{500}$ (5.9$'$ for $kT=2.3$ keV).

\paragraph{3C~449} is located at a Galactic latitude of -16 deg and at the
outskirts of a bright \iras\ 100 $\mu$m feature across several degrees,
which should explain the enhanced absorption.
It is one of the two groups with an entropy bump detected ($\S$8.1).

\paragraph{NGC~533} was studied by Piffaretti et al. (2005) (\xmm) and G07
(\chandra\ + \xmm). The explored radial range in spectral analysis is
similar in all three work (up to 240 - 260 kpc).
Our temperature profile agrees well with that derived by G07. Our $c_{500}$
(4.58$^{+3.90}_{-2.34}$) is still consistent with the results by Piffaretti
et al. (2005) (8.6$\pm$0.7) and G07 (9.0$\pm$0.7).

\paragraph{MKW4} is bright enough that the group emission can be traced to
the very edge of the \chandra\ field. We are able to separate the group emission
and the soft background emission on the S1 spectrum because of the prominent
$\sim$ 0.6 keV hump in the soft background emission and the iron L hump in
the group emission. The same \chandra\ data had been analyzed by V05 and V06.
Our temperature profile agrees well with V05's, as well as properties at $r_{2500}$
(mass and gas fraction) with V06's. However, our results at $r_{500}$ differ from V06's.
The difference should lie on the modeling of the density profile. V06 
derived a very steep density profile at large radii of MKW4 ($\beta_{\rm eff, 500}$
= 0.92, Table 2 of V06), while we derived a value of $\sim$ 0.6, more similar
to A262 and A1991 (Table 2 of V06). We notice that Vikhlinin et al. (1999)
derived $\beta_{\rm outer}$ = 0.67$\pm$0.06 for MKW4, using the same PSPC
data.
G07 used the same \chandra\ data and also analyzed an \xmm\ observation of
MKW4. Our $r_{500}$ and $M_{500}$ are consistent with G07's values, but our
$c_{500}$ lies between the values of V06 and G07. G07 assumed an NFW profile for
the dark matter halo and only derived gas properties to 322 $h_{73}^{-1}$ kpc.
The properties at $r_{500}$ thus rely on the assumption of NFW profile and
extrapolation. 
On the other hand, the \chandra\ data at the outermost
bin only covers 10\% of the area in that annulus. Better constraints on the
properties of this nearby system require more coverage at large radii.

\paragraph{NGC~5129} is at the edge of the NPS so the local R45 value is very
high. We indeed found a high local soft X-ray excess (Table 2). It is also near
an extended feature on the \iras\ 100 $\mu$m map, which should explain the enhanced
absorption. G07 presented results based on an \xmm\ observation. Our temperature
profile is consistent with that from G07. However, our $c_{500}$ is smaller
than G07's (3.43$^{+1.72}_{-1.22}$ vs. 7.7$\pm$1.3).

\paragraph{UGC 2755} is one of the faintest and most gas poor systems
in our sample ($f_{\rm gas, 2500} = 0.030\pm0.005$). It is one of the two groups with
an entropy bump detected ($\S$8.1). 

\paragraph{NGC~4325} has a luminous cooling core. 
NGC~4325 was studied by G07, who also analyzed an \xmm\ observation.
The explored radial range in spectral analysis is similar in both work
(228 kpc vs. 232 kpc in our work).
Our temperature profile agrees well with that derived by G07. Our $c_{500}$,
$M_{2500}$ and $f_{\rm gas, 2500}$ are consistent with G07's within 1 $\sigma$ errors.

\paragraph{3C~442A} has the most peaky temperature profile in this sample.
The sharp
reconstructed temperature peak at 75 - 150 kpc, and the steepening of the
surface brightness at $\sim$ 120 kpc, may best be explained by the second
most luminous radio source in our sample (only after Abell~2462). From the
NVSS image, the two radio lobes of 3C~442A extend to $\sim 5'$ (or 152 kpc) in radius.
Shock heating by the radio source may explain the high temperature peak.
Because of this high temperature peak, the total mass density profile is not
physically meaningful within the central $\sim$160 kpc, which casts doubt on
the assumption of hydrostatic equilibrium within the central 160 kpc. Therefore,
$c_{500}$ cannot be constrained reliably.

\paragraph{ESO~552-020} was also studied by G07. Our temperature profile is
consistent with G07's. Our $c_{500}$ is also consistent with G07's.

\paragraph{IC~1262} has rich substructure within its core.
However, beyond the central 60 kpc radius, it appears symmetrical and relaxed.
Its surface brightness profile shows a sharp break at $\sim$ 200 kpc.
Unlike 3C~442A and A2462, its radio source is faint. However,
this may be caused by past AGN activity, which
was also suggested to explain the rich structures in the group core by
Trinchieri et al. (2007). The sharp temperature decline beyond 200 kpc
also supports this scenario, as the regions between 80 and 200 kpc radius
may have been recently heated (Fig. 4). The adiabatic sound speed in $\sim$ 1.9 keV gas
is $\sim$ 700 km/s. Thus, a Mach 1.5 shock will travel to the current position
in $\sim$ 200 Myr, which is consistent with the typical duty cycle of radio AGN.

\paragraph{ESO~306-017} has been studied before (Sun et al. 2004; G07).
There is an adjacent \chandra\ pointing targeted at the $z=0.64$ cluster
RDCS~J0542-4100 (ObsID 914) that we used to constrain the local soft CXB.
Our temperature profile covers a wider radial range than G07's. In the overlapping
region, our temperature profile is consistent with G07's. Our results on $r_{500}$
and $M_{500}$ are also consistent with G07's.

\paragraph{NGC~5098} was also studied by G07. The \iras\ $100\mu$m map shows the
presence of Galactic cirrus around the group and we indeed find extra absorption
(Table 2). There is a second group in the field (Mahdavi et al. 2005) and the
region around it is excluded in our analysis. Our temperature profile and
results on $r_{500}$, $M_{500}$ and $c_{500}$ are consistent with G07's.

\paragraph{UGC~842} was also studied by G07. Our temperature profile and
results on $r_{500}$, $M_{500}$ and $c_{500}$ are consistent with G07's.

\paragraph{A2717} was studied by Pratt \& Arnaud (2005), G07 and Snowden et al. (2008)
with the \xmm\ data. Our temperature profile is consistent with the profiles
derived by G07 and Snowden et al. (2008).
Our derived $r_{500}$ and $M_{500}$ are consistent with
those derived by G07, while our $c_{500}$ (2.15$^{+0.36}_{-0.32}$) is close to G07's value,
3.0$\pm$0.2 and the result by Pratt \& Arnaud (2005), 2.8$\pm$0.2 (converted from
their $c_{200}$ assuming an NFW profile).

\paragraph{AS1101} (or S\'{e}rsic 159-03) is the most gas-rich system in our sample.
Its enclosed gas
fraction at 0.1$r_{500}$ ($\sim$0.06) is almost 3 times the average of other groups
in the sample. Its gas fraction at $r_{500}$ (0.114$^{+0.021}_{-0.020}$, extrapolated)
is comparable to those of 5 - 7 keV clusters. The \chandra\
exposure is short but the best-fit values of our temperature profile
agree well with the \xmm\ results by Snowden et al. (2008).
The derived $c_{500}$ in this work (5.05$^{+2.37}_{-1.34}$) is consistent with
the result by Piffaretti et al. (2005), 4.33$\pm$0.51.

\paragraph{A1991} was also studied by V05 and V06. Our derived system properties
at $r_{500}$ and $r_{2500}$ agree very well
with those of V05 and V06. The temperature decline at large radii in this system
was also found from the \xmm\ data by Snowden et al. (2008) with consistent values.

\paragraph{A2462} hosts a small corona (with a radius of $\lsim$ 4 kpc)
at the center, without a large cool core. This is common for BCGs
(Sun et al. 2007; also see $\S$8.2). The \chandra\ surface brightness
profile shows a significant break at $\sim$ 180 kpc, which is about the size
of the central radio source from the NVSS image. The central radio source in A2462 is the most
luminous one in our sample. It may have heated the group core, as shown by the
high temperature and entropy beyond the central corona.

\paragraph{RXJ~1159+5531} has been studied by V05, V06 and G07. Our temperature
profile agrees well with both V05 and G07. The derived gas fraction and total mass
at $r_{2500}$ are well consistent with those from V06. $r_{500}$ is also consistent.
Our $c_{500}$ (2.95$^{+1.16}_{-0.90}$) lies between the results from V06 (1.70$\pm$0.29)
and from G07 (5.6$\pm$1.5). Our $r_{500}$ and $M_{500}$ are consistent with G07's.

\paragraph{A2550} is in a large filamentary structure that connects with A2554
($z=0.111$, 17.5$'$ on the northeast), while A2556 ($z=0.087$) is 21$'$ to the east.
Based on the derived surface brightness profiles centered on each system, we exclude
regions within 13.7$'$ and 14.1$'$ (in radius) of A2554 and A2556 respectively,
which are about 1.4 times $r_{500}$ of each system.
There is also an X-ray clump south of A2550's core that is excised.

\end{appendix}

\clearpage

\begin{table} 
\begin{center}
\caption{The group sample and the \chandra\ observations}
{\scriptsize
\begin{tabular}{ccccccccc} \hline \hline
Group &  $z$\tablenotemark{a} & D\tablenotemark{b} & ObsID\tablenotemark{c} & Date & Exposure\tablenotemark{d} & PSPC\tablenotemark{e} & $L_{\rm Ks}$\tablenotemark{f} & $L_{\rm 1.4 GHz}$\tablenotemark{g} \\ \hline
 
NGC~1550 & 0.0124 & 51.4 & 3186 & 2002-01-08 & 9.7 & & 11.29 & 21.72 \\
         &        &      & 3187 & 2002-01-08 & 9.5 & & & \\
         &        &      & 5800 & 2005-10-22 & 43.0 (44.0) & & & \\
         &        &      & 5801 & 2005-10-24 & 44.0 (44.3) & & & \\
NGC~3402 & 0.0153 & 63.5 & 3243 & 2002-11-05 & 22.9 (29.5) & & 11.40 & 22.18 \\
A262     & 0.0163 & 67.8 & 2215 & 2001-08-03 & 28.7 (28.7) & 7.6 & 11.60 & 22.56 \\
         &        &      & 7921 & 2006-11-20 & 110.5 (110.5) & & & \\  
NGC~383 (3C~31) & 0.0170 & 70.8 & 2147 & 2000-11-06 & 41.0 (44.3) & 24.0 & 11.67 & 24.46 \\
3C~449   & 0.0171 & 71.2 & 4057 & 2003-09-18 & 14.3 (26.4) & 8.9 & 11.13 & 24.36 \\
NGC~533  & 0.0185 & 77.1 & 2880 & 2002-07-28 & 29.7 (37.1) & 11.5 & 11.76 & 22.31 \\
NGC~741  & 0.0185 & 77.3 & 2223 & 2001-01-28 & 28.7 (30.2) & 12.4 & 11.82 & 23.85 \\
MKW4     & 0.0200 & 83.4 & 3234 & 2002-11-24 & 27.6 (29.7) & 9.4 & 11.81 & 22.15 \\
A3581    & 0.0230 & 96.2 & 1650 & 2001-06-07 & 7.2 (7.2) & & 11.51 & 23.85 \\
NGC~5129 & 0.0230 & 96.2 & 6944 & 2006-04-13 & 20.5 (20.5) & 5.5 & 11.63 & 21.90 \\
         &        &      & 7325 & 2006-05-14 & 25.6 (25.6) & & & \\
NGC~1132 & 0.0233 & 97.5 & 3576 & 2003-11-16 & 20.5 (37.1) & & 11.64 & 21.79 \\
UGC~2755 & 0.0245 & 102 & 2189 & 2001-02-07 & 7.2 (15.6) & 16.4 & 11.49 & 24.26 \\
NGC~4325 & 0.0257 & 108 & 3232 & 2003-02-04 & 25.6 (29.9) & 4.8 & 11.31 & $<$21.32 \\
HCG~51   & 0.0258 & 108 & 4989 & 2004-02-15 & 18.4 (18.9) & & 11.47 & $<$21.32 \\
         &        &     & 5304 & 2004-02-16 & 12.2 (12.7) & & & \\
3C~442A  & 0.0263 & 110 & 5635 & 2005-07-27 & 26.6 & & 11.51, 11.40 & 24.70 \\
         &        &     & 6353 & 2005-07-28 & 13.6 & & & \\
         &        &     & 6359 & 2005-10-07 & 19.4 & & & \\
         &        &     & 6392 & 2006-01-12 & 31.7 & & & \\
UGC~5088 & 0.0274 & 115 & 3227 & 2002-03-10 & 33.8 & & 11.26 & 21.04 \\
NGC~6338 & 0.0274 & 115 & 4194 & 2003-09-17 & 39.9 & 3.5 & 11.73 & 22.95 \\
NGC~4104 & 0.0282 & 118 & 6939 & 2006-02-16 & 35.8 (35.8) & 13.6 & 11.85 & 21.60 \\
RBS~461  & 0.0296 & 124 & 4182 & 2003-03-11 & 22.0 & & 11.41 & 22.57 \\
ESO~552-020 & 0.0314 & 132 & 3206 & 2002-10-14 & 18.7 & & 11.92 & $<$21.72 \\
A1177    & 0.0316 & 133 & 6940 & 2006-12-27 & 32.8 (33.5) & & 11.73 & $<$21.50 \\
IC~1262  & 0.0326 & 138 & 2018 & 2001-08-23 & 25.6 (30.5) & & 11.49 & 23.22 \\
         &        &     & 6949 & 2006-04-17 & 38.1 & & & \\
         &        &     & 7321 & 2006-04-19 & 36.8 & & & \\
         &        &     & 7322 & 2006-04-22 & 37.4 & & & \\
NGC~6269 & 0.0348 & 147 & 4972 & 2003-12-29 & 38.6 & 10.1 & 11.94 & 23.11 \\
ESO~306-017 & 0.0358 & 151 & 3188 & 2002-03-08 & 13.6 & & 11.91 & 22.58 \\
            &        &     & 3189 & 2002-03-09 & 13.8 & & &  \\
            &        &     &  914* & 2000-07-26 & 50.4 & & &  \\
NGC~5098 & 0.0368 & 156 & 2231 & 2001-08-04 & 10.0 & & 11.44, 11.36 & 23.38 \\
         &        &     & 6941 & 2005-11-01 & 37.9 (38.6) & & & \\
A1139 & 0.0398 & 169 & 9387 & 2008-03-28 & 10.0 & & 11.50 & 23.16 \\
A160   & 0.0447 & 190 & 3219 & 2002-10-18 & 54.5 & & 11.69 & 24.65 \\
UGC~842 & 0.0452 & 192 & 4963 & 2005-02-13 & 38.9 (39.2) & & 11.77 & 22.24 \\
A2717   & 0.0498 & 213 & 6973 & 2006-08-17 & 46.1 & 8.5 & 11.82 & 24.52 \\ 
        &        &     & 6974 & 2006-04-10 & 18.9 & & & \\
RXJ~1022+3830 & 0.0543 & 233 & 6942 & 2006-10-14 & 40.9 (41.4) & 10.3 & 11.69, - & $<$21.98 \\
AS1101 & 0.0564 & 242 & 1668 & 2001-08-13 & 9.2 (9.4) & 11.9 & 11.89 & 24.27 \\
ESO~351-021 & 0.0571 & 245 & 5784 & 2005-04-24 & 34.8 (35.1) & & 11.95 & 22.71 \\
A3880 & 0.0581 & 250 & 5798 & 2004-12-23 & 19.7 & & 11.93 & 24.22 \\
A1991 & 0.0587 & 253 & 3193 & 2002-12-16 & 36.8 (36.3) & 20.0 & 11.86 & 23.47 \\
A1275 & 0.0637 & 275 & 6945 & 2006-02-05 & 49.1 (48.4) & & 11.31 & $<$22.13 \\
A2092 & 0.0669 & 290 & 9384 & 2007-11-13 & 9.7 & & 11.44 & $<$22.17 \\
RXJ~1206-0744 & 0.0680 & 295 & 9388 & 2007-11-15 & 10.0 & & 11.75 & $<$22.19 \\
A1238 & 0.0720 & 313 & 4991 & 2004-03-28 & 18.2 & & 11.70 & 24.44 \\
A744  & 0.0729 & 317 & 6947 & 2006-10-22 & 36.3 & & 11.80 & 22.24 \\
A2462 & 0.0733 & 319 & 4159 & 2002-11-19 & 37.9 (38.6) & 3.4 & 11.82 & 25.36 \\
RXJ~1159+5531 & 0.0808 & 354 & 4964 & 2004-02-11 & 69.6 (74.2) & & 11.97 & $<$22.34 \\
A1692 & 0.0848 & 372 & 4990 & 2004-08-12 & 21.5 & & 11.88 & 22.33 \\
      &        &     & 6930* & 2006-03-06 & 76.0 &   &    &       \\
      &        &     & 7289* & 2006-03-09 & 75.0 &   &    &       \\
A2550 & 0.122  & 550 & 2225 & 2001-09-03 & 57.3 (58.8) & & 11.62 & 23.18 \\

\hline \hline
\end{tabular}}
\begin{flushleft}
\leftskip 35pt
\tablenotetext{a}{The group redshift is extracted from NASA/IPAC Extragalactic Database (NED)}
\tablenotetext{b}{The luminosity distance of the group derived from its redshift}
\tablenotetext{c}{The ObsIDs with * are observations that happened to be close to the interested groups. We used these observations to constrain the local soft CXB.}
\tablenotetext{d}{Effective exposure after excluding time intervals of background flares. For observations with ACIS-S3 at the optical axis, two exposure values are listed, with the one in the brackets as the exposure for the FI chips.} \tablenotetext{e}{\rosat\ PSPC clean exposure time (in ksec) if the pointed PSPC observations exist.}
\tablenotetext{f}{2MASS $K_{\rm s}$ band luminosity of the cD galaxy as shown as log(L$_{\rm Ks}$/L$_{\odot}$), $M_{K\odot}$ = 3.39 mag. There are three groups (3C~442A, NGC~5098 and RXJ~1022+3830)
with two BCGs at the center. The $K_{\rm s}$ band magnitude of one central galaxy in RXJ~1022+3830
with two BCGs at the center. The $K_{\rm s}$ band magnitude of one central galaxy in RXJ~1022+3830
is unknown.}
\tablenotetext{g}{1.4 GHz luminosity of the cD galaxy as shown as log(L$_{\rm 1.4 GHz}$/W Hz$^{-1}$) from the NRAO VLA Sky Survey (NVSS) or the Sydney University Molonglo Sky Survey (SUMSS), assuming a spectral index of -0.8, unless it can be derived from NED.}
\end{flushleft}
\end{center}
\end{table}

\begin{table}
\begin{center}
\caption{Absorption, radial range of the analysis and the local CXB}
{\tiny
\begin{tabular}{cccccccccc} \hline \hline
Group & $N_{H}$\tablenotemark{a} & R45\tablenotemark{b} & outermost bin\tablenotemark{c} & r$_{\rm det, spe}$\tablenotemark{d} & r$_{\rm det, sur}$\tablenotemark{e} & $kT_{\rm hot}$\tablenotemark{f} & $f_{\rm 0.47 - 1.21 keV}$\tablenotemark{g} & $f_{\rm 2 - 8 keV}$\tablenotemark{h} & $f_{\rm 2 - 8 keV, expected}$ ($f_{\rm limit}$)\tablenotemark{i} \\
      & (10$^{20}$ cm$^{-2}$) & & (\%) & (kpc) & (kpc) & (keV) &  & & \\ \hline

NGC~1550 & 12.5 (10.0) & 125 & 11 & 355 & 364 & 0.20$^{+0.04}_{-0.07}$ & 4.5$^{+0.2}_{-0.4}$ & 13.5$^{+4.2}_{-4.5}$, 7.4$^{+4.5}_{-4.0}$ & $\sim$12.6 ($\sim$2.1), $\sim$8.5 ($\sim$0.56) \\
NGC~3402 & 11.0 (4.0) & 119 & 31 & 239 & 239 & (0.25) & 5.0$^{+0.2}_{-0.6}$ & 6.5$^{+2.5}_{-3.5}$ & $\sim$6.8 ($\sim$ 0.27) \\
A262     & 8.1 (5.8) & 152 & 19 & 363 & 387 (800) & 0.27$\pm$0.03 & 2.7$^{+0.6}_{-0.5}$ & 10$^{+4.0}_{-4.5}$, 8.4$^{+3.4}_{-4.0}$ & $\sim$9.4 (0.51-1.1), $\sim$7.7 (0.31-0.65) \\
NGC~383  & 5.3 & 100 & 21 & 289 & 350 (500) & (0.25) & 3.2$^{+0.4}_{-0.9}$ & 9.3$^{+3.2}_{-2.6}$ & $\sim$8.4 (0.47 - 0.61) \\
3C 449   & 13.3 (9.0) & 161 & 32 & 230 & 230 (380) & (0.25) & 6.7$^{+0.8}_{-0.3}$ & 10.0$^{+5.0}_{-2.6}$ & $\sim$8.4 ($\sim$0.47) \\
NGC~533  & 5.7 (3.0) & 103 & 26 & 238 & 340 (380) & 0.31$\pm$0.05 & 3.9$^{+0.9}_{-0.5}$ & 7.1$^{+2.5}_{-2.2}$ & $\sim$ 7.2 ($\sim$ 0.33) \\
NGC~741  & 5.9 (4.3) & 100 & 33 & 271 & 360 (420) & 0.23$^{+0.05}_{-0.03}$ & 3.7$^{+0.5}_{-0.3}$ & 6.7$^{+3.0}_{-3.5}$ & $\sim$ 7.5 ($\sim$ 0.36) \\
MKW4     & 3.1 (1.8) & 117 & 10 & 490 & 490 (720) & (0.25) & 4.7$^{+0.6}_{-0.3}$ & 7.5$^{+3.7}_{-3.9}$ & $\sim$8.4  (0.34 - 0.78) \\
A3581    & 5.8 (4.5) & 296 & 25 & 322 & 450 & 0.27$^{+0.03}_{-0.02}$ & 12.5$^{+0.4}_{-0.2}$ & 10.6$^{+2.3}_{-5.5}$ & $\sim$10.1 ($\sim$0.98)\\
NGC~5129 & 5.5 (1.7) & 298 & 45 & 214 & 222 (270) & 0.29$^{+0.02}_{-0.01}$ & 12.3$^{+0.7}_{-0.3}$ & 7.0$^{+3.9}_{-4.8}$ & $\sim$7.0 (0.17-0.41) \\
NGC~1132 & 7.8 (5.5) & 92 & 33 & 284 & 310 & 0.28$\pm$0.04 & 2.5$^{+0.7}_{-0.3}$ & 7.3$^{+2.9}_{-2.5}$ & $\sim$6.5 ($\sim$0.22) \\
UGC~2755 & 13.6 & 93 & 52 & 193 & 193 (205) & 0.26$^{+0.06}_{-0.04}$ & 3.3$^{+0.6}_{-0.5}$ & 10.4$^{+3.0}_{-3.7}$ & $\sim$ 8.7 (0.55-0.73) \\
NGC~4325 & 2.4 & 143 & 23 & 232 & 232 (250) & (0.25) & 4.5$^{+1.2}_{-1.3}$ & 8.1$^{+5.4}_{-5.0}$ & $\sim$ 7.0 ($\sim$ 0.30) \\
HCG~51   & 5.3 (1.1) & 118 & 30 & 329 & 360 & (0.25) & 2.2$^{+0.6}_{-0.3}$ & 6.8$^{+2.5}_{-3.0}$ & $\sim$ 7.2 ($\sim$ 0.33) \\
3C~442A  & 6.1 (4.8) & 85 & 68 & 396 & 530 & (0.25) & 2.0$^{+0.5}_{-0.6}$ & 6.6$\pm$3.4 & $\sim$7.6 (0.25-0.7) \\
UGC~5088 & 1.2 & 90 & 63 & 317 & 320 & (0.25) & 2.5$^{+0.3}_{-0.4}$  & 6.5$^{+2.4}_{-2.2}$ & $\sim$7.0 ($\sim$0.27) \\
NGC~6338 & 2.3 & 133 & 43 & 349 & 360 (510) & (0.25) & 2.0$^{+0.5}_{-0.6}$ & 9.2$^{+4.6}_{-2.8}$ & $\sim$7.0 ($\sim$0.31) \\
NGC~4104 & 4.4 (1.8) & 113 & 33 & 407 & 420 (550) & 0.23$^{+0.07}_{-0.05}$ & 2.7$^{+0.4}_{-0.2}$ & 8.1$^{+5.0}_{-5.8}$ & $\sim$7.0 ($\sim$0.30) \\
RBS~461  & 17.7 (15.0) & 65 & 54 & 359 & 515 & (0.25) & 3.6$^{+0.3}_{-0.6}$ & 10.0$^{+6.0}_{-3.7}$ & $\sim$8.6 ($\sim$ 0.56) \\
ESO~552-020 & 3.9 & 119 & 51 & 380 & 480 & (0.25) & 2.4$^{+0.4}_{-0.7}$ & 8.8$^{+3.4}_{-3.5}$ & $\sim$7.8 ($\sim$ 0.46) \\
A1177 & 4.6 (1.1) & 103 & 35 & 418 & 420 & 0.15$\pm$0.02\tablenotemark{j} & 2.0$^{+0.4}_{-0.2}$ & 6.5$^{+2.8}_{-3.4}$ & $\sim$ 7.3 ($\sim$ 0.34) \\
IC~1262  & 3.4 (1.8) & 162 & 74 & 375 & 390 & 0.32$\pm$0.04 & 5.4$^{+0.4}_{-1.0}$ & 8.1$^{+3.1}_{-2.2}$ & 6.0 - 7.3 (0.15 - 0.31) \\
NGC~6269 & 5.2 & 106 & 48 & 439 & 510 (740) & (0.25) & 2.5$\pm$0.4 & 8.2$^{+2.3}_{-3.3}$ & $\sim$ 7.0 ($\sim$ 0.30) \\
ESO~306-017 & 3.0 & 105 & 77 & 484 & 490 & 0.23$^{+0.04}_{-0.03}$ & 4.9$^{+0.6}_{-0.8}$ & 11.0$^{+4.9}_{-2.6}$ & $\sim$ 9.7 ($\sim$ 0.85) \\
NGC~5098 & 6.8 (1.3) & 120 & 81 & 421 & 480 & 0.25$^{+0.05}_{-0.04}$ & 2.6$^{+0.4}_{-0.5}$ & 5.9$^{+3.7}_{-3.9}$, 9.5$^{+4.8}_{-4.2}$ & $\sim$ 10.1 ($\sim$ 0.95), $\sim$ 6.9 ($\sim$ 0.26) \\
A1139 & 3.2 & 93 & 94 & 318 & 370 & (0.25) & 2.6$^{+0.4}_{-0.5}$ & 10.8$^{+5.8}_{-4.3}$ & $\sim$ 9.6 ($\sim$ 0.87) \\
A160 & 4.0 & 100 & 62 & 507 & 640 & (0.25) & 3.2$\pm$0.5 & 7.8$^{+3.7}_{-2.9}$ & $\sim$ 6.7 ($\sim$ 0.22) \\
UGC~842 & 4.0 & 99 & 25 & 410 & 510 & 0.26$\pm$0.03 & 4.4$^{+0.5}_{-0.6}$ & 7.5$^{+3.5}_{-2.6}$ & $\sim$6.8 ($\sim$ 0.22)\\
A2717 & 1.2 & 122 & 37 & 730 & 800 (760) & 0.31$^{+0.07}_{-0.06}$ & 3.6$\pm$0.6 & 7.1$^{+3.2}_{-2.9}$, 9.4$^{+4.1}_{-3.2}$ & $\sim$ 6.3 ($\sim$ 0.20), $\sim$ 8.6 ($\sim$ 0.58) \\
RXJ~1022+3830 & 3.3 (1.5) & 116 & 32 & 548 & 610 (670) & 0.30$\pm$0.06 & 2.5$\pm$0.6 & 8.8$^{+5.2}_{-3.8}$ & $\sim$ 7.0 ($\sim$ 0.27)\\
AS1101 & 1.1 & 130 & 23 & 568 & 650 (850) & 0.28$\pm$0.07 & 3.5$\pm$0.9 & 13.0$^{+6.5}_{-5.0}$ & $\sim$10.1 ($\sim$ 0.95) \\
ESO~351-021 & 2.4 & 129 & 34 & 511 & 515 & 0.28$^{+0.07}_{-0.04}$ & 2.5$^{+0.4}_{-0.6}$ & 7.2$^{+3.7}_{-2.9}$ & $\sim$6.8 ($\sim$ 0.26)\\
A3880 & 1.2 & 149 & 37 & 779 & 810 & (0.25) & 3.5$^{+1.0}_{-0.6}$ & 9.2$^{+3.5}_{-3.9}$ & $\sim$8.0 ($\sim$ 0.43)\\
A1991 & 2.5 & 280 & 31 & 655 & 820 (750) & 0.33$\pm$0.02 & 14.4$\pm$0.6 & 6.7$^{+2.9}_{-2.5}$ & $\sim$7.0 ($\sim$ 0.29)\\
A1275 & 6.2 (2.0) & 91 & 42 & 495 & 530 & 0.21$^{+0.06}_{-0.09}$ & 2.1$^{+0.5}_{-0.4}$ & 7.7$^{+3.8}_{-3.4}$ & $\sim$6.3 ($\sim$ 0.16) \\
A2092 & 2.2 & 130 & 79 & 555 & 600 & (0.25) & 3.1$^{+0.3}_{-0.4}$ & 6.3$^{+1.9}_{-1.8}$ & $\sim$ 9.2 ($\sim$ 0.79) \\
RXJ~1206-0744 & 2.3 & 134 & 92 & 451 & 580 & (0.25) & 2.1$^{+0.6}_{-1.2}$ & 9.2$^{+4.3}_{-3.0}$ & $\sim$ 9.4 ($\sim$ 0.84) \\
A1238 & 3.5 & 103 & 81 & 673 & 690 & 0.22$^{+0.09}_{-0.05}$ & 2.4$^{+0.4}_{-0.7}$ & 8.3$^{+4.2}_{-3.3}$ & $\sim$ 8.2 ($\sim$ 0.49) \\
A744 & 3.5 & 104 & 92 & 433 & 600 & 0.20$^{+0.30}_{-0.06}$ & 2.7$^{+0.4}_{-0.5}$ & 6.4$^{+3.3}_{-2.6}$ & $\sim$ 6.8 ($\sim$ 0.24) \\
A2462 & 3.1 & 158 & 50 & 564 & 650 (800) & 0.26$\pm$0.02 & 7.8$\pm$0.6 & 7.6$^{+5.0}_{-2.1}$ & $\sim$6.4 ($\sim$ 0.19) \\
RXJ~1159 & 1.2 & 134 & 56 & 529 & 560 & 0.27$\pm$0.03 & 2.7$\pm$0.5 & 6.0$^{+3.1}_{-2.5}$ & $\sim$5.5 (0.08-0.14) \\
A1692 & 1.8 & 158 & 80 & 766 & 770 & 0.26$^{+0.05}_{-0.04}$ & 4.2$\pm$0.5 & 7.3$^{+3.9}_{-3.0}$ & $\sim$8.0 ($\sim$ 0.39)\\
A2550 & 1.9 & 130 & 75 & 597 & 740 & 0.28$\pm$0.03 & 4.8$\pm$0.7 & 7.8$^{+3.4}_{-3.1}$ & $\sim$ 5.9 ($\sim$ 0.13) \\

\hline \hline
\end{tabular}}
\vspace{-1.2cm}
\begin{flushleft}
\leftskip 35pt
\tablenotetext{a}{The absorption column density in our analysis. If the value from
our spectral analysis ($\S$3.3) is consistent with the Galactic value from the
Leiden/Argentine/Bonn (LAB) HI survey (Kalberla et al. 2005), the LAB value is used.
Both values are listed if they are significantly different, with the number in
brackets is the LAB value.}
\tablenotetext{b}{The \rosat\ All-Sky Survey R45 flux (Snowden et al. 1997), in a unit of
10$^{-6}$ cts/s/arcmin$^{2}$, measured from an annulus centered on the source. The inner
radius of the annulus is 0.4 - 0.8 deg (depending on the source size), while the outer
radius is the inner radius + 0.4 deg. In a few cases, we have to use partial apertures
to exclude the bright sources near our targets (e.g., A1692 and A2550).}
\tablenotetext{c}{The fraction of the outermost radial bin for the spectral analysis covered
by the \chandra\ data, compared with the full annulus. The median is 37\%. Note
this fraction is always less than one because of point sources and chip gaps.}
\tablenotetext{d}{The outermost radius for the \chandra\ spectral analysis}
\tablenotetext{e}{The radius where X-ray surface brightness is detected at $> 2\sigma$.
$r_{out}/r_{in}$ = 1.06-1.1. Note that our estimate of local background is conservative so our 2$\sigma$ range is smaller than V06's 3$\sigma$ range for A1991 and RXJ~1159+5531. The value in brackets is for PSPC if available.}
\tablenotetext{f}{The temperature of the hotter component of the local soft CXB}
\tablenotetext{g}{The 0.47 - 1.21 keV observed flux of the local soft CXB
(in unit of 10$^{-12}$ ergs cm$^{-2}$ s$^{-1}$ deg$^{-2}$). The energy band is
chosen to match that of the RASS R45 band.}
\tablenotetext{h}{The 2 - 8 keV unabsorbed flux of the unresolved hard CXB
(in unit of 10$^{-12}$ ergs cm$^{-2}$ s$^{-1}$ deg$^{-2}$)}
\tablenotetext{i}{The expected 2 - 8 keV unabsorbed flux of the unresolved hard X-ray
CXB (in unit of 10$^{-12}$ ergs cm$^{-2}$ s$^{-1}$ deg$^{-2}$), estimated
based on the limiting flux (shown in brackets, in unit of 10$^{-14}$ ergs
cm$^{-2}$ s$^{-1}$) and from the derived average relation in K07.}
\tablenotetext{j}{The 0.1 keV component has zero normalization in this case.}
\end{flushleft}
\end{center}
\end{table}

\begin{table}
\begin{center}
\caption{Derived properties of groups (I: temperature, mass and gas fraction)}
\vspace{0.3cm}
{\tiny
\begin{tabular}{cccccccccc} \hline \hline
Group & $T_{500}$\tablenotemark{a} & $T_{2500}$ & $r_{500}$\tablenotemark{a} & $r_{2500}$ & $M_{500}$\tablenotemark{a} & $f_{\rm gas, 500}$\tablenotemark{a} & $f_{\rm gas, 2500}$ & $f_{\rm gas, 2500-500}$\tablenotemark{a} & $c_{500}$\tablenotemark{b} \\
      & (keV) & (keV) & (kpc) & (kpc) & (10$^{13}$ M$_{\odot}$) & & & & \\ \hline

NGC~1550 & 1.06$\pm0.02$* & 1.18$\pm0.02$ & 465$^{+15}_{-19}$* & 222$\pm6$ & 3.18$^{+0.32}_{-0.37}$* & 0.097$^{+0.012}_{-0.009}$* & 0.056$\pm$0.003 & 0.149$^{+0.033}_{-0.024}$* & 4.93$^{+0.50}_{-0.46}$ (10) \\
NGC~3402 & (0.74$\pm0.03$)& 0.80$\pm0.02$ & (380) & 205$^{+62}_{-22}$ & & & 0.032$^{+0.008}_{-0.014}$ & & \\
A262     & (1.94$^{+0.11}_{-0.15}$) & 2.18$^{+0.06}_{-0.07}$ & (644) & 288$^{+17}_{-16}$ & & & 0.064$^{+0.005}_{-0.007}$ & & 3.48$^{+0.49}_{-0.45}$ (9) \\
NGC~383  & (1.67$^{+0.13}_{-0.11}$) & 1.89$^{+0.17}_{-0.10}$ & (593) & 252$^{+35}_{-20}$ & & & 0.031$^{+0.003}_{-0.004}$ & & 3.09$^{+1.84}_{-1.30}$ (6) \\
3C 449   & (0.97$^{+0.04}_{-0.05}$) & 1.08$\pm0.04$ & (437) & 211$^{+11}_{-21}$ & & & 0.041$^{+0.006}_{-0.004}$ & & 2.97$^{+2.60}_{-1.64}$ (5) \\
NGC~533  & (1.06$^{+0.08}_{-0.04}$) & 1.21$^{+0.06}_{-0.07}$ & (461) & 207$^{+18}_{-29}$ & & & 0.031$^{+0.005}_{-0.004}$ & & 4.58$^{+3.90}_{-2.34}$ (5) \\
NGC~741  & (1.27$^{+0.08}_{-0.12}$) & 1.37$^{+0.08}_{-0.12}$ & (510) & 214$^{+20}_{-14}$ & & & 0.026$^{+0.004}_{-0.003}$ & & 3.05$^{+1.68}_{-1.22}$ (6) \\
MKW4     & 1.58$\pm$0.09 & 1.75$^{+0.05}_{-0.04}$ & 538$^{+24}_{-29}$ & 259$^{+12}_{-8}$ & 4.85$^{+0.71}_{-0.68}$ & 0.086$\pm$0.009 & 0.047$^{+0.002}_{-0.003}$ & 0.134$^{+0.38}_{-0.26}$ & 3.93$^{+1.16}_{-0.78}$ (7) \\
A3581    & (1.68$^{+0.10}_{-0.09}$) & 1.85$^{+0.12}_{-0.07}$ & (593) & 259$^{+22}_{-17}$ & & & 0.067$^{+0.006}_{-0.007}$ & & 7.43$^{+2.06}_{-1.40}$ (6) \\
NGC~5129 & (0.76$\pm0.03$) & 0.83$\pm0.02$ & (384) & 174$^{+7}_{-9}$ & & & 0.035$^{+0.004}_{-0.003}$  & & 3.43$^{+1.72}_{-1.22}$ (5) \\
NGC~1132 & (0.99$\pm0.04$) & 1.08$\pm0.03$ & (442) & 215$\pm$11 & & & 0.039$^{+0.005}_{-0.002}$ & & 1.77$^{+0.70}_{-0.58}$ (6) \\
UGC~2755 & (0.76$\pm0.05$) & 0.83$^{+0.05}_{-0.05}$ & (384) & 188$^{+11}_{-12}$ & & & 0.031$\pm$0.005 & & \\
NGC~4325 & (0.89$\pm0.03$) & 0.97$\pm0.03$ & (418) & 212$\pm$31 & & & 0.037$^{+0.014}_{-0.007}$ & & 5.19$^{+3.68}_{-2.36}$ (7) \\
HCG~51   & (1.06$^{+0.04}_{-0.03}$) & 1.15$^{+0.04}_{-0.03}$ & (460) & 275$^{+13}_{-14}$ & & & 0.028$\pm$0.003 & & 2.11$^{+0.89}_{-0.56}$ (6) \\
3C~442A  & 1.34$\pm0.04$* & 1.61$^{+0.05}_{-0.06}$ & 495$^{+12}_{-16}$* & 277$^{+16}_{-15}$ & 3.90$^{+0.22}_{-0.40}$* & 0.068$^{+0.006}_{-0.003}$* & 0.028$^{+0.003}_{-0.002}$ & 0.230$^{+0.210}_{-0.060}$* & \\
UGC~5088 & 0.81$\pm0.03$* & 0.83$^{+0.02}_{-0.03}$ & 364$^{+27}_{-19}$* & 163$^{+10}_{-8}$ & 1.48$^{+0.36}_{-0.24}$* & 0.049$^{+0.008}_{-0.010}$* & 0.029$\pm$0.003 & 0.067$^{+0.021}_{-0.013}$* & 4.15$^{+1.66}_{-1.16}$ (5) \\
NGC~6338 & (1.92$^{+0.06}_{-0.07}$) & 2.14$\pm$0.05 & (636) & 288$^{+13}_{-11}$ & & & 0.053$\pm$0.003 & & 5.27$^{+2.66}_{-1.80}$ (8) \\
NGC~4104 & 1.41$^{+0.09}_{-0.06}$* & 1.64$^{+0.05}_{-0.08}$ & 535$^{+19}_{-20}$* & 274$\pm$12 & 4.85$^{+0.55}_{-0.53}$* & 0.069$^{+0.009}_{-0.006}$* & 0.036$\pm$0.003 & 0.137$^{+0.043}_{-0.025}$* & 4.29$^{+1.58}_{-1.12}$ (7) \\
RBS~461  & (1.93$\pm0.10$) & 2.17$\pm0.06$ & (637) & 318$^{+25}_{-27}$ & & & 0.059$\pm$0.006 & & 5.45$^{+1.32}_{-1.16}$ (9) \\
ESO~552-020 & (1.72$^{+0.13}_{-0.09}$) & 1.97$\pm0.09$ & (598) & 297$^{+27}_{-24}$ & & & 0.039$\pm$0.005 & & 4.27$^{+1.84}_{-1.22}$ (5) \\
A1177    & 1.37$^{+0.06}_{-0.07}$* & 1.48$^{+0.06}_{-0.07}$ & 550$^{+29}_{-27}$* & 264$^{+20}_{-25}$ & 5.28$^{+0.84}_{-0.73}$* & 0.060$^{+0.009}_{-0.007}$* & 0.037$\pm$0.004 & 0.091$^{+0.033}_{-0.020}$* & 5.26$^{+4.68}_{-2.61}$ (6)  \\
IC~1262  & (1.73$^{+0.06}_{-0.05}$) & 1.91$^{+0.06}_{-0.05}$ & (600) & 330$^{+20}_{-40}$ & & & 0.045$^{+0.004}_{-0.010}$ & & \\
NGC~6269 & 1.72$^{+0.12}_{-0.11}$* & 2.16$^{+0.08}_{-0.10}$ & 645$^{+50}_{-52}$* & 232$^{+27}_{-21}$ & 8.49$^{+1.97}_{-2.01}$* & 0.076$^{+0.011}_{-0.010}$* & 0.044$^{+0.003}_{-0.004}$ & 0.087$^{+0.027}_{-0.015}$* & \\
ESO~306-017 & 2.37$^{+0.12}_{-0.14}$* & 2.54$^{+0.08}_{-0.10}$ & 690$^{+44}_{-30}$* & 337$^{+16}_{-14}$ & 10.3$^{+2.1}_{-1.3}$* & 0.081$^{+0.010}_{-0.011}$* & 0.052$^{+0.003}_{-0.004}$ & 0.119$^{+0.031}_{-0.026}$* & \\
NGC~5098 & 0.96$\pm0.04$ & 1.05$^{+0.02}_{-0.03}$ & 398$^{+17}_{-33}$ & 195$^{+17}_{-13}$ & 2.00$^{+0.28}_{-0.46}$ & 0.108$^{+0.021}_{-0.012}$ & 0.048$^{+0.006}_{-0.004}$ & 0.205$^{+0.131}_{-0.060}$ & 4.29$^{+1.46}_{-1.06}$ (7) \\
A1139 & (2.01$^{+0.33}_{-0.34}$) & 2.20$^{+0.35}_{-0.33}$ & (650) & 298$^{+37}_{-36}$ & & & 0.026$^{+0.016}_{-0.008}$ & & \\
A160  & 1.68$^{+0.10}_{-0.10}$ & 2.05$^{+0.07}_{-0.06}$ & 626$^{+27}_{-31}$ & 286$^{+25}_{-23}$ & 7.90$^{+1.06}_{-1.10}$ & 0.085$^{+0.009}_{-0.008}$ & 0.043$^{+0.003}_{-0.004}$ & 0.121$^{+0.038}_{-0.023}$ & 2.73$^{+0.70}_{-0.60}$ (8) \\
UGC~842  & 1.54$^{+0.14}_{-0.12}$* & 1.78$\pm0.09$ & 570$^{+84}_{-45}$* & 276$^{+23}_{-29}$ & 5.60$^{+2.60}_{-1.10}$* & 0.056$^{+0.012}_{-0.014}$* & 0.032$\pm$0.003 & 0.089$^{+0.064}_{-0.041}$* & 6.06$^{+3.31}_{-2.19}$ (6) \\
A2717 & 2.43$^{+0.13}_{-0.12}$ & 2.60$^{+0.08}_{-0.09}$ & 732$^{+49}_{-32}$ & 342$^{+13}_{-11}$ & 12.9$^{+2.7}_{-1.7}$ & 0.076$\pm$0.010 & 0.053$\pm$0.003 & 0.098$^{+0.029}_{-0.025}$ & 2.15$^{+0.36}_{-0.32}$ (9) \\
RXJ~1022+3830 & 1.94$^{+0.20}_{-0.14}$ & 2.36$^{+0.17}_{-0.12}$ & 631$^{+32}_{-41}$ & 326$^{+25}_{-23}$ & 8.00$^{+1.31}_{-1.40}$ & 0.075$^{+0.007}_{-0.013}$ & 0.038$^{+0.004}_{-0.005}$ & 0.134$^{+0.074}_{-0.046}$ & 4.03$^{+1.24}_{-0.80}$ (6) \\
AS1101   & 2.57$^{+0.14}_{-0.11}$* & 2.69$^{+0.10}_{-0.09}$ & 768$^{+90}_{-65}$* & 362$^{+21}_{-18}$ & 14.1$^{+5.0}_{-3.4}$* & 0.114$^{+0.021}_{-0.020}$* & 0.085$\pm$0.008 & 0.129$^{+0.088}_{-0.041}$* & 5.05$^{+2.37}_{-1.34}$ (10) \\
ESO~351-021 & 1.14$^{+0.07}_{-0.04}$ & 1.34$^{+0.07}_{-0.03}$ & 437$^{+78}_{-38}$ & 194$^{+28}_{-35}$ & 3.22$^{+1.80}_{-0.90}$ & 0.074$^{+0.013}_{-0.018}$ & 0.036$^{+0.007}_{-0.005}$ & 0.119$^{+0.066}_{-0.050}$ & 2.42$^{+1.20}_{-0.76}$ (6) \\
A3880 & 2.49$^{+0.14}_{-0.12}$ & 2.75$\pm0.11$ & 799$^{+89}_{-68}$ & 309$^{+24}_{-12}$ & 14.9$^{+5.0}_{-3.5}$ & 0.088$^{+0.016}_{-0.021}$ & 0.074$^{+0.005}_{-0.007}$ & 0.090$^{+0.047}_{-0.022}$ & 4.03$^{+1.54}_{-1.26}$ (8) \\
A1991 & 2.68$^{+0.10}_{-0.08}$ & 2.86$\pm0.07$ & 749$^{+45}_{-35}$ & 348$^{+13}_{-10}$ & 13.4$^{+2.5}_{-1.9}$ & 0.094$^{+0.010}_{-0.012}$ & 0.066$\pm$0.004 & 0.115$^{+0.038}_{-0.022}$ & 4.69$^{+0.76}_{-0.70}$ (9) \\
A1275 & 1.46$^{+0.08}_{-0.07}$* & 1.63$^{+0.06}_{-0.07}$ & 592$^{+87}_{-50}$* & 202$^{+12}_{-9}$ & 6.90$^{+3.00}_{-1.67}$* & 0.094$^{+0.018}_{-0.031}$* & 0.069$\pm$0.004 & 0.100$^{+0.032}_{-0.035}$* & 4.10$^{+1.69}_{-1.27}$ (7) \\
A2092 & 1.67$^{+0.13}_{-0.12}$* & 2.14$^{+0.17}_{-0.21}$ & 659$^{+45}_{-40}$* & 283$^{+34}_{-24}$ & 8.95$^{+1.81}_{-1.62}$* & 0.078$\pm$0.013* & 0.050$^{+0.006}_{-0.007}$ & 0.096$^{+0.027}_{-0.020}$* & \\
RXJ~1206-0744 & (1.91$^{+0.20}_{-0.21}$) & 2.14$^{+0.19}_{-0.21}$ & (624) & 248$^{+28}_{-27}$ & & & 0.053$^{+0.009}_{-0.007}$ & & \\
A1238    & (2.51$^{+0.27}_{-0.32}$)& 2.88$^{+0.35}_{-0.36}$ & (725) & 272$^{+63}_{-59}$ & & & 0.035$\pm$0.008 & & \\
A744     & (2.24$^{+0.19}_{-0.16}$) & 2.49$\pm$0.16 & (681) & 325$^{+30}_{-24}$ & & & 0.046$^{+0.005}_{-0.006}$ & & 3.33$^{+1.66}_{-1.04}$ (6) \\
A2462    & 2.32$^{+0.12}_{-0.10}$ & 2.62$\pm$0.09 & 646$\pm$30 & 327$^{+13}_{-12}$ & 8.80$^{+1.29}_{-1.19}$ & 0.099$^{+0.011}_{-0.009}$ & 0.052$\pm$0.003 & 0.180$^{+0.058}_{-0.034}$ & 3.41$^{+1.56}_{-0.84}$ (6) \\
RXJ~1159 & 1.84$^{+0.14}_{-0.08}$* & 2.12$^{+0.09}_{-0.10}$ & 630$^{+77}_{-30}$* & 273$^{+15}_{-13}$ & 8.30$^{+3.10}_{-1.12}$* & 0.065$^{+0.007}_{-0.012}$* & 0.042$\pm$0.003 & 0.080$^{+0.019}_{-0.023}$* & 2.95$^{+1.16}_{-0.90}$ (6) \\
A1692    & 2.61$^{+0.16}_{-0.24}$ & 3.06$^{+0.25}_{-0.22}$ & 658$^{+64}_{-47}$ & 356$^{+42}_{-32}$ & 9.70$^{+2.99}_{-1.91}$ & 0.090$^{+0.014}_{-0.020}$ & 0.043$\pm$0.006 & 0.216$^{+0.155}_{-0.087}$ & 5.46$^{+1.77}_{-1.19}$ (7) \\
A2550    & 1.95$\pm$0.10 & 2.05$\pm$0.08 & 617$^{+74}_{-27}$ & 286$\pm$9 & 7.90$^{+2.90}_{-1.00}$ & 0.093$^{+0.011}_{-0.016}$ & 0.067$^{+0.003}_{-0.004}$ & 0.119$^{+0.027}_{-0.038}$ & 4.57$^{+0.92}_{-0.76}$ (7) \\
\hline \hline
\end{tabular}}
\vspace{-1.2cm}
\begin{flushleft}
\leftskip 35pt
\tablenotetext{a}{The $T_{500}$ in parenthesis are from the empirical relation between $T_{1500}$
and $T_{500}$ ($\S$4). The $r_{500}$ in  parenthesis are estimated from the $M_{500} - T_{500}$
relation in this work. The values with an asterisk are for tier 2 groups from extrapolation.}
\tablenotetext{b}{The value in parenthesis is the number of radial points between 40 kpc
and $r_{\rm det, spe}$ (including 40 kpc and $r_{\rm det, spe}$) are used in the NFW fit ($\S$7.3).}
\end{flushleft}
\end{center}
\end{table}

\begin{table}
\begin{center}
\caption{Derived properties of groups (II: entropy)}
{\scriptsize
\begin{tabular}{ccccccc} \hline \hline
Group & $K_{500}$\tablenotemark{a} & $K_{1000}$\tablenotemark{a} & $K_{1500}$\tablenotemark{a} & $K_{2500}$ & $K_{\rm 0.15 r500}$ & $K_{\rm 30 kpc}$ \\ \hline

NGC~1550 & (297$^{+46}_{-42}$) & 253$\pm$23 & 228$\pm$15 & 198$^{+9}_{-8}$ & 94$\pm$3 & 49$\pm$1 \\
NGC~3402 & & & (278$^{+218}_{-88}$) & 226$^{+100}_{-52}$ & 66$\pm$4 & 34$\pm$2 \\
A262 & & (385$^{+87}_{-67}$) & 383$^{+73}_{-46}$ & 354$^{+43}_{-29}$ & 161$^{+9}_{-7}$ & 69$\pm$2 \\
NGC~383 & & & (495$^{+83}_{-46}$) & 448$^{+71}_{-39}$ & 327$^{+23}_{-24}$ & 214$^{+28}_{-25}$ \\
3C 449 & & & (217$^{+28}_{-37}$) & 199$^{+22}_{-27}$ & 158$^{+11}_{-12}$ & 96$\pm$6 \\
NGC~533 & & (336$^{+52}_{-76}$) & 317$^{+45}_{-66}$ & 289$^{+44}_{-50}$ & 171$^{+13}_{-11}$ & 103$\pm$6 \\
NGC~741 & & (460$^{+64}_{-114}$) & 428$^{+57}_{-86}$ & 388$^{+53}_{-66}$ & 243$^{+36}_{-29}$ & 146$^{+15}_{-17}$ \\
MKW4 & 574$^{+54}_{-96}$ & 454$^{+44}_{-59}$ & 388$^{+42}_{-39}$ & 332$^{+31}_{-24}$ & 161$\pm$6 & 80$^{+3}_{-4}$ \\
A3581 & & (384$^{+84}_{-78}$) & 344$^{+72}_{-59}$ & 298$^{+49}_{-41}$ & 124$\pm$7 & 38$\pm$1 \\
NGC~5129 & & (262$^{+32}_{-30}$) & 229$^{+23}_{-22}$ & 193$^{+18}_{-17}$ & 100$\pm$6 & 75$^{+6}_{-4}$ \\
NGC~1132 & & & 336$^{+49}_{-45}$ & 238$\pm$26 & 97$^{+7}_{-6}$ & 68$^{+9}_{-7}$ \\
UGC~2755 & & & (260$^{+65}_{-72}$) & 211$^{+50}_{-45}$ & 120$^{+15}_{-14}$ & 113$^{+24}_{-18}$ \\
NGC~4325 & & & (195$^{+86}_{-70}$) & 179$^{+46}_{-36}$ & 84$^{+7}_{-6}$ & 39$\pm$2 \\
HCG~51 & & (371$^{+196}_{-166}$) & 348$^{+120}_{-101}$ & 303$^{+49}_{-47}$ & 126$\pm$5 & 82$\pm$4 \\
3C~442A & (455$^{+49}_{-77}$) & 438$^{+34}_{-49}$ & 431$^{+29}_{-40}$ & 419$^{+33}_{-37}$ & 210$^{+27}_{-26}$ & 72$\pm$8 \\
UGC~5088 & (414$^{+91}_{-191}$) & 326$^{+48}_{-49}$ & 288$^{+32}_{-33}$ & 230$\pm$23 & 84$^{+8}_{-7}$ & 51$^{+6}_{-4}$ \\
NGC~6338 & & (401$^{+46}_{-45}$) & 371$^{+36}_{-35}$ & 334$^{+28}_{-23}$ & 239$^{+14}_{-11}$ & 108$\pm$5 \\
NGC~4104 & (571$^{+74}_{-85}$) & 471$^{+42}_{-50}$ & 421$^{+32}_{-36}$ & 360$^{+26}_{-29}$ & 221$^{+16}_{-15}$ & 148$^{+12}_{-11}$ \\
RBS~461 & & & (446$^{+59}_{-56}$) & 370$^{+43}_{-41}$ & 165$\pm$8 & 93$\pm$5 \\
ESO~552-020 & & (580$^{+107}_{-124}$) & 550$^{+79}_{-87}$ & 460$^{+58}_{-61}$ & 190$\pm$9 & 97$^{+6}_{-5}$ \\
A1177 & (648$^{+130}_{-111}$) & 480$^{+74}_{-70}$ & 397$^{+48}_{-47}$ & 317$^{+28}_{-45}$ & 179$^{+11}_{-12}$ & 138$^{+13}_{-11}$ \\
IC~1262 & & & (489$^{+63}_{-99}$) & 411$^{+70}_{-63}$ & 134$\pm$4 & 47$\pm$2 \\
NGC~6269 & (390$^{+189}_{-183}$) & 440$^{+56}_{-68}$ & 434$^{+54}_{-50}$ & 406$^{+38}_{-40}$ & 272$^{+18}_{-21}$ & 127$^{+7}_{-8}$ \\
ESO~306-017 & (914$^{+203}_{-192}$) & 700$^{+108}_{-81}$ & 596$^{+74}_{-59}$ & 475$^{+44}_{-38}$ & 191$\pm$12 & 86$\pm$6 \\
NGC~5098 & 217$\pm$43 & 210$^{+28}_{-26}$ & 205$\pm$21 & 194$^{+16}_{-17}$ & 84$^{+7}_{-8}$ & 45$\pm$2 \\
A1139 & & & (486$^{+429}_{-206}$) & 431$^{+112}_{-100}$ & 368$^{+69}_{-48}$ & 261$^{+69}_{-43}$ \\
A160 & 477$^{+86}_{-90}$ & 429$^{+34}_{-36}$ & 407$^{+29}_{-28}$ & 379$^{+33}_{-25}$ & 268$^{+17}_{-16}$ & 172$^{+31}_{-26}$ \\
UGC~842 & (652$^{+231}_{-149}$) & 522$^{+90}_{-77}$ & 459$^{+60}_{-58}$ & 406$^{+48}_{-39}$ & 238$^{+22}_{-20}$ & 122$^{+11}_{-9}$ \\
A2717 & 1022$^{+163}_{-181}$ & 817$^{+106}_{-102}$ & 673$^{+71}_{-77}$ & 502$^{+40}_{-41}$ & 186$\pm$11 & 75$\pm$5 \\
RXJ~1022+3830 & 655$^{+89}_{-94}$ & 587$^{+69}_{-64}$ & 548$^{+62}_{-51}$ & 488$^{+72}_{-46}$ & 223$^{+33}_{-30}$ & 77$^{+6}_{-7}$ \\
AS1101 & (476$^{+457}_{-218}$) & 465$^{+142}_{-91}$ & 441$^{+73}_{-53}$ & 386$^{+39}_{-36}$ & 120$^{+15}_{-13}$ & 29$\pm$1 \\
ESO~351-021 & 309$^{+105}_{-84}$ & 315$^{+85}_{-40}$ & 310$^{+58}_{-34}$ & 289$^{+35}_{-38}$ & 175$^{+19}_{-20}$ & 57$\pm$4 \\
A3880 & 624$^{+198}_{-236}$ & 681$^{+103}_{-98}$ & 560$^{+88}_{-52}$ & 420$^{+44}_{-35}$ & 176$\pm$20 & 50$\pm$3 \\
A1991 & 752$^{+252}_{-275}$ & 653$^{+95}_{-142}$ & 568$^{+68}_{-94}$ & 463$^{+43}_{-63}$ & 166$^{+11}_{-23}$ & 39$\pm$1 \\
A1275 & (464$^{+278}_{-98}$) & 324$^{+66}_{-44}$ & 284$^{+23}_{-30}$ & 231$^{+18}_{-17}$ & 130$^{+16}_{-14}$ & 59$\pm$3 \\
A2092 & (423$^{+192}_{-140}$) & 449$^{+90}_{-66}$ & 439$^{+85}_{-69}$ & 401$^{+69}_{-57}$ & 202$^{+24}_{-21}$ & 200$^{+57}_{-34}$ \\
RXJ~1206-0744 & & 432$^{+97}_{-104}$ & 402$^{+88}_{-80}$ & 354$^{+68}_{-67}$ & 201$^{+34}_{-41}$ & 102$^{+32}_{-35}$ \\
A1238 &         & 805$^{+347}_{-165}$ & 676$^{+167}_{-113}$ & 584$^{+114}_{-91}$ & 464$^{+129}_{-76}$ & 341$^{+200}_{-78}$ \\
A744 &             & (695$^{+249}_{-220}$) & 580$^{+134}_{-115}$ & 470$^{+72}_{-63}$ & 215$^{+16}_{-15}$ & 115$^{+12}_{-11}$ \\
A2462 & 627$^{+102}_{-118}$ & 537$^{+79}_{-77}$ & 488$^{+62}_{-60}$ & 422$^{+48}_{-40}$ & 247$^{+12}_{-13}$ & 172$^{+20}_{-21}$ \\
RXJ~1159 & (800$^{+200}_{-121}$) & 580$^{+78}_{-56}$ & 491$^{+49}_{-40}$ & 410$^{+35}_{-28}$ & 230$^{+16}_{-17}$ & 92$\pm$6 \\
A1692 & 622$^{+202}_{-176}$ & 644$^{+186}_{-167}$ & 636$^{+174}_{-156}$ & 576$^{+126}_{-120}$ & 252$\pm$29 & 149$^{+30}_{-24}$ \\
A2550 & 704$^{+132}_{-128}$ & 530$^{+65}_{-67}$ & 431$^{+46}_{-45}$ & 331$^{+27}_{-26}$ & 114$^{+11}_{-8}$ & 46$\pm$3 \\

\hline \hline
\end{tabular}}
\vspace{-1.2cm}
\begin{flushleft}
\leftskip 35pt
\tablenotetext{a}{The values in parentheses are from extrapolation (see $\S$4).}
\end{flushleft}
\end{center}
\end{table}

\begin{table}
\begin{center}
\caption{$K - T$ relations\tablenotemark{a}}
\begin{tabular}{lcc} \hline \hline
Relation & $K_{1}$ & $\alpha$  \\
       & (keV cm$^{2}$) & \\ \hline

$K_{500} - T_{500}$ (tier 1) & 286$\pm$60 & 1.12$\pm$0.28 \\
$K_{500} - T_{500}$ (tier 1+2) & 356$\pm$45 & 0.825$\pm$0.200 \\
$K_{1000} - T_{500}$ (tier 1+2+3) & 310$\pm$19 & 0.832$\pm$0.100 \\
$K_{1500} - T_{500}$ (all groups) & 288$\pm$14 & 0.790$\pm$0.078 \\
$K_{2500} - T_{2500}$ (all groups) & 230$\pm$10 & 0.760$\pm$0.061 \\
$K_{0.15 r500} - T_{2500}$ (all groups) & 114$\pm$9 & 0.778$\pm$0.124 \\
$K_{30 \rm kpc} - T_{2500}$ (all groups) & 68$\pm$9 & 0.459$\pm$0.229 \\
\hline
$K_{500} - T_{500}$ (tier 1 + clusters) & 286$\pm$23 & 1.08$\pm$0.05 \\
$K_{500} - T_{500}$ (tier 1+2 + clusters) & 329$\pm$25 & 0.994$\pm$0.054 \\
$K_{1000} - T_{500}$ (tier 1+2+3 + clusters) & 303$\pm$13 & 0.887$\pm$0.034 \\
$K_{2500} - T_{500}$ (all groups + clusters) & 252$\pm$8 & 0.740$\pm$0.027 \\
$K_{0.15 r500} - T_{500}$ (all groups + clusters) & 137$\pm$9 & 0.494$\pm$0.047 \\

\hline \hline
\end{tabular}
\vspace{-1cm}
\tablenotetext{a}{$E(z)^{4/3} K = K_{1}$ ($T$ / 1 keV)$^{\alpha}$,
where $K_{1}$ is the corresponding entropy at 1 keV. The last four relations
include 14 clusters from V06 and V08. There is little correction between $K_{30 \rm kpc}$
and $T_{2500}$, but we still list it for the studies of intrinsic scatter (Fig. 11).
We used the BCES (Y$|$X) regression (Akritas \& Bershady 1996) as the temperature
errors are smaller than the entropy errors.}
\end{center}
\end{table}

\begin{table}
\begin{center}
\caption{$M_{500} - T_{500}$ relation\tablenotemark{a}}
\begin{tabular}{cccc} \hline \hline
Sample & $M_{3}$ & $\alpha$ & $r_{3}$ \\
       & (10$^{14} h^{-1} M_{\odot}$) &  & ($h^{-1}$ Mpc) \\ \hline

Tier 1 & 1.17$\pm$0.21 & 1.64$\pm$0.21 & 0.587$\pm$0.035 \\
Tier 1 + clusters & 1.21$\pm$0.08 & 1.68$\pm$0.04 & 0.593$\pm$0.014 \\
Tier 1+2 & 1.27$\pm$0.12 & 1.67$\pm$0.15 & 0.602$\pm$0.020 \\
Tier 1+2 + clusters & 1.26$\pm$0.07 & 1.65$\pm$0.04 & 0.600$\pm$0.011 \\
Tier 1+2 ($>$ 1 keV) + clusters & 1.27$\pm$0.06 & 1.60$\pm$0.03 & 0.602$\pm$0.010 \\

\hline \hline
\end{tabular}
\vspace{-1cm}
\tablenotetext{a}{$E(z) M_{500} = M_{3}$ ($T_{500}$ / 3 keV)$^{\alpha}$,
where $T_{500}$ is spectroscopic temperature. $r_{3}$ is the corresponding
scale in the $r_{500} - T_{500}$ relation ($E(z) r_{500} = r_{3}$ ($T_{500}$ /
3 keV)$^{\alpha/3}$). The cluster sample includes fourteen
$T_{500} > 3.7$ keV systems from V08. We used the BCES orthogonal regression
(Akritas \& Bershady 1996).}
\end{center}
\end{table}

\begin{table}
\begin{center}
\caption{$M_{500} - Y_{\rm X, 500}$ relation\tablenotemark{a}}
\begin{tabular}{ccc} \hline \hline
Sample & $M_{\rm C}$ & $\alpha$ \\
       & (10$^{14} h^{-1} M_{\odot}$) &  \\ \hline

Tier 1 & 1.12$\pm$0.16 & 0.605$\pm$0.050  \\
Tier 1 + clusters & 1.10$\pm$0.07 & 0.588$\pm$0.012  \\
Tier 1+2 & 1.12$\pm$0.10 & 0.564$\pm$0.031 \\
Tier 1+2 + clusters & 1.14$\pm$0.05 & 0.571$\pm$0.010  \\

\hline \hline
\end{tabular}
\vspace{-1cm}
\tablenotetext{a}{$E(z)^{2/5} M_{500} = M_{\rm C}$ ($Y_{\rm X, 500}$ / 4$\times10^{13}$ keV $M_{\odot}$)$^{\alpha}$.
The cluster sample includes fourteen $T_{500} > 3.7$ keV systems from V08.
We used the BCES orthogonal regression (Akritas \& Bershady 1996).}
\end{center}
\end{table}

\clearpage

\begin{figure}
\centerline{\includegraphics[height=0.4\linewidth]{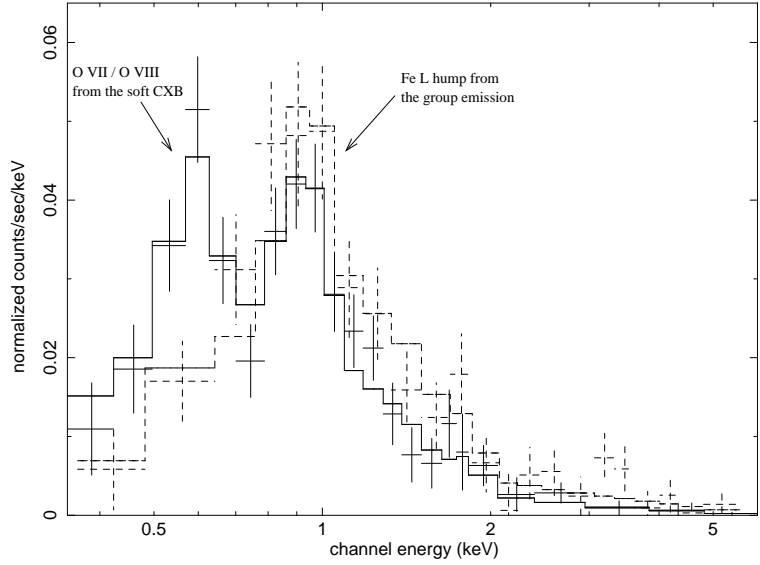}}
  \caption{The spectra of the NGC~1550 in the outermost radial bin that is to the edge
of the \chandra\ FOV (one from the S1 chip of the ObsID 5800 in the solid line and
another from the S2 chip of the ObsID 3187 in the dashed line). Besides the still
significant iron L hump from the group emission, the O hump from the soft CXB is also
strong in the S1 spectrum, which allows a robust separation of these two components.
}
\end{figure}

\begin{figure}
\centerline{\includegraphics[height=0.5\linewidth,angle=270]{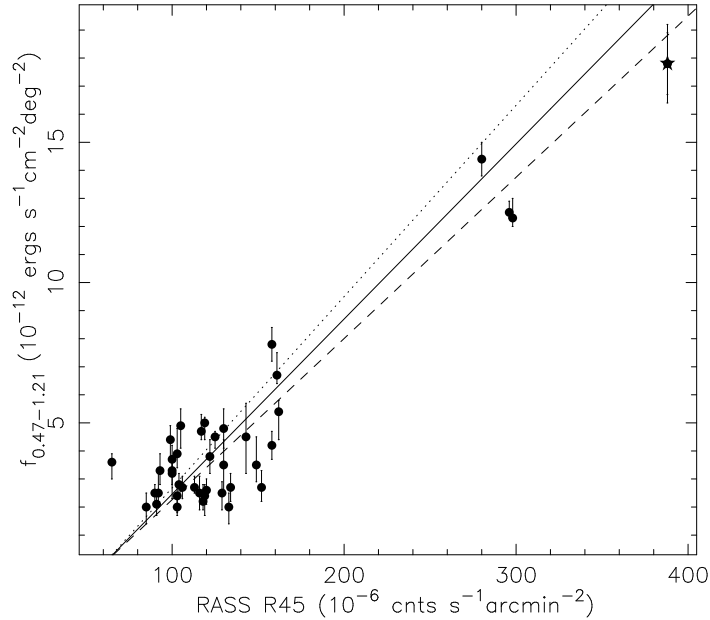}}
  \caption{The 0.47 - 1.21 keV observed flux surface density of the soft
CXB from the \chandra\ data vs. the RASS R45 flux (in RASS channels of 13 - 18)
measured in an annulus around the target (inner radius of 0.4-0.8 deg and the
outer radius is the inner radius + 0.4 deg). The three lines are the expected
conversions between two fluxes, with the assumed two thermal components for
the soft CXB (all with $T_{\rm cool}$ = 0.1 keV). The solid line is for:
$T_{\rm hot}$ = 0.25 keV and NORM$_{\rm hot}$ / NORM$_{\rm cool}$ = 0.5. The
dotted line is for the $T_{\rm hot}$ = 0.2 keV and NORM$_{\rm hot}$ /
NORM$_{\rm cool}$ = 0.5. The dashed line is for: $T_{\rm hot}$ = 0.3 keV and
NORM$_{\rm hot}$ / NORM$_{\rm cool}$ = 1. The
assumed absorption is 4$\times10^{20}$ cm$^{-2}$. The total hard CXB
(1.74$\times10^{-11}$ ergs s$^{-1}$ cm$^{-2}$ deg$^{-2}$ in the 2-8 keV band
from K07) is added in the conversion. The good agreement on average can be seen.
One should be aware that the RASS R45 flux is extracted in a much larger area
surrounding the interested group range and some of the soft CXB may come from
the time-variable SWCX emission. We also include 3C~296 ($T_{\rm hot} = 0.38 \pm 0.03$)
in the high CXB flux end (star). It is an 1 keV group but was left out
as the high local soft CXB (on NPS) prohibits deriving gas properties to $r_{2500}$. 
}
\end{figure}

\begin{figure}
\centerline{\includegraphics[height=1.25\linewidth]{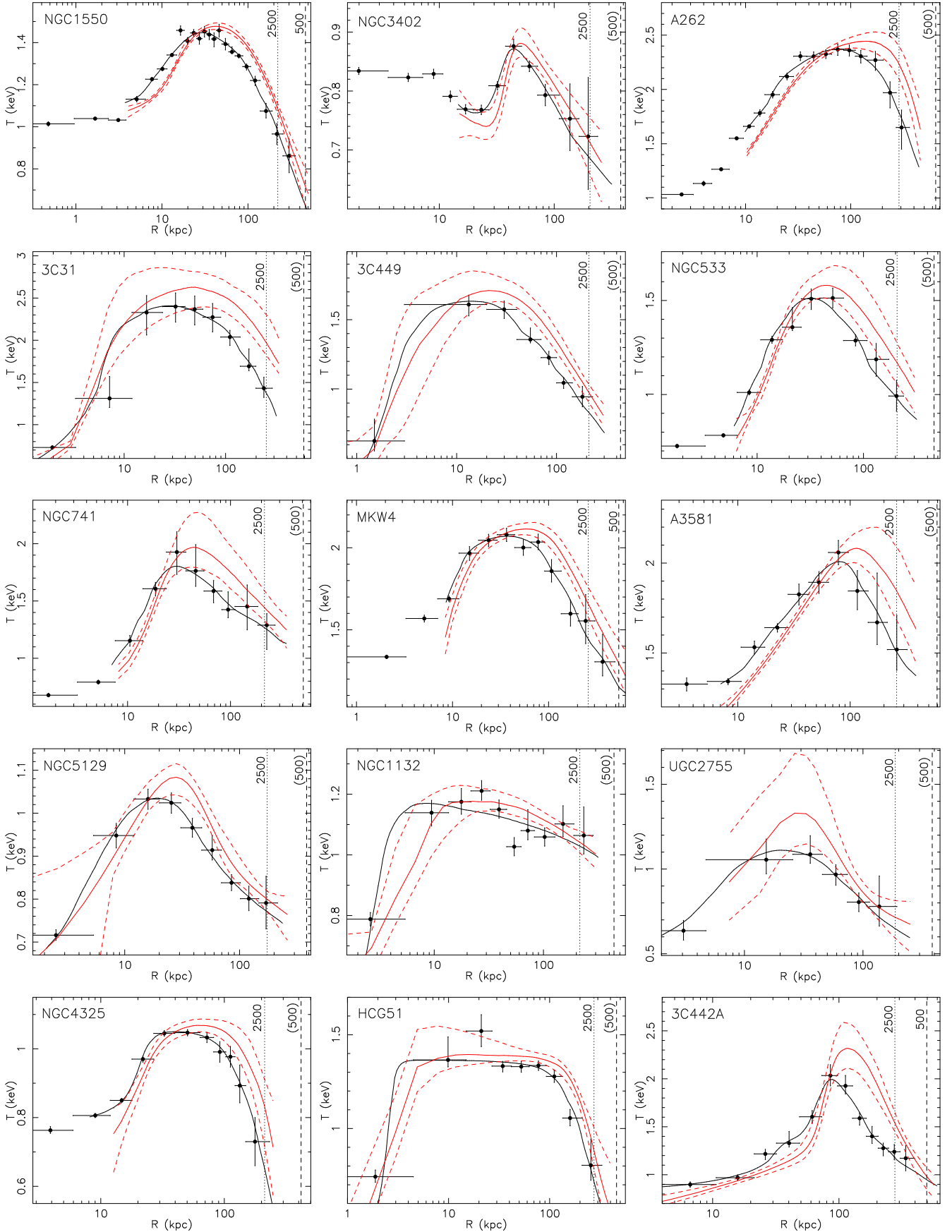}}
  \caption{The temperature profiles of 43 groups in our sample. The red lines
are the reconstructed 3D temperature profile with the 1 $\sigma$ errors,
derived from 1000 simulations. The black line is the best-fit projected
temperature profile. The effective radius of each bin for the projected temperature
profile is derived by weighting the projected emissivity profile with the actual
spatial coverage of the each bin. $r_{2500}$ and $r_{500}$ are marked. The $r_{500}$
in parentheses is estimated from the $M_{500} - T_{500}$ relation.
}
\end{figure}
\clearpage

\begin{figure}
\centerline{\includegraphics[height=1.25\linewidth]{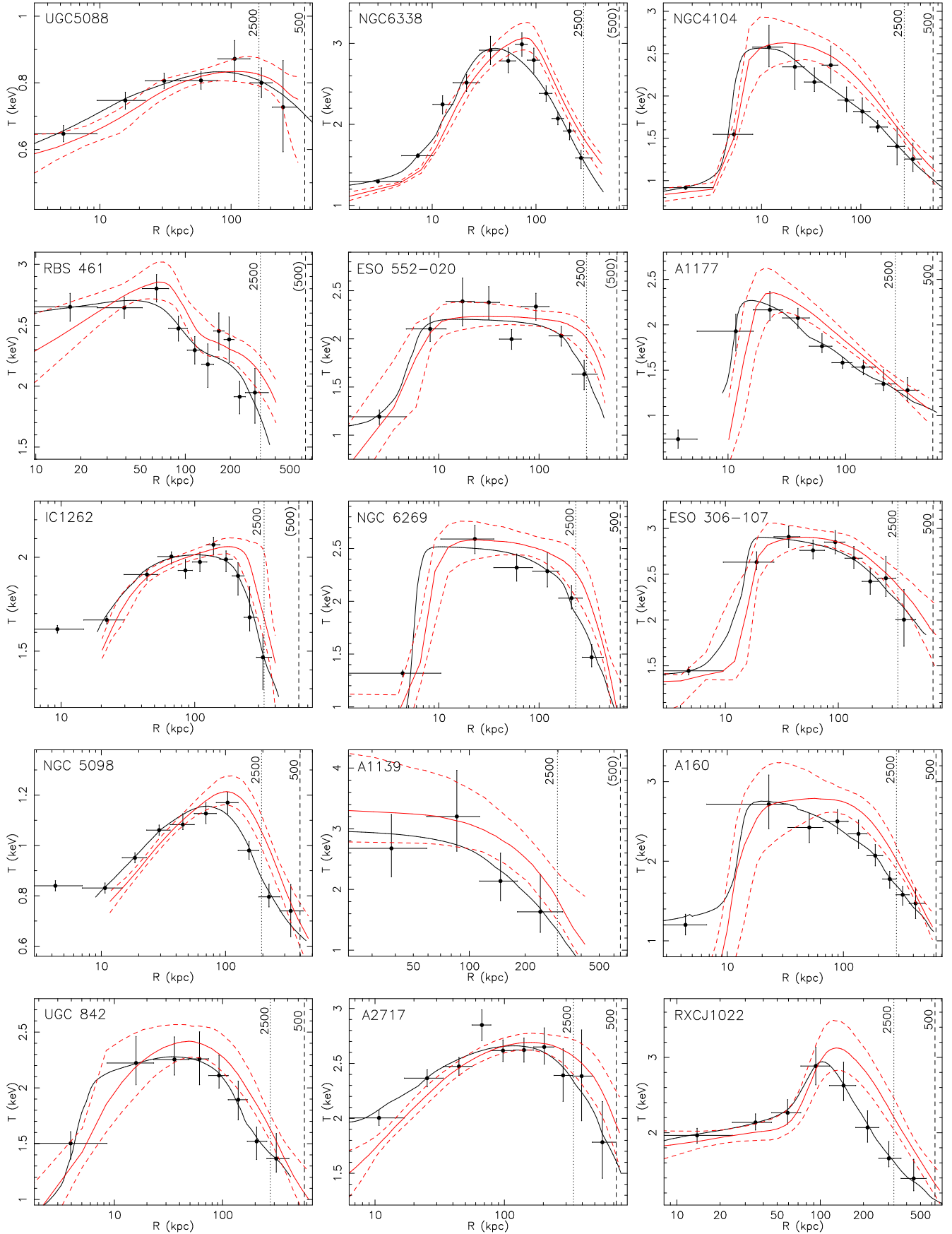}}
  \caption{Continue from Fig. 3.
}
\end{figure}
\clearpage

\begin{figure}
\centerline{\includegraphics[height=1.25\linewidth]{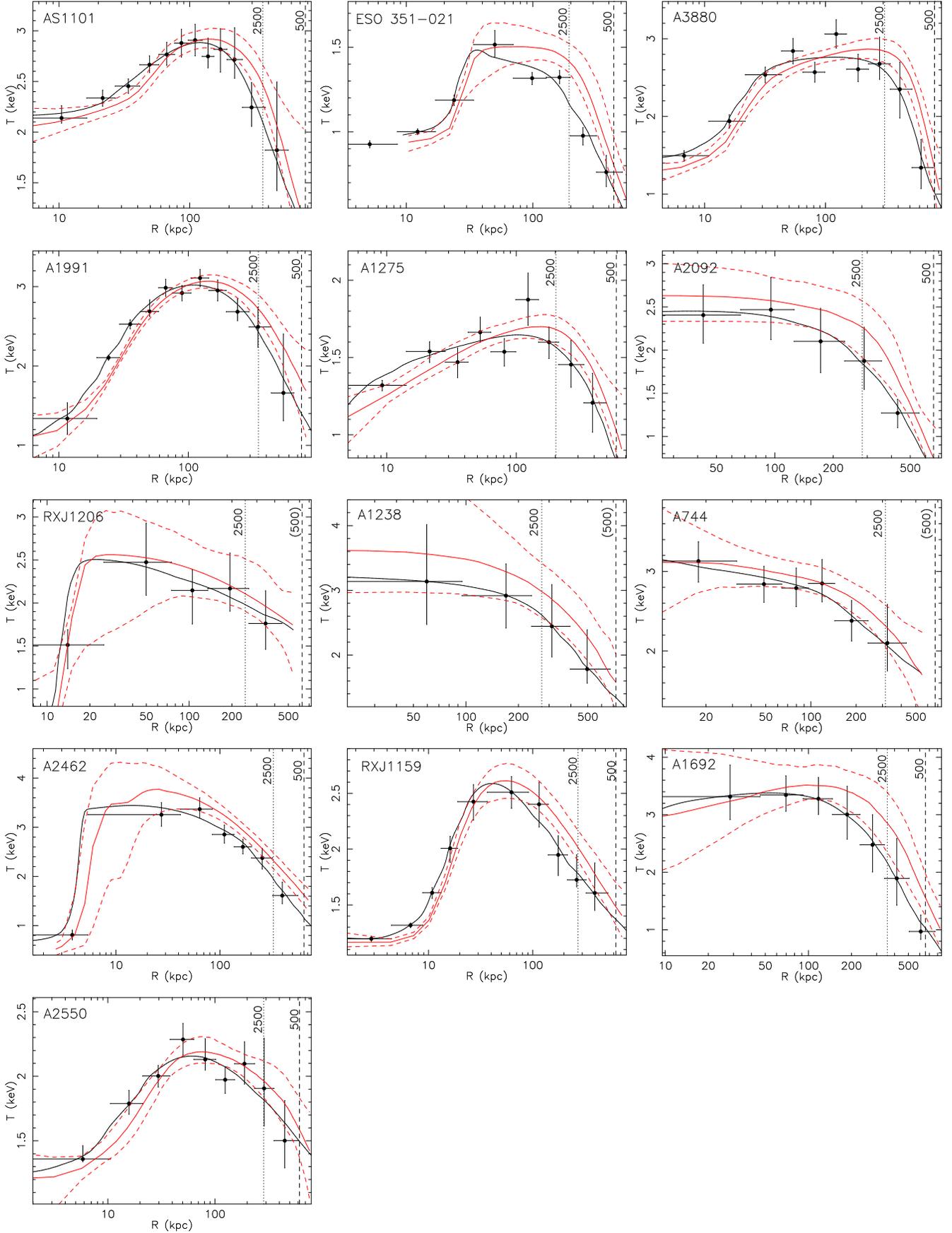}}
  \caption{Continue from Fig. 4.
}
\end{figure}
\clearpage 

\begin{figure}
\centerline{\includegraphics[height=0.5\linewidth,angle=270]{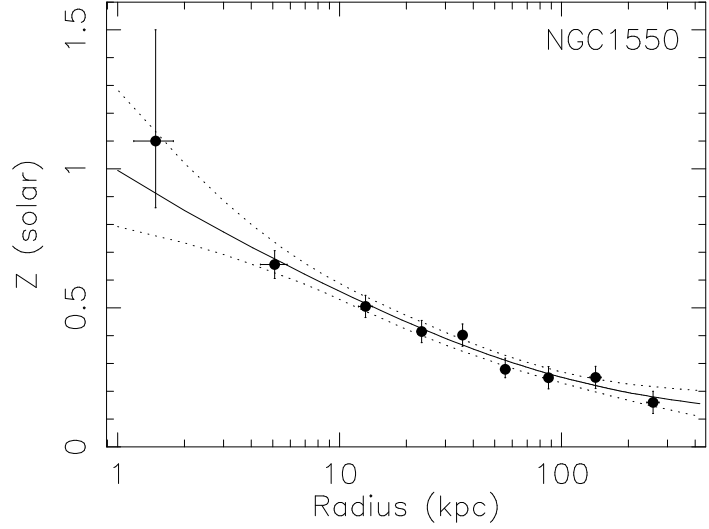}}
  \caption{The best constrained 3D abundance profile in this sample with the best
parametric fit ($\S$3.3) and 1 $\sigma$ errors from 1000 simulations. 
The 1000 simulated abundance profiles are all
used to derive 1000 simulated temperature and density profiles.
As errors of other abundance profiles in this sample are larger and there are
fewer bins, our simple 3D abundance model ($\S$3.3) always fits well
beyond the central 10 kpc.
}
\end{figure}

\begin{figure}
\centerline{\includegraphics[height=0.5\linewidth,angle=270]{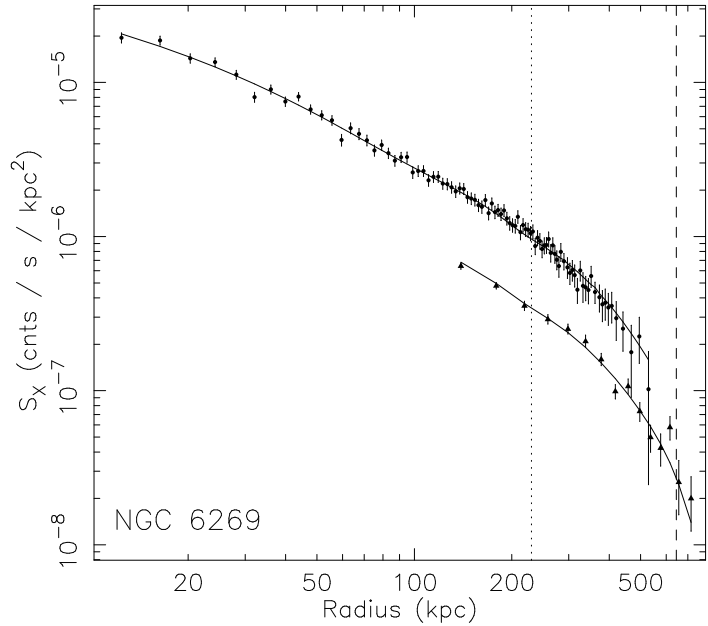}}
  \caption{One example of the surface brightness profiles (\chandra\ +
PSPC) with the best fits derived from the best-fit 3D temperature and abundance profiles.
The \chandra\ profile is the upper one, while the PSPC profile is the lower one.
We generated response files for each \chandra\ radial bin.
Note that the density errors are derived from 1000 Monte Carlo simulations
with 1000 simulated 3D temperature and abundance profiles folded in.
The dashed and dotted lines mark $r_{2500}$ and $r_{500}$ (see Fig. 4 for NGC~6269).
}
\end{figure}
\clearpage

\begin{figure}
\centerline{\includegraphics[height=0.47\linewidth]{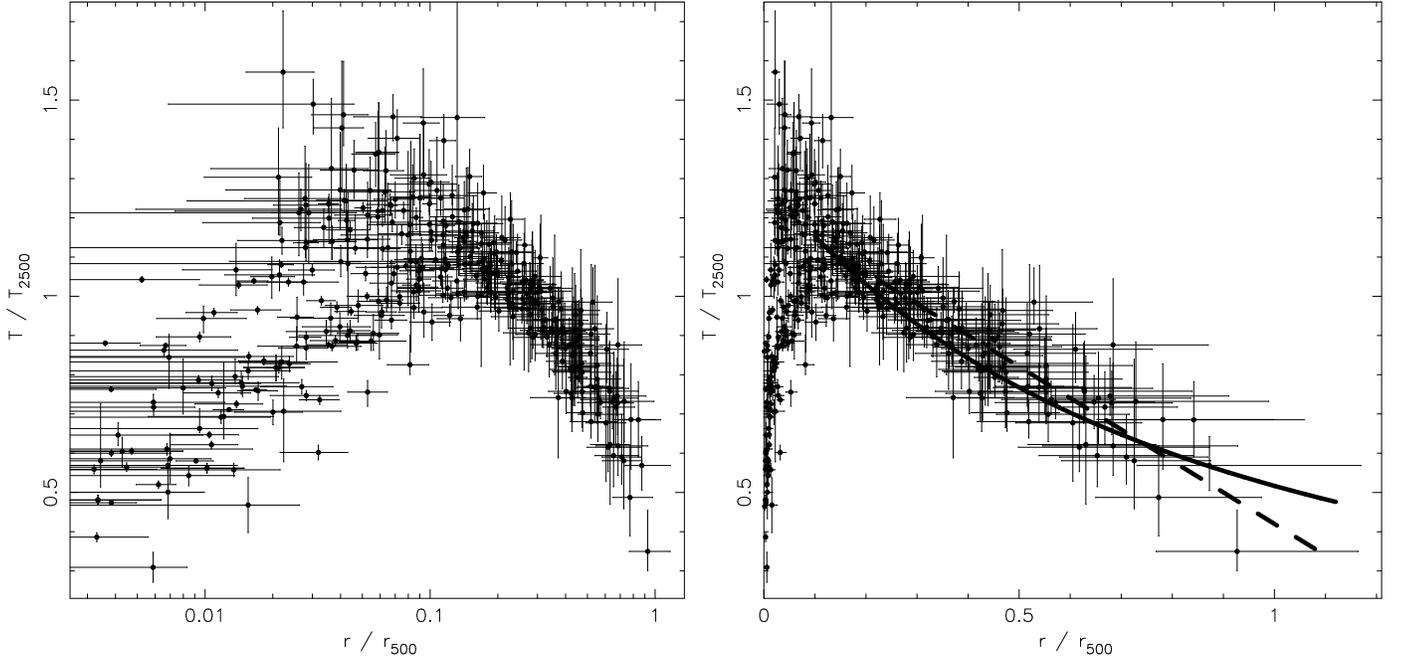}}
  \caption{Temperature profiles in the logarithmic and linear scales of $r_{500}$.
Temperatures are scaled by $T_{2500}$. Despite the large scatter at small radii,
the temperature profiles outside of 0.2 $r_{500}$ are generally similar. The thick
solid line in the linear plot is the universal temperature profile (also projected)
derived from the simulations in Loken et al. (2002). We simply used $T_{2500}$ to replace
$T_{0}$ in Loken et al. (2002). Good agreement can be seen even though the normalization
is not adjusted. The thick dashed line is a simple linear fit to the data (see $\S$5).
}
\end{figure}

\begin{figure}
\vspace{-1.2cm}
\centerline{\includegraphics[height=0.5\linewidth,angle=270]{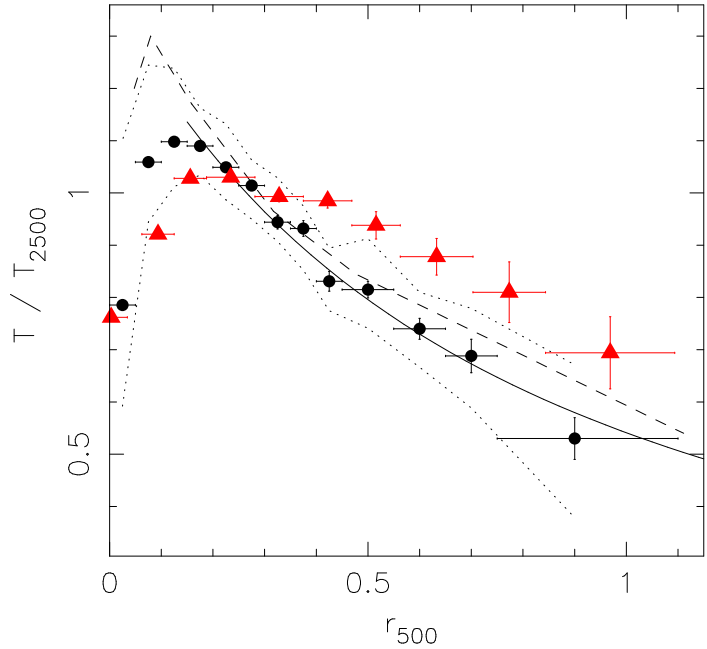}}
  \caption{Mean temperature profile of groups (black circles) and the 1-$\sigma$ scatter
in dotted lines. The solid line is the best-fit from equ. 6. The dashed
line is the mean temperature profile of 1-3 keV systems from Borgani et al.
(2004) simulations. The data points in red triangles are the mean temperature
profile from LM08 on 48 $kT >$ 3.3 keV clusters
at $z$ = 0.1 - 0.3. Note the mean temperature $T_{\rm M}$ defined in
LM08 is computed by fitting the profile with a constant after excluding the
central 0.1 $r_{180}$ region. It should be smaller than $T_{2500}$ as
$T_{2500}$ is emission-weighted within $r_{2500}$.
Nevertheless, it is clear that the group temperature profiles are more
peaky than those of clusters around the center.
}
\end{figure}
\clearpage

\begin{figure}
\centerline{\includegraphics[height=0.4\linewidth,angle=270]{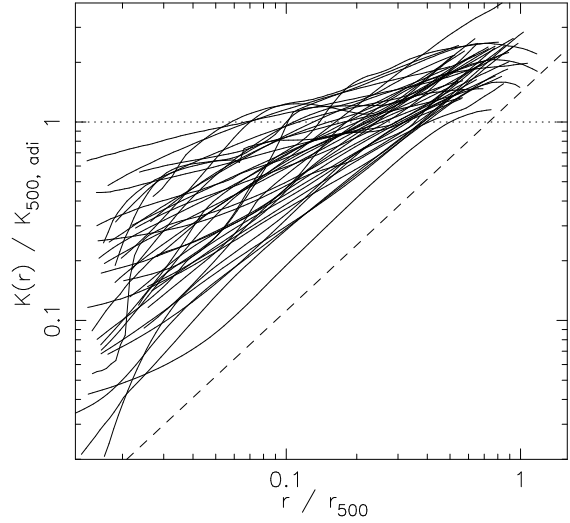}}
  \caption{The entropy profiles scaled by $K_{\rm 500, adi}$ ($\S$6).
The dashed line represents the baseline entropy profile derived by VKB05
with a power index of 1.1.
The observed entropy profiles all lie above the baseline.
}
\end{figure}

\begin{figure}
\vspace{-5cm}
\centerline{\includegraphics[height=1.27\linewidth]{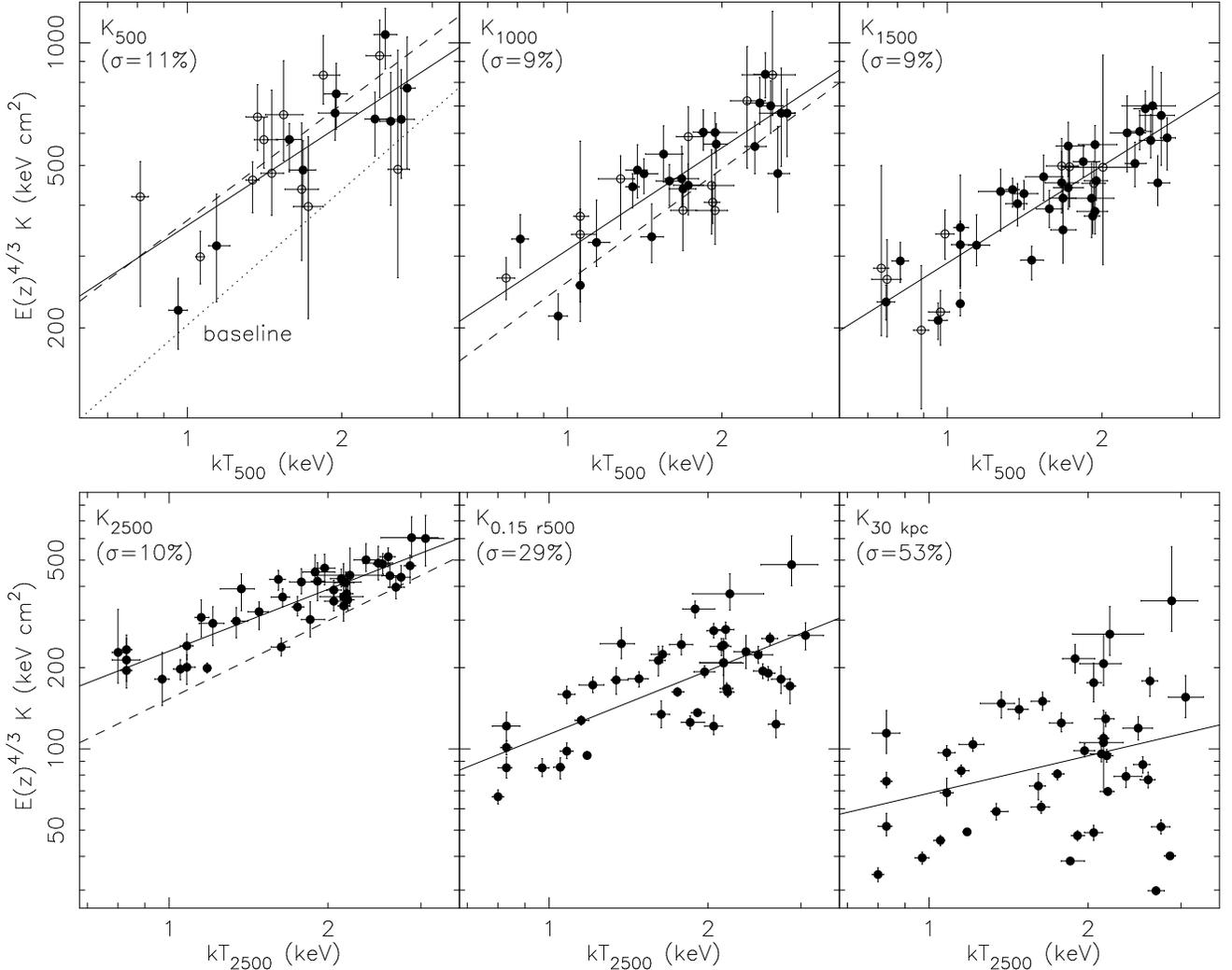}}
\vspace{-4.8cm} 
  \caption{Entropy values at $r_{500}$, $r_{1000}$, $r_{1500}$, $r_{2500}$,
0.15 $r_{500}$ and 30 kpc vs. system temperature ($T_{500}$ or $T_{2500}$).
The open data points (in the upper three panels) are based on extrapolation of
the temperature and density profiles (see $\S$4 for details).
The solid lines are the best-fits to our data from the BCES (Y$|$X) estimator (Table 5),
while the dashed lines are the best-fits from the NKV07 simulations.
The entropy excess above the NKV07 simulations (with cooling and SF) is
significant at $r_{2500}$, while the agreement is better at larger radii.
The dotted line in the $K_{500}$ plot represents the base-line entropy by
VKB05 (or 1.40 $K_{\rm 500, adi}$, $\S$6). Note that the baseline has a slope
of 1.1 rather than 1.0 as the $M_{500} - T_{500}$ relation used (the fourth row of
Table 6) has a slope of 1.65 (rather than 1.5).
It is also clear that the entropy values at 0.15 $r_{500}$ and 30 kpc
show large intrinsic scatter. The measured intrinsic scatter in the $K - T$
relations decreases with radius and stays the same from $r_{2500}$ to $r_{500}$ at
$\sim$ 10\%.
}
\end{figure}

\begin{figure}
\centerline{\includegraphics[height=0.7\linewidth,angle=270]{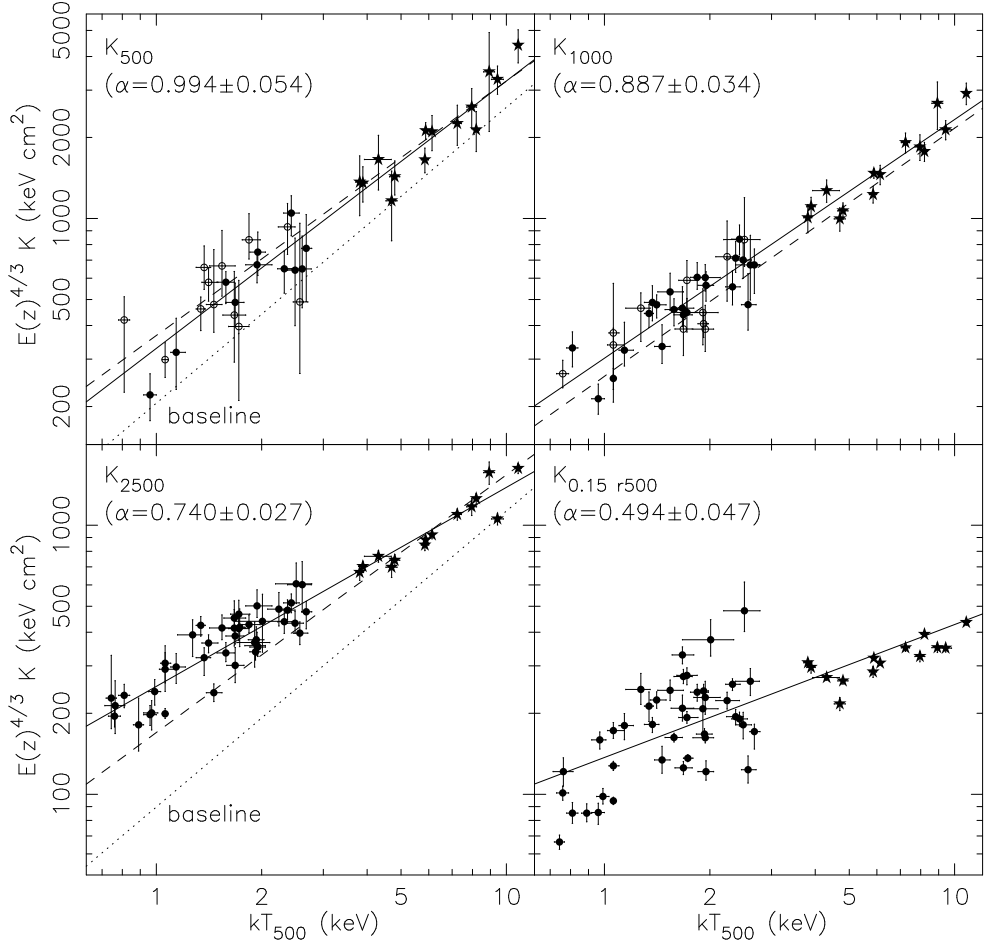}}
  \caption{$K - T_{500}$ relations at $r_{500}$, $r_{1000}$, $r_{2500}$ and 0.15 $r_{500}$
for groups in our sample and 14 clusters from V08. The solid lines are the best-fits of
the data from the BCES (Y$|$X) estimator (Table 5), while the dashed lines are
the best-fits from the NKV07 simulations. The dotted lines in the $K_{500}$ and $K_{2500}$
plots represent the base-line entropy by VKB05.
The best-fit slopes from observations are
also shown. The agreement between observations and the NKV07 simulations becomes better
with increasing radius. At $r_{2500}$, the observed entropy values are on average 56\% - 22\%
higher than those from the NKV07 simulations at 0.8 - 2.5 keV.
At $r_{1000}$, the difference is 18\% - 8\% from 0.8 - 10 keV.
At $r_{500}$, the NKV07 line is basically the same as our best-fit, which also
has a slope expected from the self-similar relation (1.0).
The $K_{0.15 r500} - T_{500}$ relation has significant scatter.
Almost all clusters in the V08 sample have dense cool cores. Inclusion of
non-cool-core clusters may steepen the $K - T$ relation at 0.15 $r_{500}$.
}
\end{figure}

\vspace{-1cm}
\begin{figure}
\centerline{\includegraphics[height=0.96\linewidth,angle=270]{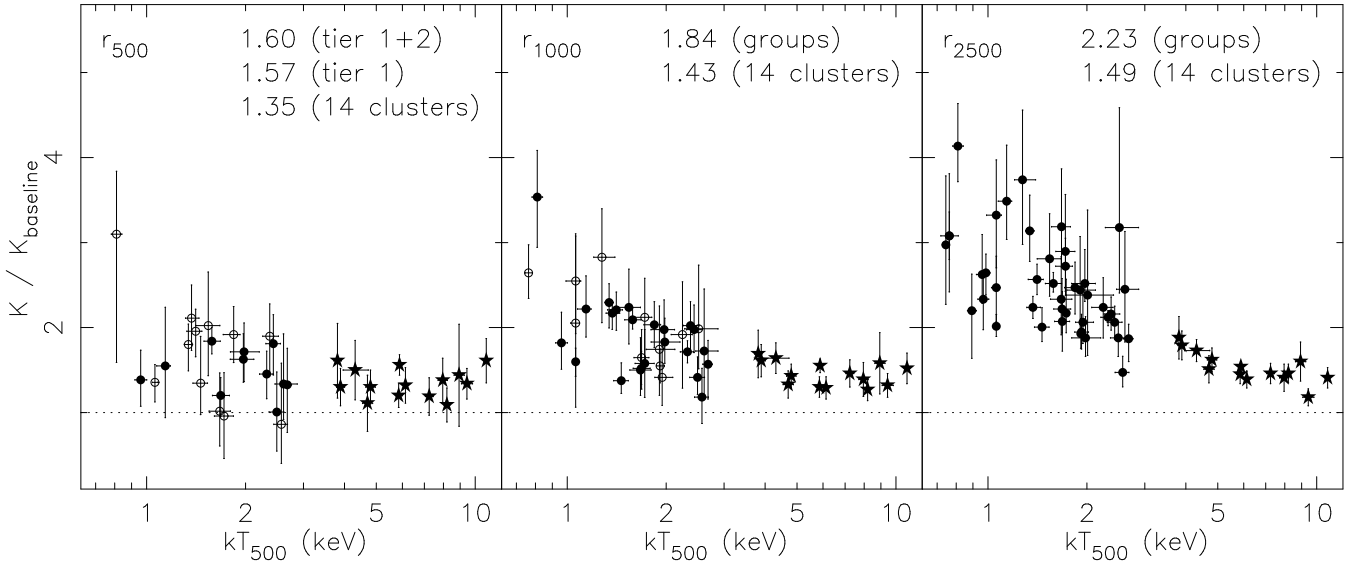}}
  \caption{The ratios of the observed entropy values to the entropy baseline from
VKB05 at $r_{500}$, $r_{1000}$ and $r_{2500}$. Fourteen clusters from V06 and V08 are included
for comparison. The open data points are from extrapolation. We also show weighted
means at each radius for the group and cluster samples. The observed entropy values are
always larger than or comparable to the baseline at all radii, but the average ratios
decrease with radius for both clusters and groups. The decrease
is more dramatic in groups.
}
\end{figure}

\begin{figure}
\centerline{\includegraphics[height=0.8\linewidth,angle=270]{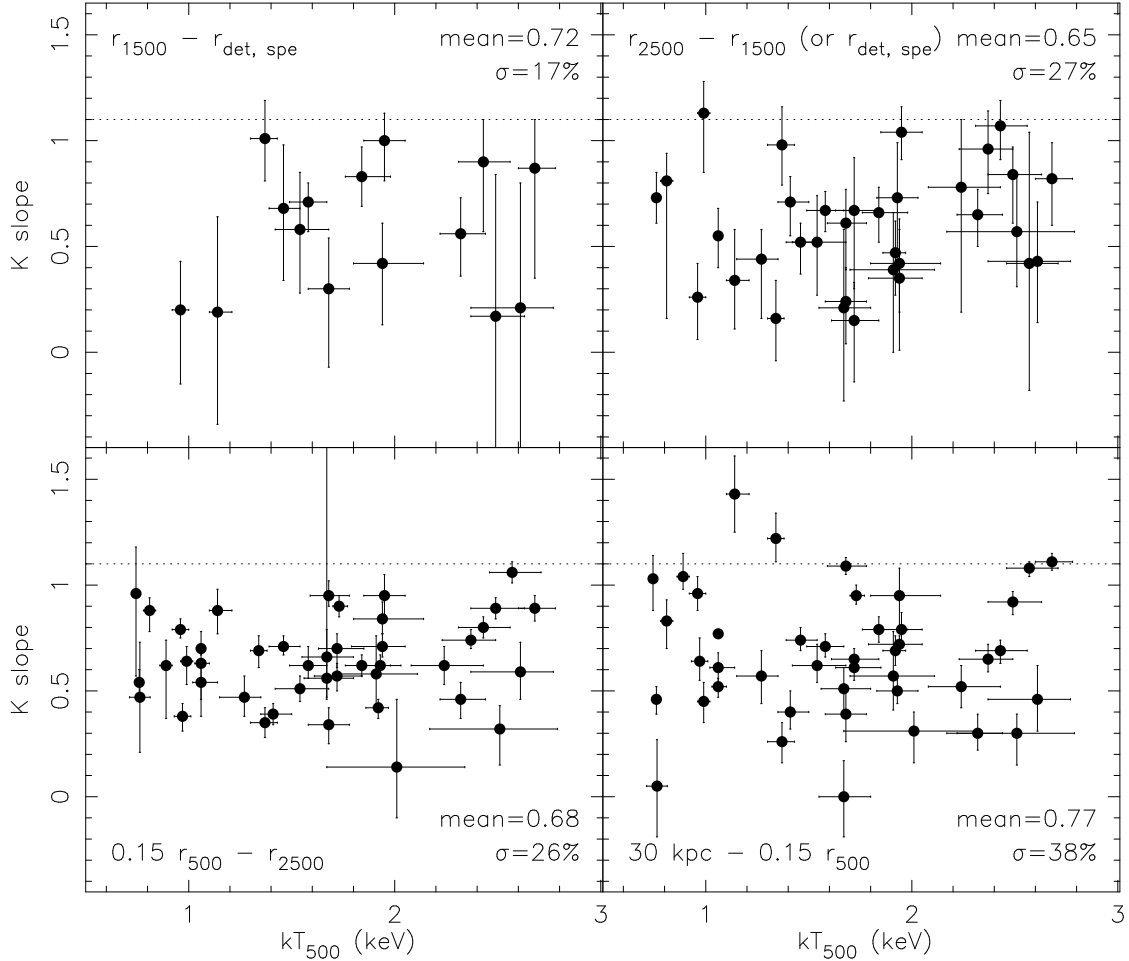}}
  \caption{The entropy slopes between $r_{1500}$ - $r_{\rm det, spe}$, $r_{2500}$ - $r_{1500}$
(or $r_{\rm det, spe}$), 0.15 $r_{500}$ - $r_{2500}$ and 30 kpc - 0.15 $r_{500}$ vs. $T_{500}$.
Beyond 0.15 $r_{500}$, the slopes are always shallower than 1.1 and the weighted averages
are all around 0.7.
}
\end{figure}

\begin{figure}
\centerline{\includegraphics[height=0.42\linewidth]{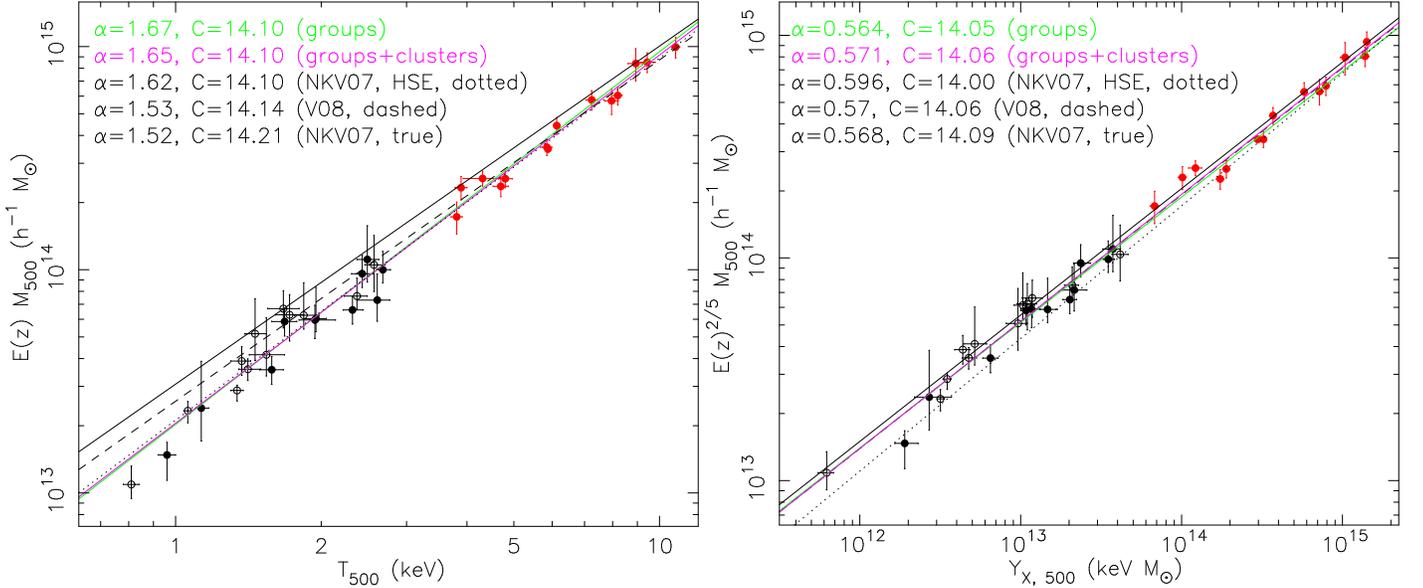}}
  \caption{$M_{500} - T_{500}$ (left panel) and $M_{500} - Y_{\rm X, 500}$ (right panel) relations, combining the
results from this work and V08. The open data points are tier 2 groups. The solid
lines are the relations for the real mass in the NKV07 simulations, while the dotted
lines are the relations for the mass derived under the assumption of hydrostatic
equilibrium in the NKV07 simulations. The dashed lines are the best-fit relations from
V08. The $M_{500} - T_{500}$ relation can be well described by a power law down to at
least $M_{500}$ of 2 $\times 10^{13}$ h$^{-1}$ M$_{\odot}$, although the HSE mass
may be systematically lower than the real mass. The $M_{500} - Y_{\rm X, 500}$ relation
has a smaller scatter and the agreement with the NKV07 simulations is much better.
Note as $Y_{\rm X, 500} \propto h^{-2.5}$, the $h$-dependence of $M_{500}$ should be
$h^{1/2}$ for the self-similar relation, $M_{500} \propto Y_{\rm X, 500}^{3/5}$.
We still use the $h^{-1}$ dependence to directly compare with the left panel
}
\end{figure}

\begin{figure}
\centerline{\includegraphics[height=0.45\linewidth,angle=270]{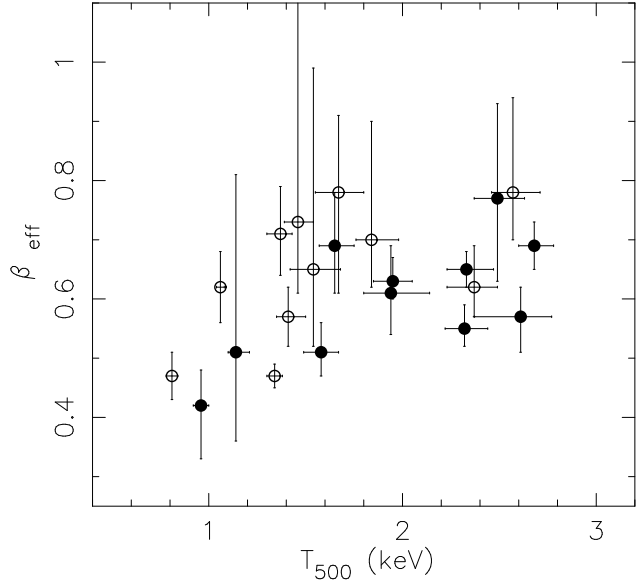}}
  \caption{The density gradient at $r_{500}$ in terms of $\beta_{\rm eff}$
(see the appendix of V06) for tier 1 and 2 groups that $M_{500}$ is derived.
Most groups have $\beta_{\rm eff}$ of 0.55 - 0.75.
}
\end{figure}

\begin{figure} 
\centerline{\includegraphics[height=0.45\linewidth,angle=270]{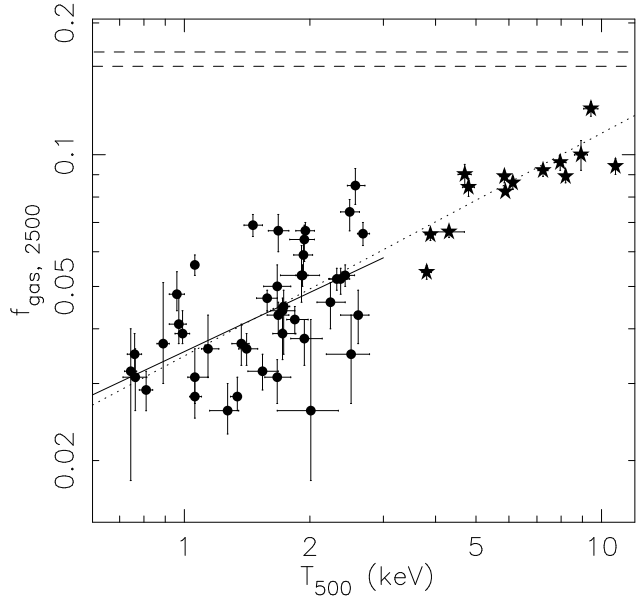}}
  \caption{The enclosed gas fraction within $r_{2500}$ vs. $T_{500}$ (groups +
14 clusters from V06 and V08). The solid line is the BCES fit to the group
sample (0.0355$\pm$0.0018 ($T_{500}$/1 keV)$^{0.449\pm0.096}$),
while the dotted line is the BCES fit to the group + cluster sample
(0.0347$\pm$0.0016 ($T_{500}$/1 keV)$^{0.509\pm0.034}$). The intrinsic scatter on the
$f_{\rm gas, 2500} - T_{500}$ relation is 22\%.
Two dashed lines enclose the 1 $\sigma$ region of the universal baryon fraction
derived from the \wmap\ 5-year data combined with the data of the Type Ia supernovae
and the Baryon Acoustic Oscillations (0.1669$\pm$0.0063, Komatsu et al. 2008).
}
\end{figure}

\begin{figure}
\centerline{\includegraphics[height=0.95\linewidth,angle=270]{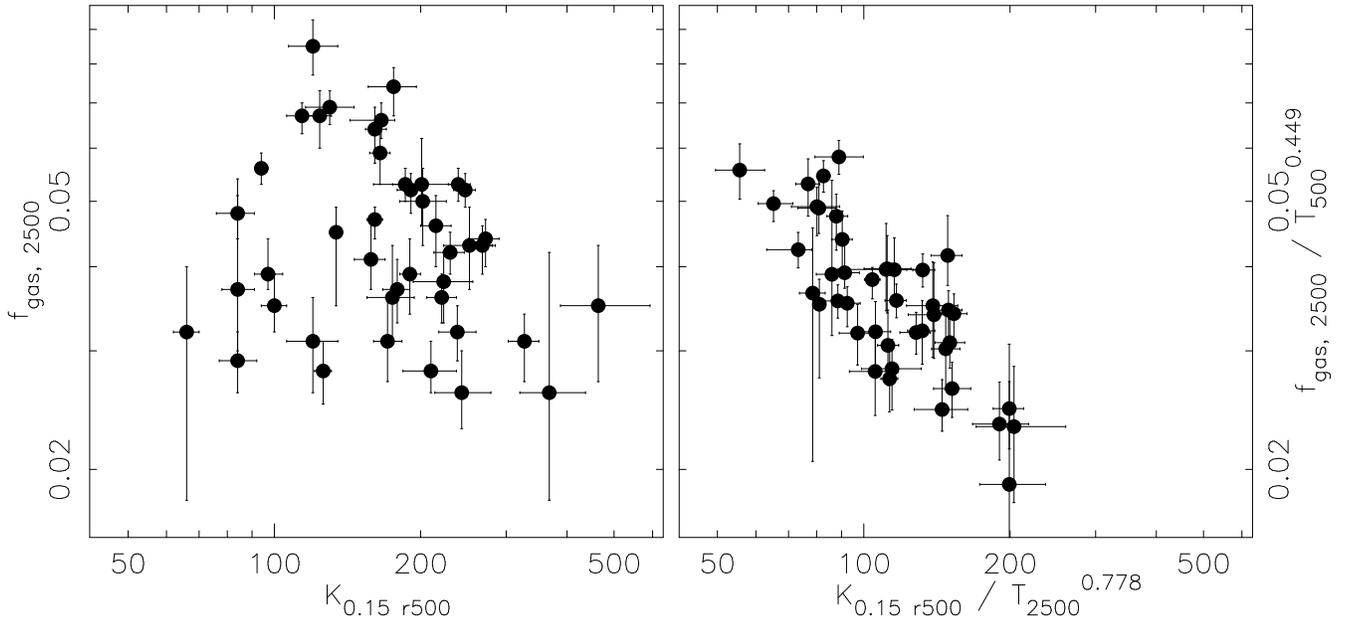}}
  \caption{The enclosed gas fraction within $r_{2500}$ is correlated with
the entropy at 0.15 $r_{500}$, after the temperature dependence on both variables
are removed (right panel). The slope is about -0.7. The intrinsic scatter on
the scaled $f_{\rm gas, 2500}$ and the scaled
$K_{0.15 r500}$ is 11\% and 14\% respectively, compared with 22\% and
29\% intrinsic scatter in their relations to temperature.
This is primarily driven by density - density correlation, but quantitatively
shows their connection.
}
\end{figure}

\begin{figure}
\centerline{\includegraphics[height=1.0\linewidth,angle=270]{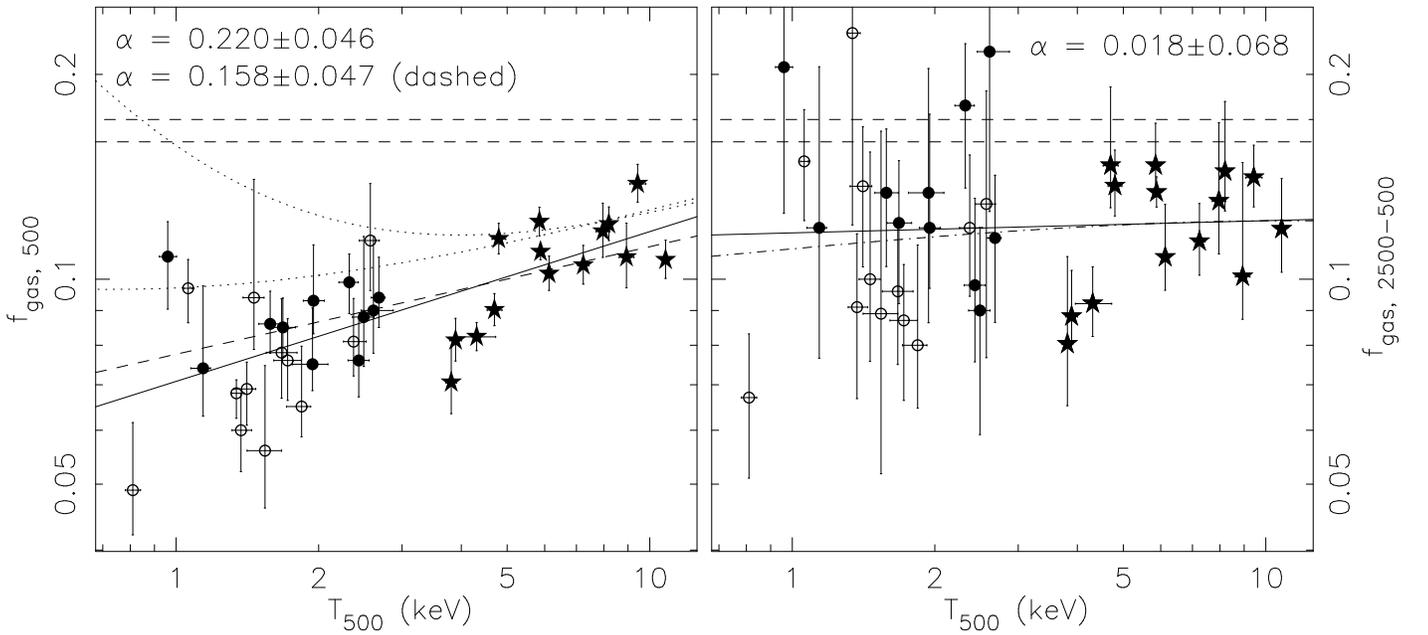}}
  \caption{The enclosed gas fraction within $r_{500}$ (left panel) and between
$r_{2500}$ - $r_{500}$ (right panel) vs. $T_{500}$. The open data points are tier 2 groups.
The solid lines are fits to all 23 groups and 14 V08 clusters, 0.0708$\pm$0.0046
($T_{500}$/1 keV)$^{0.220\pm0.046}$, while the dashed line in the left panel is
the fit excluding 12 tier 2 groups, 0.0776$\pm$0.0057 ($T_{500}$/1 keV)$^{0.158\pm0.047}$.
The listed power law index can be compared with that in NKV07 simulations
(0.152). Two horizontal dashed lines enclose the 1 $\sigma$ region of the
universal baryon fraction (see the caption of Fig. 17). We also estimate
the total baryon fraction within $r_{500}$ by adding the $f_{\rm gas, 500} - T_{500}$
relation from 23 groups and 14 V08 clusters and the relation for stellar fraction
from Lin et al. (2003) and Gonzalez et al. (2007), as shown by the two dotted
lines. The upper one uses the Gonzalez et al. relation, while the lower one
uses the Lin et al. relation (see $\S$7.4, note that the  Gonzalez et al. relation
includes the intracluster light). It appears that gas fraction between
$r_{2500}$ - $r_{500}$ has no temperature dependence ($f_{\rm gas} \sim$ 0.12)
on average, although there is still scatter. The solid line is the fit to all
23 groups and 14 clusters with a slope of 0.018$\pm$0.068. The dashed-dotted line
is the expected average $f_{\rm gas, 2500-500}$ expected from the best-fit
$f_{\rm gas, 500} - T_{500}$ and $f_{\rm gas, 2500} - T_{500}$
scaling relations in this work (see $\S$7.2).
}
\end{figure}

\begin{figure}
\centerline{\includegraphics[height=0.4\linewidth,angle=270]{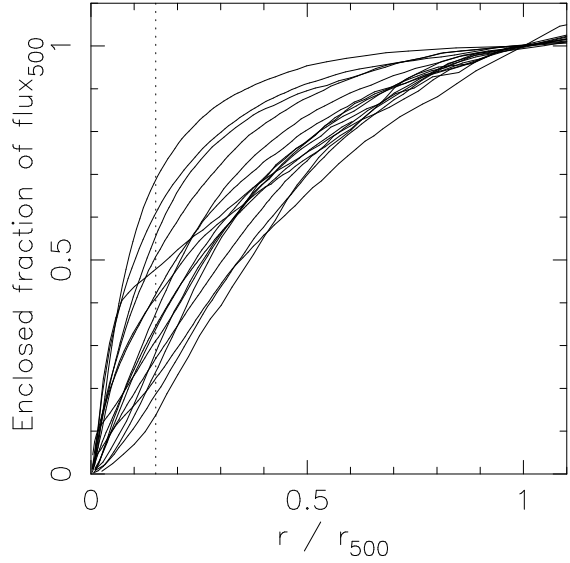}}
  \caption{The enclosed fraction of the count flux for 17 groups in our
sample that $r_{500}$ is reached by \chandra\ or (and) PSPC. The dotted line
marks the position of 0.15 $r_{500}$. There is large scatter for the
enclosed fractions within 0.15 $r_{500}$ and $r_{2500}$ ($\sim$ 0.465 $r_{500}$ on
average in this sample), mainly depending on the existence of a central cool core.
}
\end{figure}

\begin{figure}
\centerline{\includegraphics[height=0.45\linewidth,angle=270]{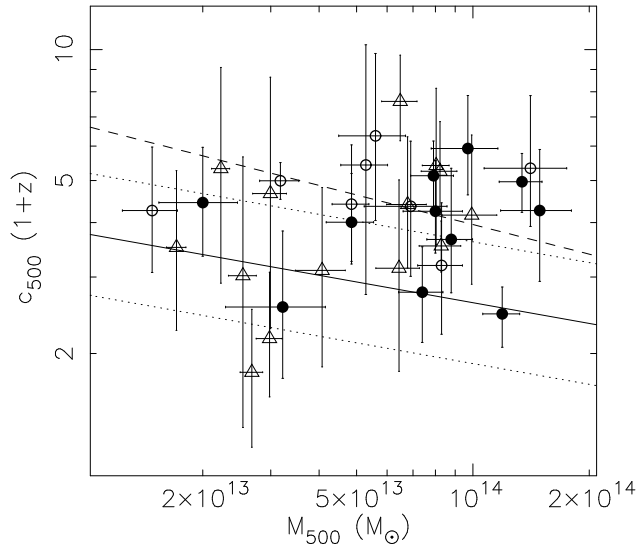}}
  \caption{$c_{500} - M_{500}$ relation from this work. The filled and open
circles are for tier 1 and 2 groups respectively. The open triangles are
for tier 3 and 4 groups with mass estimated from the $M_{500} - T_{500}$
relation derived in this work. The solid line is the median relation from the
model of Bullock et al. (2001) with parameters F=0.001 and K=2.8, while
the two dotted lines enclose the 1$\sigma$ region. We use:
$\sigma_{8}$=0.817, $\Omega$$_{\rm M}$=0.279, $n_{s}$=0.96 (tilt) from
Komatsu et al. (2008). Note that $\sigma_{8}$ affects the predicted $c_{500}$
significantly (see Buote et al. 2007). The dashed line is the best-fit from
G07. Our results show no significant mass dependence of $c_{500}$ in this
narrow mass range and are generally consistent with the prediction under the current
value of $\sigma_{8}$. The weighted mean of $c_{500}$ is 4.2.
}
\end{figure}

\begin{figure}
\centerline{\includegraphics[height=0.45\linewidth,angle=270]{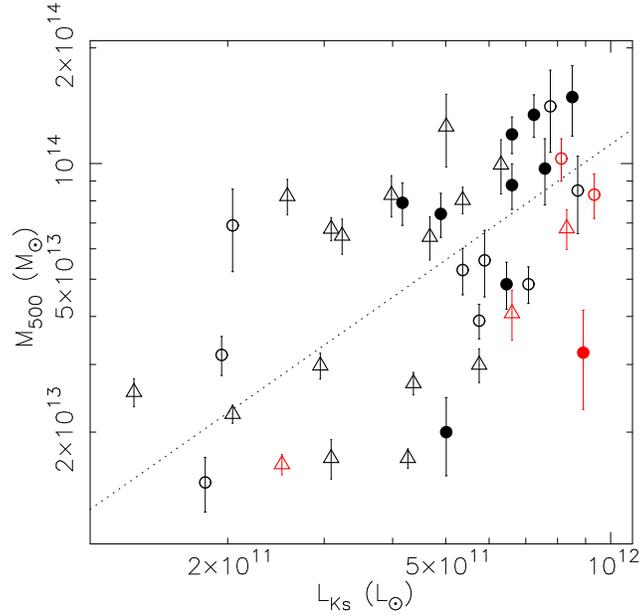}}
  \caption{2MASS $K_{\rm s}$ band luminosity of the cD galaxy vs. $M_{500}$.
The filled and open circles are for tier 1 and 2 groups respectively. The
open triangles are for tier 3 and 4 groups with mass estimated from the
$M_{500} - T_{500}$ relation derived in this work. The red points are fossil
groups identified in this work. There is a general trend that more massive groups
host more massive central galaxies. The dotted line represents a constant
$M_{500} / L_{\rm Ks, cD}$ line. For the same system mass, the fossil groups
host more luminous (or more massive) cDs than non-fossil groups.
}
\end{figure}

\begin{figure}
\centerline{\includegraphics[height=0.9\linewidth,angle=270]{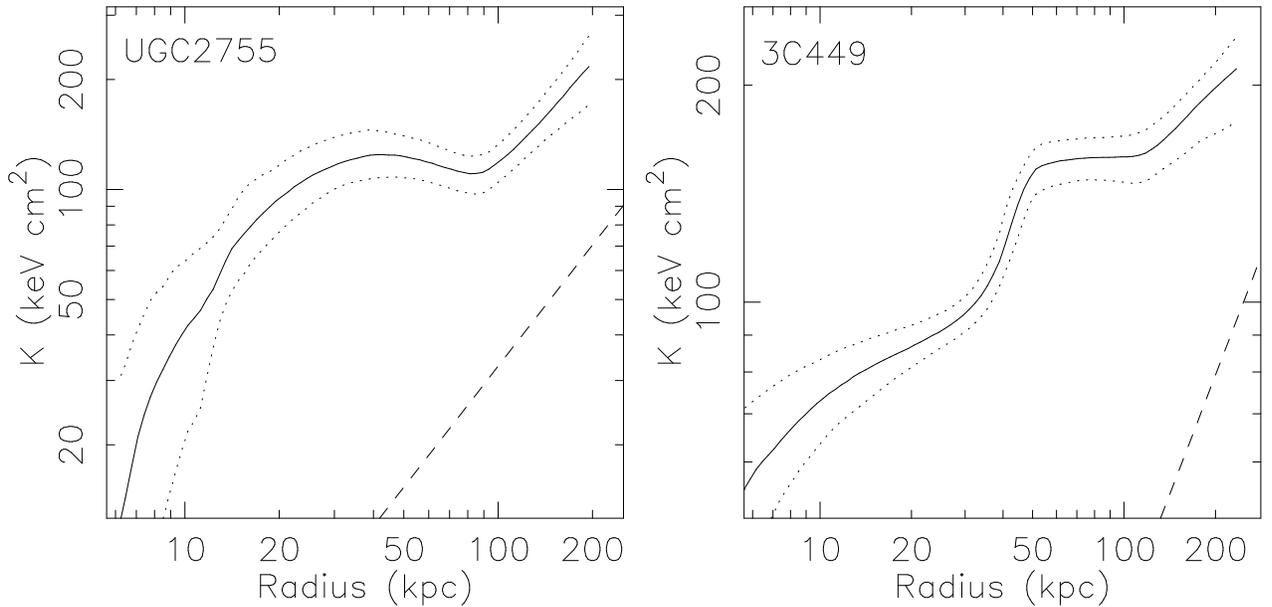}}
  \caption{Two groups with a significant entropy bump: UGC~2755 and 3C~449.
Both hosts an active FR I radio source (see $\S$8.1).
The dotted lines enclose the 1$\sigma$ error region, while the dashed lines
represent the baseline entropy profile (estimated from the group's mass)
with a slope of 1.1.
}
\end{figure}

\begin{figure}
\centerline{\includegraphics[height=0.45\linewidth,angle=270]{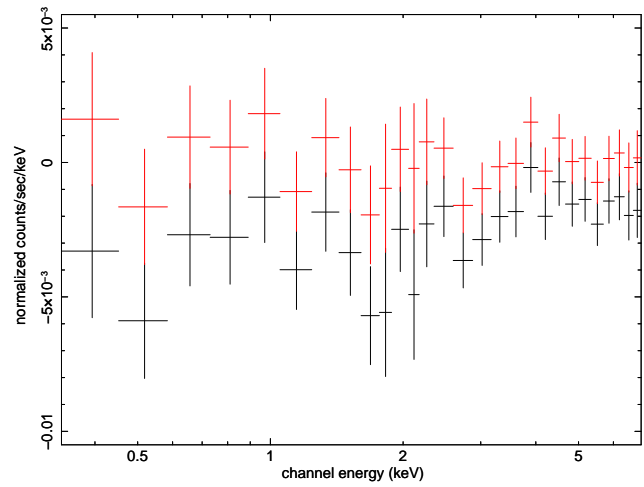}}
  \caption{The spectral residual after the period D stowed background subtracted by the
period E stowed background (in black). Their 9.5 - 12 keV fluxes are scaled to be the same.
The residual is flat and is only $\sim$ 6\%. The red data points are
the residual ($\sim$ 0\%) if the period D stowed background is scaled up by 5.7\%.
Clearly after this special scaling, the 9.5 - 12 keV fluxes of two background are not the same,
which implies a small spectral change for the BI PB from the period D to E.
}
\end{figure}

\end{document}